\pdfoutput=1
\documentclass[12pt,a4paper]{article}

\usepackage{ifthen} 
\newboolean{pdflatex}
\setboolean{pdflatex}{true} 

\newboolean{articletitles}
\setboolean{articletitles}{true} 

\newboolean{uprightparticles}
\setboolean{uprightparticles}{false} 

\def\paperauthors{BESIII and LHCb collaborations} 
\def\paperasciititle{Measurement of the CKM angle gamma in B->D(->KSh'h')h decays with a novel approach} 
\def\papertitle{Measurement of the CKM angle $\gamma$ in $\decay{\Bpm}{D(\to\KS h^{\prime+}h^{\prime-})h^{\pm}}$ decays with a novel approach} 
\def\paperkeywords{{High Energy Physics}, {BESII}, {LHCb}} 

\def\papercopyright{\the\year\ CERN for the benefit of the LHCb collaboration and \\ IHEP for the benefit of the BESIII collaboration} 
\def\paperlicence{CC BY 4.0 licence}
\def\paperlicenceurl{https://creativecommons.org/licenses/by/4.0/}

\newif\ifEnableSectionTOCLinks
\EnableSectionTOCLinksfalse 

\usepackage[top=1in, bottom=1.25in, left=1in, right=1in]{geometry}

\usepackage{microtype}
\usepackage{lineno}  
\usepackage{xspace} 
\usepackage{caption} 

\usepackage{graphicx}  
\usepackage{color}
\usepackage{colortbl}
\graphicspath{{./figs/}} 

\usepackage{amsmath} 
\usepackage{amssymb}
\usepackage{amsfonts}
\usepackage{upgreek} 

\newcommand*\patchAmsMathEnvironmentForLineno[1]{%
\expandafter\let\csname old#1\expandafter\endcsname\csname #1\endcsname
\expandafter\let\csname oldend#1\expandafter\endcsname\csname
end#1\endcsname
 \renewenvironment{#1}%
   {\linenomath\csname old#1\endcsname}%
   {\csname oldend#1\endcsname\endlinenomath}%
}
\newcommand*\patchBothAmsMathEnvironmentsForLineno[1]{%
  \patchAmsMathEnvironmentForLineno{#1}%
  \patchAmsMathEnvironmentForLineno{#1*}%
}
\AtBeginDocument{%
\patchBothAmsMathEnvironmentsForLineno{equation}%
\patchBothAmsMathEnvironmentsForLineno{align}%
\patchBothAmsMathEnvironmentsForLineno{flalign}%
\patchBothAmsMathEnvironmentsForLineno{alignat}%
\patchBothAmsMathEnvironmentsForLineno{gather}%
\patchBothAmsMathEnvironmentsForLineno{multline}%
\patchBothAmsMathEnvironmentsForLineno{eqnarray}%
}

\usepackage[pdftex,
            pdfauthor={\paperauthors},
            pdftitle={\paperasciititle},
            pdfkeywords={\paperkeywords}]{hyperref}
\usepackage{hyperxmp}
\hypersetup{
    pdfcopyright={Copyright (C) \papercopyright},
    pdflicenseurl={\paperlicenceurl}
}

\usepackage[colorinlistoftodos,textsize=scriptsize]{todonotes}

\usepackage[bottom,flushmargin,hang,multiple]{footmisc}

\usepackage[all]{hypcap} 

\usepackage{multirow}
\usepackage{csquotes}
\usepackage{float}

\usepackage{xspace}
\usepackage{upgreek}

\def\lhcb   {\mbox{LHCb}\xspace}

\def\babar  {\mbox{BaBar}\xspace}
\def\belle  {\mbox{Belle}\xspace}
\def\belletwo {\mbox{Belle~II}\xspace}
\def\besiii {\mbox{BESIII}\xspace}
\def\cleo   {\mbox{CLEO}\xspace}

\def\MagUp {\mbox{\em Mag\kern -0.05em Up}\xspace}

\ifthenelse{\boolean{uprightparticles}}%
{

 \def\Pnu         {\ensuremath{\upnu}\xspace}
 
 \def\Ppi         {\ensuremath{\uppi}\xspace}

 \def\Ppsi        {\ensuremath{\uppsi}\xspace}

 \def\PDelta      {\ensuremath{\Delta}\xspace}
 \def\PXi         {\ensuremath{\Xi}\xspace}
 \def\PLambda     {\ensuremath{\Lambda}\xspace}
 \def\PSigma      {\ensuremath{\Sigma}\xspace}
 \def\POmega      {\ensuremath{\Omega}\xspace}
 \def\PUpsilon    {\ensuremath{\Upsilon}\xspace}
 \let\oldPi\Pi
 \def\PPi         {\ensuremath{\oldPi}\xspace}

 \def\PB      {\ensuremath{\mathrm{B}}\xspace}
 \def\PD      {\ensuremath{\mathrm{D}}\xspace}

 \def\PK      {\ensuremath{\mathrm{K}}\xspace}
 \def\Pb      {\ensuremath{\mathrm{b}}\xspace}
 \def\Pc      {\ensuremath{\mathrm{c}}\xspace}
 \def\Pd      {\ensuremath{\mathrm{d}}\xspace}
 \def\Pe      {\ensuremath{\mathrm{e}}\xspace}
 \def\Ps      {\ensuremath{\mathrm{s}}\xspace}
 \def\Pt      {\ensuremath{\mathrm{t}}\xspace}
 \def\Pu      {\ensuremath{\mathrm{u}}\xspace}
 \def\thebaroffset{0.0em}
}
{

 \def\Pnu         {\ensuremath{\nu}\xspace}
 
 \def\Ppi         {\ensuremath{\pi}\xspace}

 \def\Ppsi        {\ensuremath{\psi}\xspace}
 
 \mathchardef\PDelta="7101
 \mathchardef\PXi="7104
 \mathchardef\PLambda="7103
 \mathchardef\PSigma="7106
 \mathchardef\POmega="710A
 \mathchardef\PUpsilon="7107
 \mathchardef\PPi="7105
 \def\PB      {\ensuremath{B}\xspace}
 \def\PD      {\ensuremath{D}\xspace}

 \def\PK      {\ensuremath{K}\xspace}
 \def\Pb      {\ensuremath{b}\xspace}
 \def\Pc      {\ensuremath{c}\xspace}
 \def\Pd      {\ensuremath{d}\xspace}
 \def\Pe      {\ensuremath{e}\xspace}
 \def\Ps      {\ensuremath{s}\xspace}
 \def\Pt      {\ensuremath{t}\xspace}
 \def\Pu      {\ensuremath{u}\xspace}
 \def\thebaroffset{0.18em}
}
\newcommand{\offsetoverline}[2][\thebaroffset]{\kern #1\overline{\kern -#1 #2}}%

\makeatletter
\ifcase \@ptsize \relax
  \newcommand{\miniscule}{\@setfontsize\miniscule{4}{5}}
\or
  \newcommand{\miniscule}{\@setfontsize\miniscule{5}{6}}
\or
  \newcommand{\miniscule}{\@setfontsize\miniscule{5}{6}}
\fi
\makeatother

\DeclareRobustCommand{\optbar}[1]{\shortstack{{\miniscule (\rule[.5ex]{1.25em}{.18mm})}
  \\ [-.7ex] $#1$}}

\def\epem       {{\ensuremath{\Pe^+\Pe^-}}\xspace}

\def\neu        {{\ensuremath{\Pnu}}\xspace}

\def\neue       {{\ensuremath{\neu_e}}\xspace}

\def\neum       {{\ensuremath{\neu_\mu}}\xspace}

\def\neul       {{\ensuremath{\neu_\ell}}\xspace}

\def\uquark    {{\ensuremath{\Pu}}\xspace}
\def\uquarkbar {{\ensuremath{\overline \uquark}}\xspace}

\def\dquark    {{\ensuremath{\Pd}}\xspace}
\def\dquarkbar {{\ensuremath{\overline \dquark}}\xspace}

\def\squark    {{\ensuremath{\Ps}}\xspace}
\def\squarkbar {{\ensuremath{\overline \squark}}\xspace}

\def\cquark    {{\ensuremath{\Pc}}\xspace}
\def\cquarkbar {{\ensuremath{\overline \cquark}}\xspace}

\def\bquark    {{\ensuremath{\Pb}}\xspace}

\def\tquark    {{\ensuremath{\Pt}}\xspace}

\def\pion   {{\ensuremath{\Ppi}}\xspace}
\def\piz    {{\ensuremath{\pion^0}}\xspace}
\def\pip    {{\ensuremath{\pion^+}}\xspace}
\def\pim    {{\ensuremath{\pion^-}}\xspace}
\def\pipm   {{\ensuremath{\pion^\pm}}\xspace}
\def\pimp   {{\ensuremath{\pion^\mp}}\xspace}

\def\kaon    {{\ensuremath{\PK}}\xspace}

\def\KorKbar {\kern \thebaroffset\optbar{\kern -\thebaroffset \PK}{}\xspace}

\def\Kp      {{\ensuremath{\kaon^+}}\xspace}
\def\Km      {{\ensuremath{\kaon^-}}\xspace}
\def\Kpm     {{\ensuremath{\kaon^\pm}}\xspace}
\def\Kmp     {{\ensuremath{\kaon^\mp}}\xspace}
\def\KS      {{\ensuremath{\kaon^0_{\mathrm{S}}}}\xspace}

\def\KL      {{\ensuremath{\kaon^0_{\mathrm{L}}}}\xspace}

\def\Kstar   {{\ensuremath{\kaon^*}}\xspace}

\def\Dbar    {{\ensuremath{\offsetoverline{\PD}}}\xspace}
\def\D       {{\ensuremath{\PD}}\xspace}
\def\Db      {{\ensuremath{\Dbar}}\xspace}
\def\DorDbar {\kern \thebaroffset\optbar{\kern -\thebaroffset \PD}\xspace}
\def\Dz      {{\ensuremath{\D^0}}\xspace}
\def\Dzb     {{\ensuremath{\Dbar{}^0}}\xspace}
\def\Dp      {{\ensuremath{\D^+}}\xspace}
\def\Dm      {{\ensuremath{\D^-}}\xspace}

\def\DpDm    {\ensuremath{\Dp {\kern -0.16em \Dm}}\xspace}
\def\Dstar   {{\ensuremath{\D^*}}\xspace}

\def\Dstarmp {{\ensuremath{\D^{*\mp}}}\xspace}

\def\B       {{\ensuremath{\PB}}\xspace}

\def\BorBbar {\kern \thebaroffset\optbar{\kern -\thebaroffset \PB}\xspace}
\def\Bz      {{\ensuremath{\B^0}}\xspace}

\def\Bd      {{\ensuremath{\B^0}}\xspace}

\def\BdorBdbar {\kern \thebaroffset\optbar{\kern -\thebaroffset \Bd}\xspace}
\def\Bu      {{\ensuremath{\B^+}}\xspace}
\def\Bub     {{\ensuremath{\B^-}}\xspace}
\def\Bp      {{\ensuremath{\Bu}}\xspace}
\def\Bm      {{\ensuremath{\Bub}}\xspace}
\def\Bpm     {{\ensuremath{\B^\pm}}\xspace}

\def\Bs      {{\ensuremath{\B^0_\squark}}\xspace}

\def\BsorBsbar {\kern \thebaroffset\optbar{\kern -\thebaroffset \Bs}\xspace}

\def\psiprpr  {{\ensuremath{\Ppsi(3770)}}\xspace}

\def\Y#1S{\ensuremath{\PUpsilon{(#1S)}}\xspace}

\def\Lz          {{\ensuremath{\PLambda}}\xspace}

\def\LorLbar     {\kern \thebaroffset\optbar{\kern -\thebaroffset \PLambda}\xspace}

\def\Lc          {{\ensuremath{\Lz^+_\cquark}}\xspace}

\def\Lb           {{\ensuremath{\Lz^0_\bquark}}\xspace}

\def\BF         {{\ensuremath{\mathcal{B}}}\xspace}

\newcommand{\decay}[2]{\mbox{\ensuremath{#1\!\to #2}}\xspace}

\def\to                 {\ensuremath{\rightarrow}\xspace}

\def\order   {{\ensuremath{\mathcal{O}}}\xspace}

\def\CP                {{\ensuremath{C\!P}}\xspace}

\def\Vud  {{\ensuremath{V_{\uquark\dquark}^{\phantom{\ast}}}}\xspace}
\def\Vcd  {{\ensuremath{V_{\cquark\dquark}^{\phantom{\ast}}}}\xspace}
\def\Vtd  {{\ensuremath{V_{\tquark\dquark}^{\phantom{\ast}}}}\xspace}

\def\Vubs  {{\ensuremath{V_{\uquark\bquark}^\ast}}\xspace}
\def\Vcbs  {{\ensuremath{V_{\cquark\bquark}^\ast}}\xspace}
\def\Vtbs  {{\ensuremath{V_{\tquark\bquark}^\ast}}\xspace}

\def\AT#1     {\ensuremath{A_{\mathrm{T}}^{#1}}\xspace}           

\def\C#1      {\ensuremath{\mathcal{C}_{#1}}\xspace}                       
\def\Cp#1     {\ensuremath{\mathcal{C}_{#1}^{'}}\xspace}                    
\def\Ceff#1   {\ensuremath{\mathcal{C}_{#1}^{\mathrm{(eff)}}}\xspace}        
\def\Cpeff#1  {\ensuremath{\mathcal{C}_{#1}^{'\mathrm{(eff)}}}\xspace}       
\def\Ope#1    {\ensuremath{\mathcal{O}_{#1}}\xspace}                       
\def\Opep#1   {\ensuremath{\mathcal{O}_{#1}^{'}}\xspace}                    

\newcommand{\nospaceunit}[1]{\ensuremath{\text{#1}}}
\newcommand{\aunit}[1]{\ensuremath{\text{\,#1}}}

\newcommand{\tev}{\aunit{Te\kern -0.1em V}\xspace}
\newcommand{\gev}{\aunit{Ge\kern -0.1em V}\xspace}
\newcommand{\mev}{\aunit{Me\kern -0.1em V}\xspace}
\newcommand{\kev}{\aunit{ke\kern -0.1em V}\xspace}
\newcommand{\ev}{\aunit{e\kern -0.1em V}\xspace}

\newcommand{\mevc}{\ensuremath{\aunit{Me\kern -0.1em V\!/}c}\xspace}
\newcommand{\gevc}{\ensuremath{\aunit{Ge\kern -0.1em V\!/}c}\xspace}
\newcommand{\mevcc}{\ensuremath{\aunit{Me\kern -0.1em V\!/}c^2}\xspace}
\newcommand{\gevcc}{\ensuremath{\aunit{Ge\kern -0.1em V\!/}c^2}\xspace}

\def\mum  {\ensuremath{\,\upmu\nospaceunit{m}}\xspace}

\def\pb {\aunit{pb}\xspace}

\def\fb   {\ensuremath{\aunit{fb}}\xspace}
\def\invfb   {\ensuremath{\fb^{-1}}\xspace}

\def\ps   {\ensuremath{\aunit{ps}}\xspace}

\newcommand{\stat}{\aunit{(stat)}\xspace}
\newcommand{\syst}{\aunit{(syst)}\xspace}

\def\order{{\ensuremath{\mathcal{O}}}\xspace}

\def\deriv {\ensuremath{\mathrm{d}}}

\def\gsim{{~\raise.15em\hbox{$>$}\kern-.85em
          \lower.35em\hbox{$\sim$}~}\xspace}
\def\lsim{{~\raise.15em\hbox{$<$}\kern-.85em
          \lower.35em\hbox{$\sim$}~}\xspace}

\newcommand{\mean}[1]{\ensuremath{\left\langle #1 \right\rangle}} 

\def\sPlot{\mbox{\em sPlot}\xspace}

\def\sqs   {\ensuremath{\protect\sqrt{s}}\xspace}

\def\pt         {\ensuremath{p_{\mathrm{T}}}\xspace}

\def\degrees{\ensuremath{^{\circ}}\xspace}

\def\evtgen     {\mbox{\textsc{EvtGen}}\xspace}

\def\geant      {\mbox{\textsc{Geant4}}\xspace}

\def\photos     {\mbox{\textsc{Photos}}\xspace}

\def\pythia     {\mbox{\textsc{Pythia}}\xspace}

\def\tell1  {TELL1\xspace}
\def\ukl1   {UKL1\xspace}

\newcommand{\eg}{\mbox{\itshape e.g.}\xspace}

\newcommand{\vs}{\mbox{\itshape vs.}\xspace}

\newcommand{\lhcborcid}[1]{\href{https://orcid.org/#1}{\hspace*{0.1em}\raisebox{-0.45ex}{\includegraphics[width=1em]{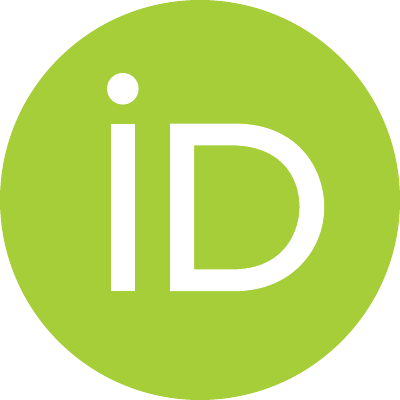}}}}

\def\rb       {\ensuremath{r_{B}}\xspace}

\def\db       {\ensuremath{\delta_{B}}\xspace}
\def\dd       {\ensuremath{\delta_{D}}\xspace}
\def\rdk      {\ensuremath{r_{B}^{DK}}\xspace}
\def\rdpi     {\ensuremath{r_{B}^{D\pi}}\xspace}
\def\rdh      {\ensuremath{r_{B}^{Dh}}\xspace}
\def\ddk      {\ensuremath{\delta_{B}^{DK}}\xspace}
\def\ddpi     {\ensuremath{\delta_{B}^{D\pi}}\xspace}
\def\ddh      {\ensuremath{\delta_{B}^{Dh}}\xspace}

\def\ampd     {\ensuremath{A_{D}}\xspace}
\def\ampdb    {\ensuremath{\overline{A}_{D}}\xspace}
\def\pd       {\ensuremath{p_{D}}\xspace}
\def\pdb      {\ensuremath{\overline{p}_{D}}\xspace}
\def\pb       {\ensuremath{p_{B}}\xspace}
\def\pbb      {\ensuremath{\overline{p}_{B}}\xspace}

\def\xpdk     {\ensuremath{x_{+}^{DK}}\xspace}
\def\xmdk     {\ensuremath{x_{-}^{DK}}\xspace}
\def\ypdk     {\ensuremath{y_{+}^{DK}}\xspace}
\def\ymdk     {\ensuremath{y_{-}^{DK}}\xspace}

\def\xpdh     {\ensuremath{x_{+}^{Dh}}\xspace}
\def\xmdh     {\ensuremath{x_{-}^{Dh}}\xspace}
\def\ypdh     {\ensuremath{y_{+}^{Dh}}\xspace}
\def\ymdh     {\ensuremath{y_{-}^{Dh}}\xspace}

\def\xpmdk    {\ensuremath{x_{\pm}^{DK}}\xspace}
\def\ypmdk    {\ensuremath{y_{\pm}^{DK}}\xspace}
\def\xpmdh    {\ensuremath{x_{\pm}^{Dh}}\xspace}
\def\ypmdh    {\ensuremath{y_{\pm}^{Dh}}\xspace}
\def\xpmdpi   {\ensuremath{x_{\pm}^{D\pi}}\xspace}
\def\ypmdpi   {\ensuremath{y_{\pm}^{D\pi}}\xspace}
\def\xxi      {\ensuremath{x_{\xi}^{D\pi}}\xspace}
\def\yxi      {\ensuremath{y_{\xi}^{D\pi}}\xspace}

\def\bz       {\ensuremath{\mathbf{z}}\xspace}
\def\bzb      {\ensuremath{\overline{\mathbf{z}}}\xspace}
\def\Cph      {\ensuremath{\mathcal{C}}\xspace}
\def\Sph      {\ensuremath{\mathcal{S}}\xspace}

\def\mkp      {\ensuremath{m^{2}_{+}}}
\def\mkm      {\ensuremath{m^{2}_{-}}}
\def\mkpm     {\ensuremath{m^{2}_{\pm}}}

\def\KSpp     {\ensuremath{\KS\pip\pim}}
\def\KLpp     {\ensuremath{\KL\pip\pim}}
\def\KSL      {\ensuremath{K^{0}_{\mathrm{S/L}}}}
\def\KSLpp    {\ensuremath{K^{0}_{\mathrm{S/L}}\pip\pim}}
\def\KSkk     {\ensuremath{\KS\Kp\Km}}
\def\KLkk     {\ensuremath{\KL\Kp\Km}}
\def\KSLkk    {\ensuremath{K^{0}_{\mathrm{S/L}}\Kp\Km}}
\def\KShh     {\ensuremath{\KS h^{\prime+}h^{\prime-}}}
\def\KLhh     {\ensuremath{\KL h^{\prime+}h^{\prime-}}}
\def\KSLhh    {\ensuremath{K^{0}_{\mathrm{S/L}}h^{\prime+}h^{\prime-}}}

\def\Nobs{\ensuremath{\mathcal{N}}}
\def\Nobsbar{\ensuremath{\smash{\overline{\mathcal{N}}}}}

\hypersetup{
  colorlinks   = true, 
  urlcolor     = blue, 
  linkcolor    = blue, 
  citecolor    = red   
}

\ifEnableSectionTOCLinks
    \usepackage[explicit]{titlesec} 
    
    \let\oldcontentsline\contentsline
    \renewcommand

    \titleformat{\section}{\normalfont\Large\bf}{\hyperlink{tocsection.\thesection}{{\thesection} \parbox[t]{\dimexpr\textwidth-1pc}{#1}}}{1pc}{}

    \titleformat{\subsection}{\normalfont\bf}{\hyperlink{tocsubsection.\thesubsection}{{\thesubsection} \parbox[t]{\dimexpr\textwidth-1pc}{#1}}}{1pc}{}

    \titleformat{name=\section,numberless}[display]{}{}{0pt}{\normalfont\Huge\bfseries #1}
\fi

\usepackage{cite} 
\usepackage{LHCb/mciteplus}

\newcommand{\BESIIIorcid}[1]{\href{https://orcid.org/#1}{\hspace*{0.1em}\raisebox{-0.45ex}{\includegraphics[width=1em]{LHCb/Figures/orcidIcon.pdf}}}}
\usepackage{booktabs, siunitx}
\begin{document}

\renewcommand{\thefootnote}{\fnsymbol{footnote}}
\setcounter{footnote}{1}


\begin{titlepage}
\pagenumbering{roman}

\vspace*{-1.5cm}
\centerline{\large EUROPEAN ORGANIZATION FOR NUCLEAR RESEARCH (CERN)}
\centerline{\large INSTITUTE OF HIGH ENERGY PHYSICS (IHEP)}
\vspace*{1.5cm}
\noindent
\begin{tabular*}{\linewidth}{lc@{\extracolsep{\fill}}r@{\extracolsep{0pt}}}
\ifthenelse{\boolean{pdflatex}}
{\vspace*{-1.5cm}\mbox{\!\!\!
\includegraphics[width=.09\textwidth,angle=90]{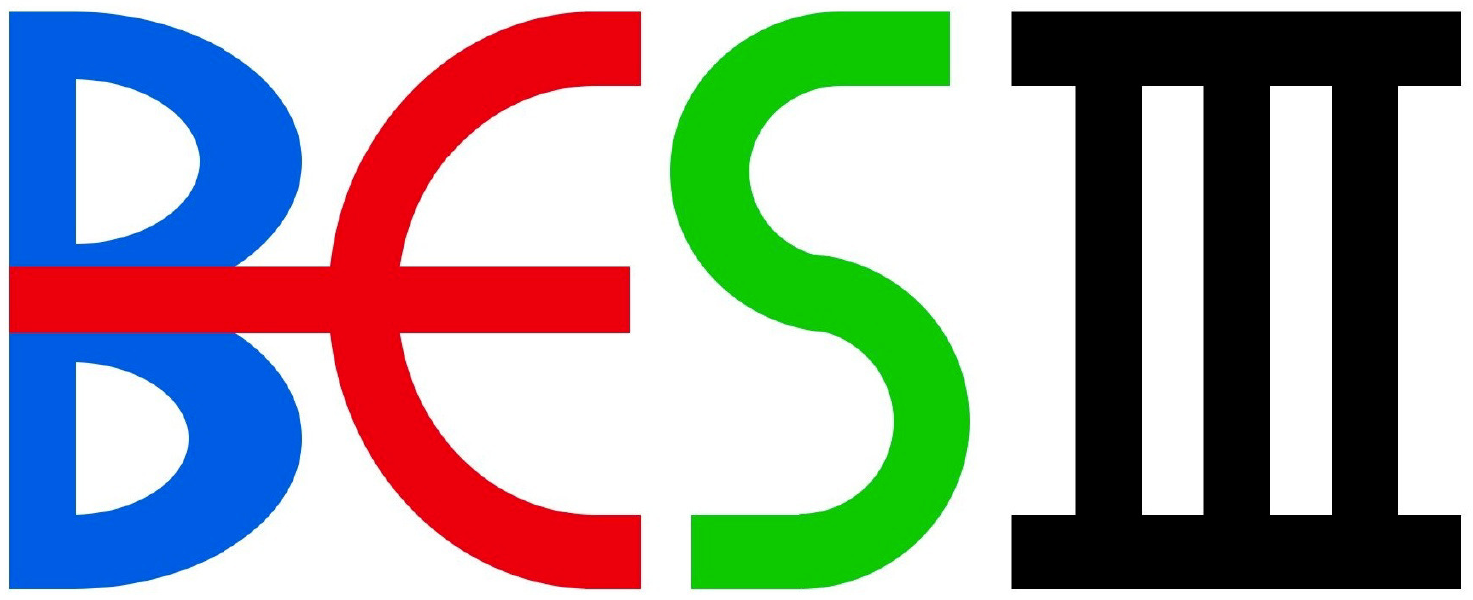}~
\includegraphics[width=.14\textwidth]{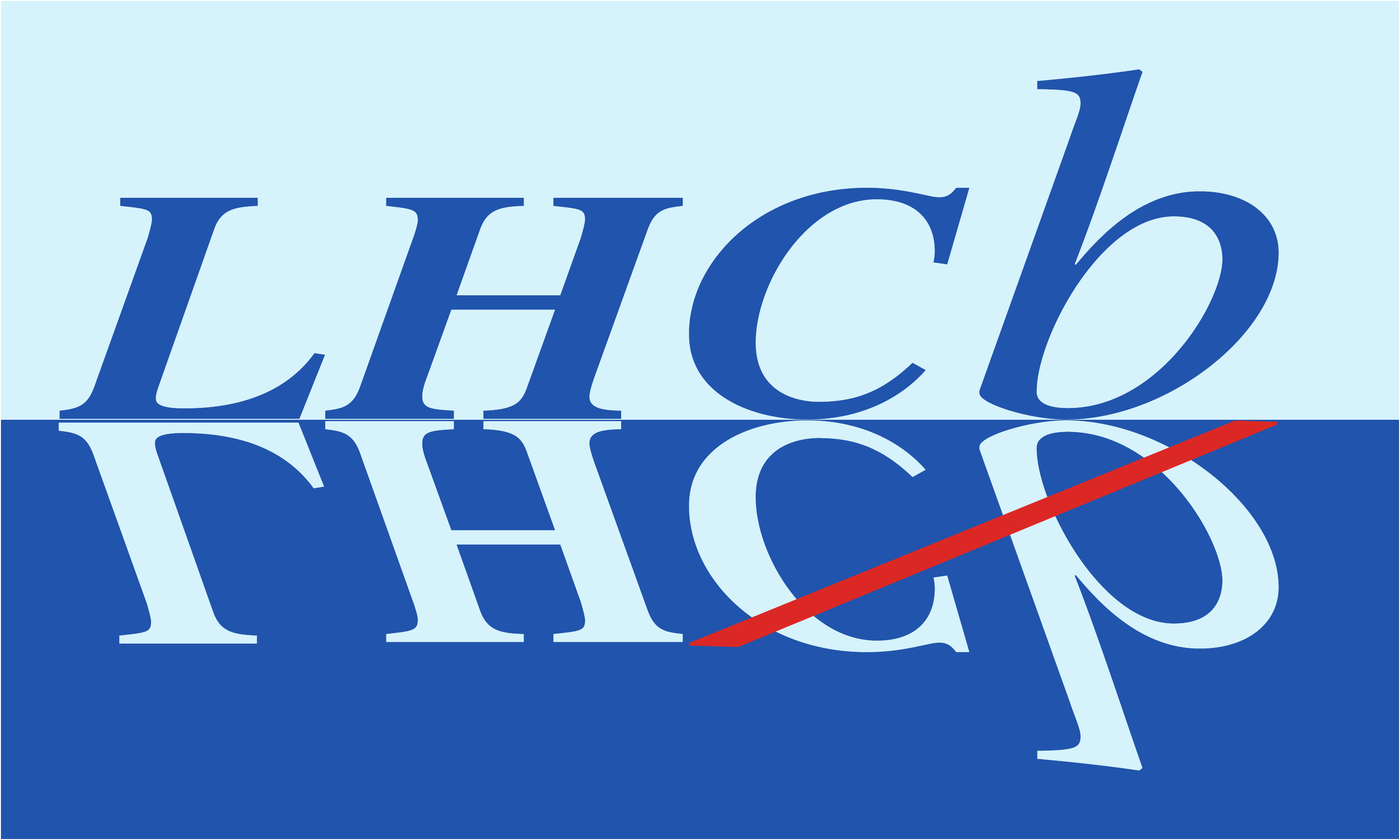}} & &}%
{\vspace*{-1.2cm}\mbox{\!\!\!
\includegraphics[width=.08\textwidth,angle=90]{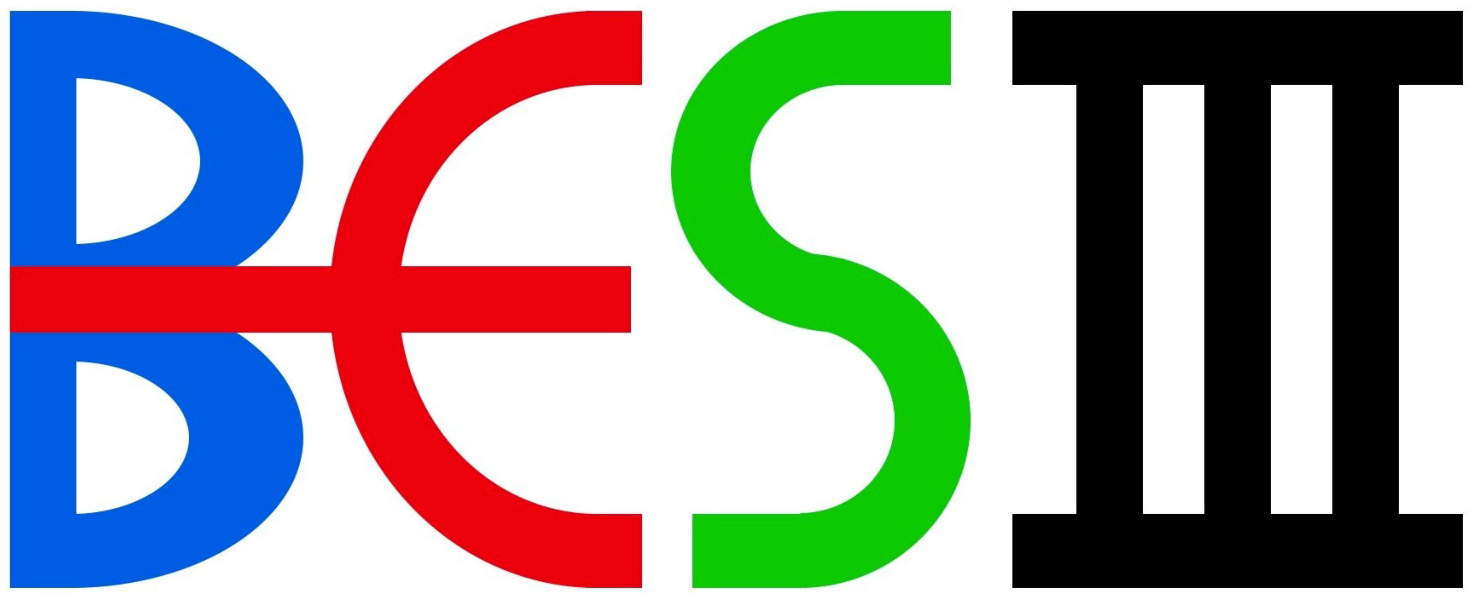}~
\includegraphics[width=.12\textwidth]{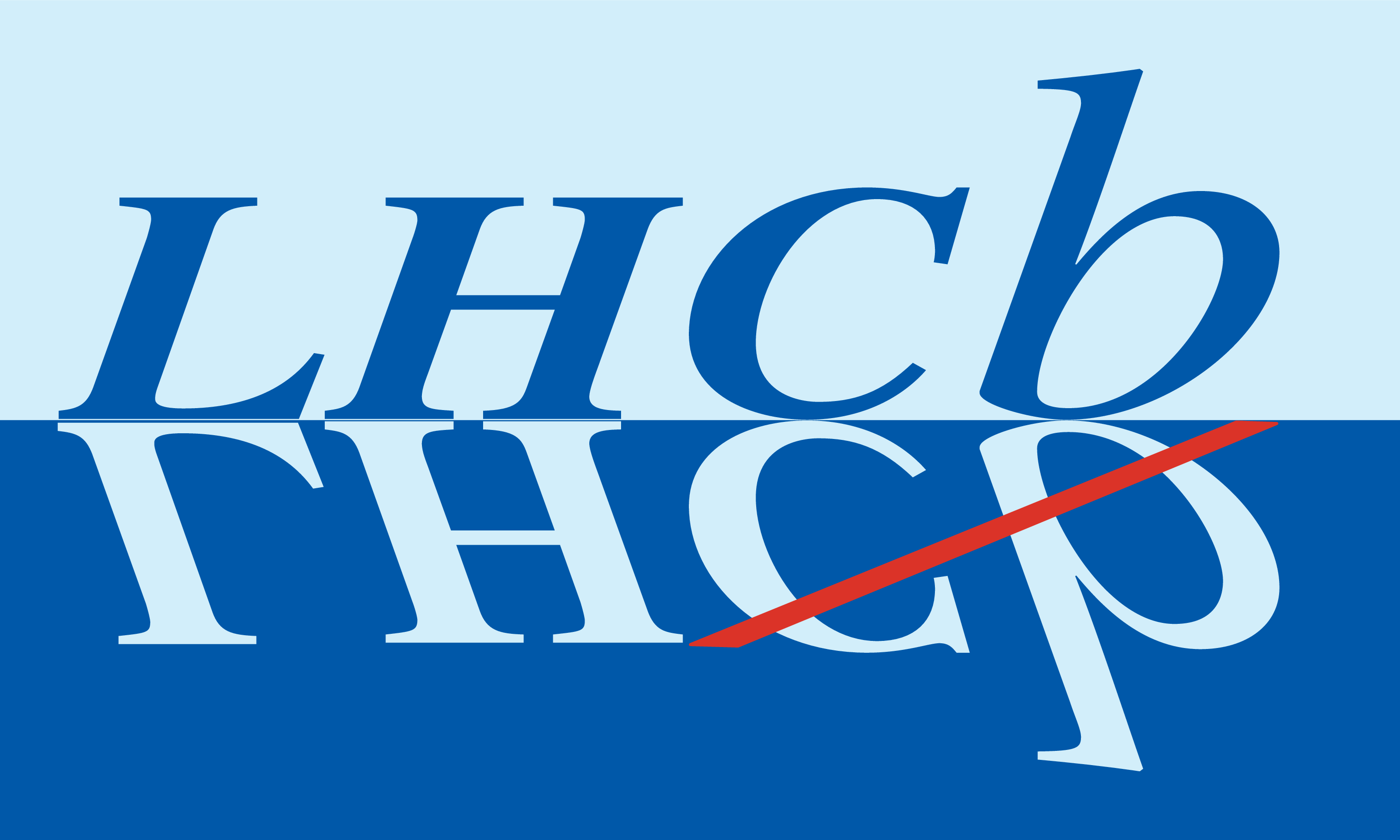}} & &}%
\\
 & & CERN-EP-2026-067 \\  
 & & LHCb-PAPER-2025-063 \\  
 & & April 7, 2026 \\
\end{tabular*}

\vspace*{2.0cm}

{\normalfont\bfseries\boldmath\huge
\begin{center}
  \papertitle 
\end{center}
}

\vspace*{1.5cm}

\begin{center}
\paperauthors\footnote{Authors are listed at the end of this paper.}
\end{center}

\vspace{\fill}

\begin{abstract}
  \noindent
    A measurement of the CKM angle $\gamma$ and related strong-phase parameters is performed using a novel, model-independent approach in $\decay{\Bpm}{D(\to\KS h^{\prime+}h^{\prime-}) h^{\pm}}$ decays, where $h^{(\prime)} \equiv \pi, K$.
  The analysis uses a joint data sample of electron-positron collisions collected by the \besiii experiment at the Beijing Electron-Positron Collider~II during 2010--2011 and 2021--2022, corresponding to an integrated luminosity of 8\invfb, and proton-proton collisions collected by the \lhcb experiment at the Large Hadron Collider during 2011--2018,
  corresponding to an integrated luminosity of 9\invfb.
  The two datasets are analyzed simultaneously by applying per-event weights based on the amplitude variation over the $D$-decay phase space to enhance the sensitivity to \CP-violating observables.
  The CKM angle $\gamma$ is determined to be $\gamma = (71.3\pm 5.0){\degrees}$, which constitutes the most precise single measurement to date.
  
\end{abstract}

\vspace*{2.0cm}

\begin{center}
  Submitted to Phys.~Rev.~D
\end{center}

\vspace{\fill}

{\footnotesize 
\begin{center}
\copyright~\papercopyright. \href{\paperlicenceurl}{\paperlicence}.
\end{center}
}
\vspace*{2mm}

\end{titlepage}


\newpage
\setcounter{page}{2}
\mbox{~}


\renewcommand{\thefootnote}{\arabic{footnote}}
\setcounter{footnote}{0}


\cleardoublepage


\pagestyle{plain} 
\setcounter{page}{1}
\pagenumbering{arabic}


\section{Introduction}
\label{sec:Introduction}
In the Standard Model of particle physics~(SM), the single irreducible complex phase in the Cabibbo--Kobayashi--Maskawa (CKM) quark-mixing matrix~\cite{Cabibbo:1963yz,*Kobayashi:1973fv} is responsible for all the \CP-violating phenomena in the quark sector.
Assuming unitarity of the CKM matrix, the relation ${\Vud\Vubs+\Vcd\Vcbs+\Vtd\Vtbs=0}$ ($V_{xy}$ represents the transition amplitude from quark $x$ to quark $y$) forms a closed triangle in the complex plane, known as the Unitarity Triangle (UT),
which has been extensively studied due to the high experimental accessibility of its sides and angles~\cite{CKMfitter2005,CKMfitter2015,UTfit-UT}. 
Among its three internal angles, ${\gamma=\phi_3\equiv\arg(-\Vud \Vubs/\Vcd \Vcbs)}$ can be measured experimentally by exploiting interference between $\decay{b}{c\uquarkbar s}$ and $\decay{b}{u\cquarkbar s}$ transitions at leading order within $\decay{B}{D^{(*)}K^{(*)}}$ decays.
The extraction of $\gamma$ in these decays has negligible theoretical uncertainty, as small as $\order(10^{-7})$~\cite{Brod:2013sga}.
The angle $\gamma$ can also be measured with charmless $B$-meson decays~\cite{Fleischer:1999pa,Fleischer:2007hj,Fleischer:2010ib,Fleischer:2022rkm,Rey-LeLorier:2011ltd,Bhattacharya:2014eca,Bhattacharya:2023pef}.
Comparison of the results measured with the two methods can provide insights into potential new-physics contributions entering  loop-mediated processes within charmless $b$-hadron decays.
In addition to direct measurements from $\decay{B}{D^{(*)}K^{(*)}}$ decays, 
indirect determinations of $\gamma$ are obtained from global fits to other CKM parameters under the assumption of unitarity~\cite{CKMfitter2005,CKMfitter2015,UTfit-UT}. 
Since some of these inputs arise from loop-level processes, discrepancies between direct and indirect determinations would indicate the existence of new physics.

The world-average values of $\gamma$ are obtained from global fits to various measurements that constrain the CKM matrix elements.
When direct measurements of $\gamma$ are excluded, the results are found to be ${\gamma = (66.29^{+0.72}_{-1.86}){\degrees}}$ from CKMfitter~\cite{CKMfitter2005,CKMfitter2015} and ${\gamma = (64.9\pm 1.4){\degrees}}$ from UTfit~\cite{UTfit-UT}.
A recent combination of direct measurements by the \lhcb collaboration yields ${\gamma=(62.8\pm2.6){\degrees}}$~\cite{LHCb-CONF-2025-003},
while a combination of \babar measurements yields $\gamma=(69^{+17}_{-16}){\degrees}$~\cite{BaBar:2013caj}, and a combination of \belle and \belletwo measurements yields $\gamma = (75.2\pm 7.6){\degrees}$~\cite{Belle:2024knt}.
These direct measurements are consistent with the indirect constraints but have larger uncertainties.
Hence, larger data samples and optimized analysis techniques are required to improve the precision on $\gamma$.

The most precise direct measurement of $\gamma$ to date is from the \lhcb experiment through the decays $\decay{\Bpm}{D(\to \KS h^{\prime+}h^{\prime-})h^{\pm}}$, where $h^{(\prime)}$ is either a $\pi$ or $K$ meson, and $D$ represents a superposition of the $\Dz$ and $\Dzb$ mesons~\cite{LHCb-PAPER-2020-019}.
The analysis employed a binned phase-space approach as proposed in Refs.~\cite{Giri:2003ty,BondarProc,Bondar:2008hh}, benefiting from the local asymmetries across the phase space.
The reported result of ${\gamma=(68.7^{+5.2}_{-5.1}){\degrees}}$ is statistically limited.
Since the binned phase-space approach only exploits about 85\% of the sensitivity to $\gamma$~\cite{CLEO:2010iul}, further optimization of the analysis method is desirable. 
A novel approach that applies per-event weights to data has been proposed as an alternative~\cite{Poluektov:2017zxp}.
This method for the determination of $\gamma$ has been found to use the phase-space information more effectively than the binned approach.

The strong-phase differences between the decay amplitudes of $\Dz$ and $\Dzb$ to the ${\KS h^{\prime+}h^{\prime-}}$ final states are essential inputs for determining $\gamma$.
However, uncertainties due to model assumptions of the decay amplitudes are difficult to quantify. 
In practice, the strong-phase differences are measured directly from quantum-correlated charm decays typically produced at the $\psi(3770)$ resonance. 
The strong-phase differences for the binned approach have been measured by the \cleo~\cite{CLEO:2010iul} and \besiii experiments~\cite{BESIII:2020hlg,BESIII:2020khq,BESIII:2020hpo,BESIII:2025nsp}, and are statistically limited. 
These uncertainties introduce a non-negligible systematic uncertainty to $\gamma$.
However, these inputs can only be used by the binned approach.

This paper, which expands upon the accompanying Letter~\cite{LHCb-PAPER-2025-064}, presents a more precise determination of the CKM angle $\gamma$ by applying the novel approach to the same \lhcb dataset used in Ref.~\cite{LHCb-PAPER-2020-019}. 
The required strong-phase parameters for this method are obtained from a joint measurement using quantum-correlated ${D\Db}$ decays produced at the \besiii experiment.
The \besiii data were collected during 2010--2011 and 2021--2022 in $\epem$ collisions at the $\psiprpr$ threshold, with an integrated luminosity of 8\invfb, where the signal $D$ mesons are reconstructed in the $\KSL h^{\prime+}h^{\prime-}$ final states.
The \lhcb data were collected during 2011--2018 in $pp$ collisions at center-of-mass energies of ${\sqs=7,8}$ and 13\tev, corresponding to an integrated luminosity of 9\invfb.
The decay channels are the same as those in Ref.~\cite{LHCb-PAPER-2020-019}, namely $\decay{\Bpm}{D\Kpm}$ and $\decay{\Bpm}{D\pipm}$, 
where the $D$ meson is reconstructed via the self-conjugate decays $\decay{D}{\KShh}$. 
In this analysis, both the \besiii and \lhcb datasets have also been analyzed using the binned phase-space approach~\cite{BESIII:2025nsp,LHCb-PAPER-2020-019} except for the $\decay{D}{\KSL\Kp\Km}$ signal in \besiii data.

The outline of this paper is as follows: the formalism of the novel approach is explained in Sect.~\ref{sec:analysis_strategy};
the \besiii and \lhcb detectors are introduced in Sect.~\ref{sec:bes_detector};
selection of candidates from data and extraction of signal yields are described in Sect.~\ref{sec:selection}, 
while Sect.~\ref{sec:coefficient} presents the method to obtain key observables from data;
the \CP observables and strong-phase parameters are obtained using the fits described in Sect.~\ref{sec:CPfit};
the corresponding systematic uncertainties are analyzed in Sect.~\ref{sec:systematic};
Section~\ref{sec:result} presents the results of the measurements of \CP observables and strong-phase parameters, together with the interpretation of \CP observables determining the CKM angle $\gamma$; and
finally, Sect.~\ref{sec:conclusions} summarizes the measurement.

\section{Analysis strategy}
\label{sec:analysis_strategy}

The amplitude of the $\decay{\Bm}{Dh^-}$ decay is a superposition of the favored $\decay{\Bm}{\Dz h^-}$ and suppressed $\decay{\Bm}{\Dzb h^-}$ decays,
\begin{equation}
    A_B \propto A_D + \rdh e^{i(\ddh-\gamma)}\overline{A}_D,\label{eq:Bmamp}
\end{equation}
where $A_D$ ($\overline{A}_D$) is the amplitude of the $\decay{\Dz(\Dzb)}{\KS h^{\prime+}h^{\prime-}}$ decay.
The amplitude of the $\decay{\Bp}{Dh^+}$ decay is obtained by substituting $-\gamma$ with $+\gamma$ and exchanging $A_D$ and $\overline{A}_D$.
The hadronic parameters $\rdh$ and $\ddh$ are the ratio and strong-phase difference between the suppressed and favored $\Bm$ decay amplitudes, respectively.
As the decay phase space has two degrees of freedom, often referred to as the Dalitz plot~\cite{Dalitz:1953cp,Fabri:1954zz}, the amplitudes are expressed as functions of the squared $\KS h^{\prime\pm}$ masses,~$\mkpm$.
Neglecting \CP violation and mixing in $D$ decays, as their impact on the phase-space distributions is expected to be well below current experimental sensitivity~\cite{Bondar:2010qs}, the amplitudes are defined as ${A_D\equiv A_D(\bz)}$ and ${\overline{A}_D\equiv\overline{A}_D(\bz) = A_D(\bzb)}$ with $\bz\equiv(\mkp, \mkm)$ and $\bzb\equiv(\mkm, \mkp)$.

The rates of the \Bm and \Bp decays, $\pb$ and $\pbb$, can be derived as
\begin{equation}
\begin{aligned}
    \pb(\bz)  &\propto \pd(\bz) + \left[(\xmdh)^2 + (\ymdh)^2\right]\pdb(\bz) + 2\left[\xmdh\Cph(\bz) + \ymdh\Sph(\bz)\right], \\
    \pbb(\bz) &\propto \pdb(\bz) + \left[(\xpdh)^2 + (\ypdh)^2\right]\pd(\bz) + 2\left[\xpdh\Cph(\bz) - \ypdh\Sph(\bz)\right],
    \end{aligned}
    \label{eq:Bpmrate}
\end{equation}
where $\optbar{p}_{\hspace*{-0.35em}D} \equiv \big|\optbar{A}_{\hspace*{-0.25em}D}\big|^2$ is the decay rate of the $\decay{\Dz(\Dzb)}{\KS h^{\prime+}h^{\prime-}}$ process.
The \CP observables, $\xpmdh$ and $\ypmdh$, are expressed as 
\begin{equation}
    \xpmdh = \rdh\cos(\ddh\pm\gamma), \quad \ypmdh=\rdh\sin(\ddh\pm\gamma).
\end{equation}
The functions $\Cph$ and $\Sph$ are related to the strong-phase difference between $\Dz$ and $\Dzb$ decays,
\begin{equation}
\begin{aligned}
    \Cph(\bz) &\equiv |\ampd(\bz)||\ampd(\bzb)|\cos\left[\delta_D(\bz)-\delta_D(\bzb)\right], \\
    \Sph(\bz) &\equiv |\ampd(\bz)||\ampd(\bzb)|\sin\left[\delta_D(\bz)-\delta_D(\bzb)\right],
\end{aligned}
\end{equation}
where $\delta_D$ is the strong phase of the $\decay{\Dz}{\KS h^{\prime+}h^{\prime-}}$ decay amplitude.

The \CP observables $\xpmdh$ and $\ypmdh$ are independent of the phase-space coordinates. Therefore, a phase-space-dependent weighting function can be applied to Eq.~\ref{eq:Bpmrate}, while $\xpmdh$ and $\ypmdh$ remain unchanged, and the weighted integrals of the decay rates can then be evaluated experimentally.
The binned phase-space approach~\cite{Giri:2003ty,BondarProc,Bondar:2008hh} is a special case in that regard, using a uniform weighting function within different regions of the Dalitz plot.

\begin{figure}[tb]
    \centering
    \includegraphics[width=0.48\linewidth]{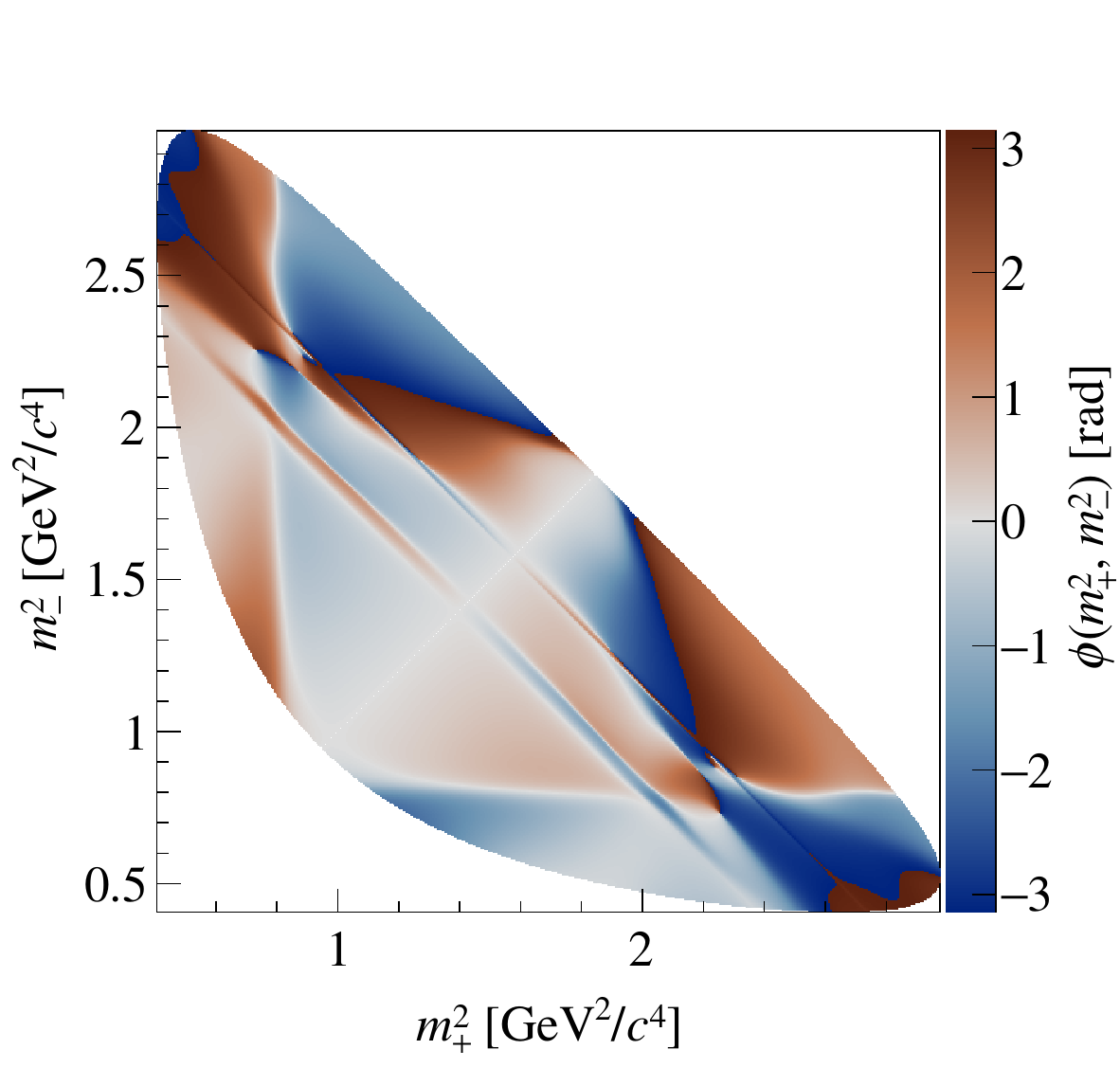}
    \includegraphics[width=0.48\linewidth]{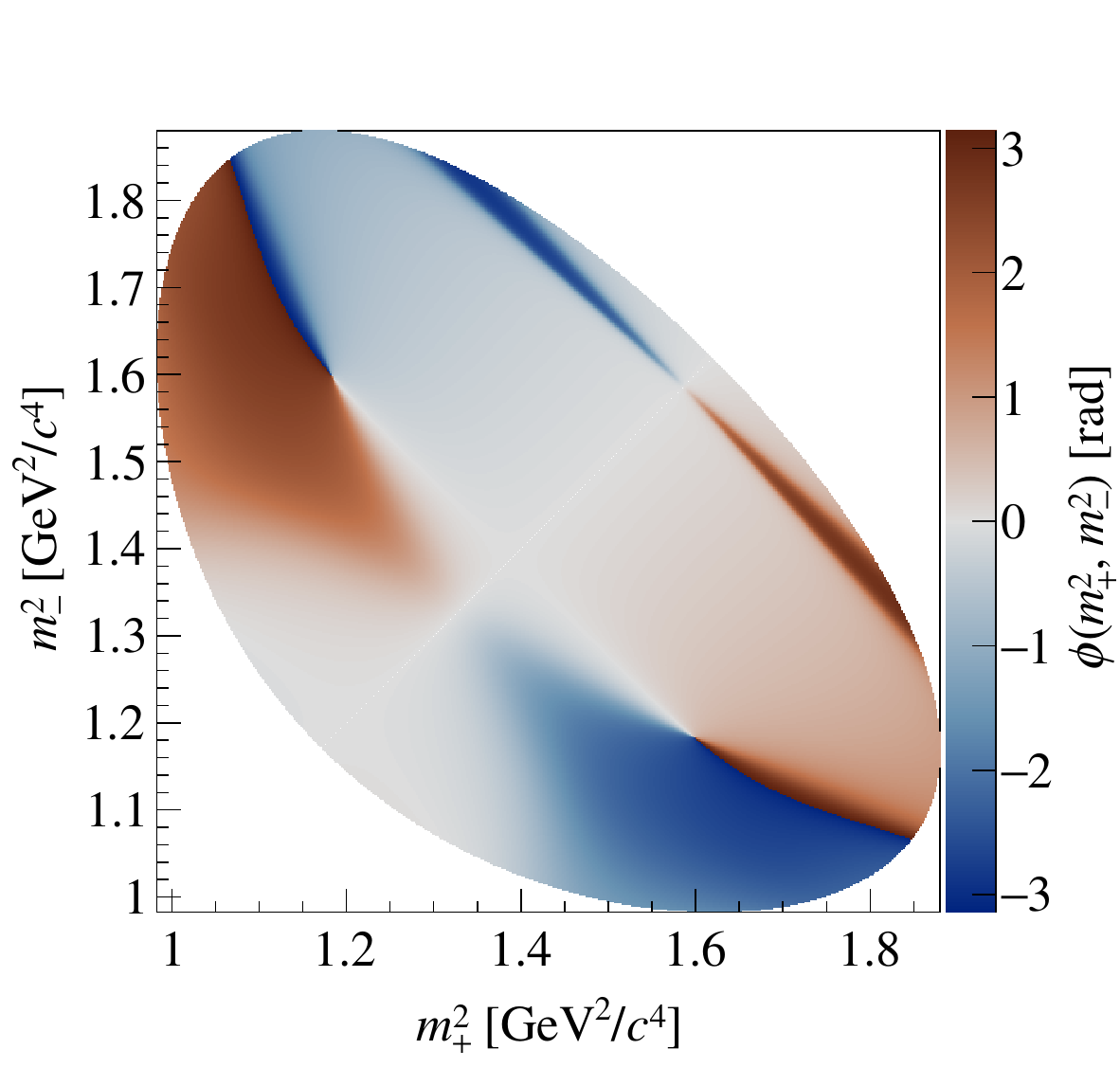}
    \caption{Distributions of the strong-phase difference $\phi(\mkp,\mkm)$ across the (left) $\decay{D}{\KS\pip\pim}$ and (right) $\decay{D}{\KS\Kp\Km}$ Dalitz plots. They are derived from amplitude models measured by the  \belle and \babar~\cite{BaBar:2018agf,BaBar:2018cka,BaBar:2010nhz} experiments.}
    \label{fig:phi}
\end{figure}

A novel approach, referred to as the \textit{Fourier split} method, has been proposed in Ref.~\cite{Poluektov:2017zxp}. 
It applies Fourier expansions of the strong-phase difference ${\phi(\bz)\equiv\dd(\bz)-\dd(\bzb)}$ to the functions in Eq.~\ref{eq:Bpmrate}, which is equivalent to applying per-event Fourier weights $\cos(k\phi)$ and $\sin(k\phi)$.
The distributions of $\phi(\bz)$, shown in Fig.~\ref{fig:phi}, are derived from amplitude models measured by the \belle and \babar experiments~\cite{BaBar:2018agf,BaBar:2018cka,BaBar:2010nhz}.
As in the case of the binned approach, the choice of model affects the sensitivity to $\gamma$ but does not bias the results.
This method achieves a lower statistical uncertainty on $\gamma$ than the binned approach by making use of the included intra-bin variation of the strong-phase difference across the phase space.

\begin{figure}[tb]
    \centering
    \includegraphics[width=0.49\linewidth]{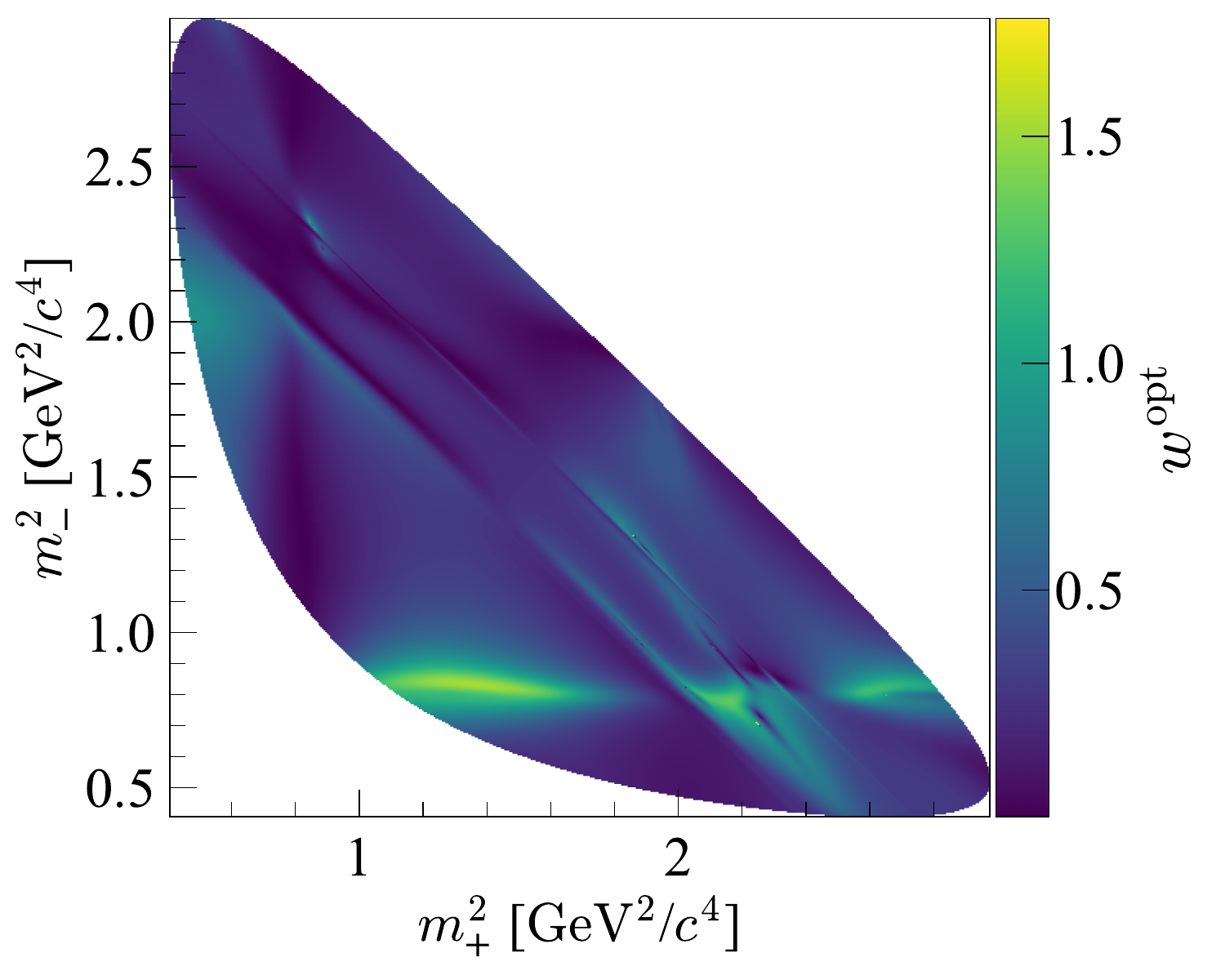}
    \includegraphics[width=0.49\linewidth]{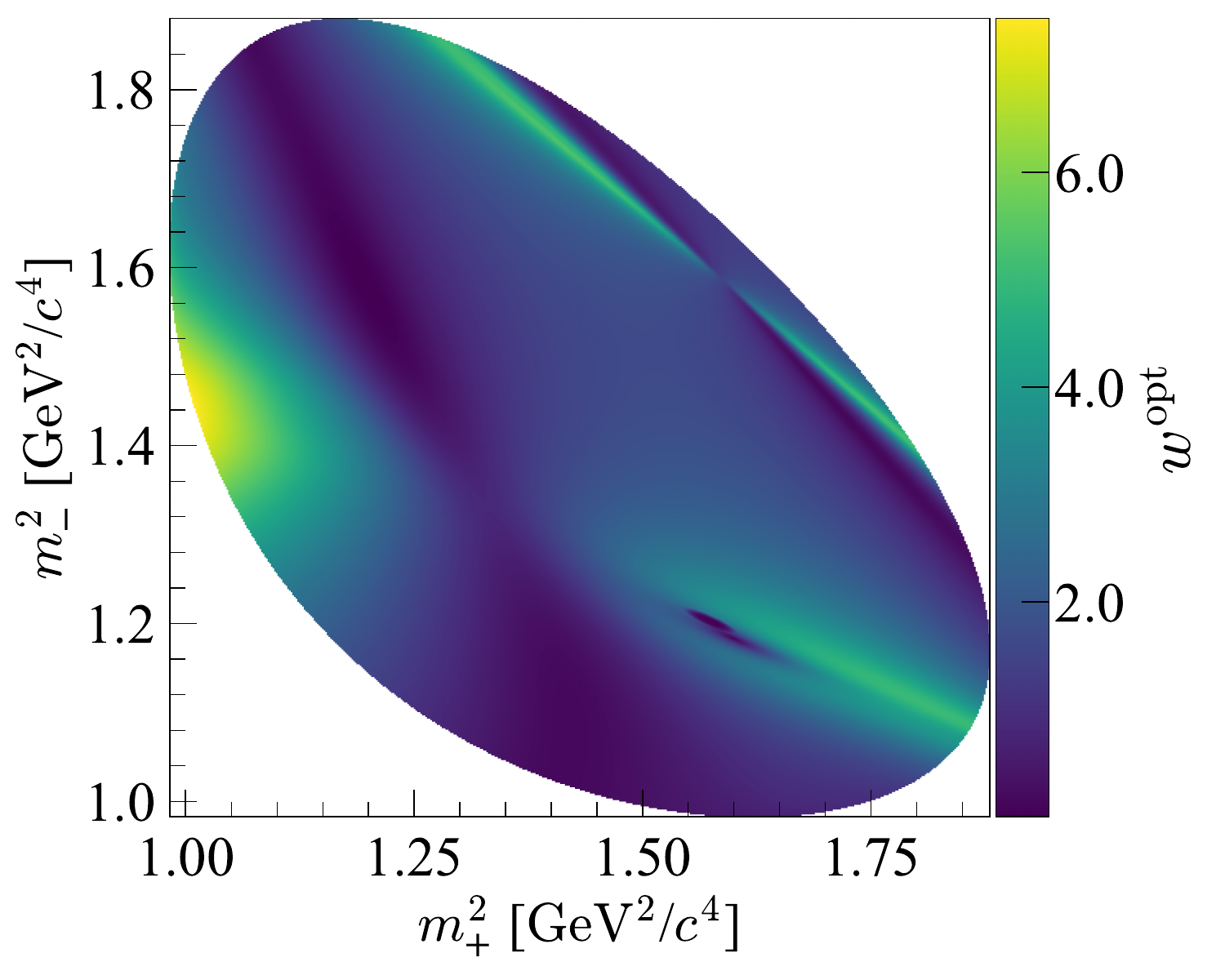}
    \caption{Distributions of the optimal weight $w^{\rm opt}$ across the (left) $\decay{D}{\KS\pip\pim}$ and (right) $\decay{D}{\KS\Kp\Km}$ Dalitz plots. They are obtained from amplitude models measured by the \belle and \babar~\cite{BaBar:2018agf,BaBar:2018cka,BaBar:2010nhz} experiments, and also take into account the effects of efficiency and background in the \lhcb measurement~\cite{LHCb-PAPER-2020-019}. The integrals of the weighting functions are normalized to unity.}
    \label{fig:opt_weights}
\end{figure}

The sensitivity to $\gamma$ also depends on the decay rate $p_D$ and the background level in data, which were not considered in Ref.~\cite{Poluektov:2017zxp}.
To improve the sensitivity to $\gamma$, an additional weighting function is introduced, which is denoted as the \emph{optimal weight}, $w^{\rm opt}$.
The optimal weights, as shown in Fig.~\ref{fig:opt_weights}, are obtained using amplitude models for the $D$ decay, which originate from the \belle and \babar measurements~\cite{BaBar:2018agf,BaBar:2018cka,BaBar:2010nhz}.
The weights also take into account the effects of efficiency and background in the \lhcb measurement~\cite{LHCb-PAPER-2020-019}.
The structure of these weights across the Dalitz plot reflects the statistical uncertainty of signal events, showing enhancement and suppression around the $\Kstar(892)^{\pm}$ ($\phi(1020)$) resonance in the $\decay{D}{\KS\pip\pim~(\KS\Kp\Km)}$ decay.
The values of these optimal weights across the Dalitz plots are provided in a \textsc{HEPData} record~\cite{hepdata}.
Analogous to the symmetric splitting of the phase space by $\mkp >(<) \mkm$ in the binned approach,
the diametrically opposed optimal weight, ${w^{\rm opt}(\bzb)}$, is also applied to improve the sensitivity.

In summary, the complete set of weighting functions combines the optimal weights $w^{\rm opt}$ with the Fourier weights proposed in Ref.~\cite{Poluektov:2017zxp},
\begin{equation}
\begin{aligned}
    w_n(\bz) &= \left\{
    \begin{array}{ll}
        w^{\rm opt}(\bz)\cos\left[k\phi(\bz)\right], & n = 2k, \\
        w^{\rm opt}(\bz)\sin\left[k\phi(\bz)\right], & n = 2k-1.
    \end{array}
    \right.
    \\
    \overline{w}_n(\bz) &= \left\{
    \begin{array}{ll}
        w^{\rm opt}(\bzb)\cos\left[k\phi(\bz)\right], & n = 2k, \\
        w^{\rm opt}(\bzb)\sin\left[k\phi(\bz)\right], & n = 2k-1.
    \end{array}
    \right.
\end{aligned}
\label{eq:wp_weights} 
\end{equation}
With a maximum Fourier order of $M$, $k$ enumerates from 0 (1) to $M$ for the cosine (sine) weights, 
and $2\times (2M+1)$ weighting functions are defined.
This novel approach is denoted as the \textit{optimal Fourier} method.

The weighted integrals of the $\Bpm$ decay rates yield $4\times(2M+1)$ parameters,
\begin{equation}
\begin{aligned}
    \Nobs_{n}^{+}    &\equiv h_{\Bp}[P_n^- + \left((\xpdh)^2 + (\ypdh)^2\right)P_n^+ + 2(-1)^n(\xpdh C_n - \ypdh S_n)],  \\
    \Nobsbar_{n}^{+} &\equiv h_{\Bp}[P_n^+ + \left((\xpdh)^2 + (\ypdh)^2\right)P_n^- + 2(-1)^n(\xpdh C_n + \ypdh S_n)],  \\
    \Nobs_{n}^{-}    &\equiv h_{\Bm}[P_n^+ + \left((\xmdh)^2 + (\ymdh)^2\right)P_n^- + 2(-1)^n(\xmdh C_n + \ymdh S_n)],  \\
    \Nobsbar_{n}^{-} &\equiv h_{\Bm}[P_n^- + \left((\xmdh)^2 + (\ymdh)^2\right)P_n^+ + 2(-1)^n(\xmdh C_n - \ymdh S_n)], \label{eq:Bfou} 
\end{aligned}
\end{equation}
where separate normalization factors $h_{\Bpm}$ for $\Bpm$ decays are applied, thereby accounting for any production and detection asymmetry in the $\Bpm$ decays.
The parameters $P_n^+$, $P_n^-$, $C_n$ and $S_n$ are $w_n$-weighted integrals of the $\pd$, $\pdb$, $\Cph$ and $\Sph$ functions, respectively. 
In addition to $h_{\Bpm}$ that are proportional to the signal yields,
a normalization over the amplitude is applied by requiring $P_0^+ + P_0^- = 1$. 
The weighted parameters are defined as
\begin{equation}
    X_n^{+\{-\}} \equiv \frac{\int f_X(\bz) w_n\{\overline{w}_n\}(\bz)\deriv\bz}{\int \left[\pd(\bz) + \pdb(\bz)\right]w^{\rm opt}(\bz)\deriv\bz},
\end{equation}
where $f_X$ denotes the functions $\pd$, $\Cph$ and $\Sph$ for $X=P$, $C$ and $S$, respectively.
Implicit definitions
\begin{equation}
    C_n \equiv C_n^+ = (-1)^n C_n^-, \quad S_n \equiv S_n^+ = (-1)^{n+1}S_n^-,
\end{equation}
are applied, because the $\Cph$ function and cosine weights are symmetric, while the $\Sph$ function and sine weights are antisymmetric across the Dalitz plot.
However, due to the asymmetric nature of $w^{\rm opt}$, weighted $\Cph$ and $\Sph$ functions are no longer symmetric or antisymmetric, respectively, which yields nonzero $C_n$ and $S_n$ terms at every sine and cosine order. This relationship is identical to that in the binned method.

By including both $\decay{\Bpm}{D\Kpm}$ and $\decay{\Bpm}{D\pipm}$ decays, the above equations contain $8\times(2M+1)$ observables when the Fourier order $k$ enumerates from 0 to $M$.
The number of free parameters is $2\times(2M+1) + 12$ when the strong-phase parameters $C_n, S_n$ are constrained by the \besiii dataset.
As a result, under the condition $M \geq 1$, these observables are sufficient to determine all the parameters.

As the $\decay{\Bpm}{D\Kpm}$ and $\decay{\Bpm}{D\pipm}$ processes have the same weak phase difference $\gamma$,
only two additional observables are needed to describe the $\decay{\Bpm}{D\pipm}$ decay. They are defined as
\begin{equation}
    \xxi \equiv \frac{\rdpi}{\rdk}\cos(\ddpi - \ddk), \quad \yxi = \frac{\rdpi}{\rdk}\sin(\ddpi - \ddk),
\end{equation}
and the \CP observables of the $\decay{\Bpm}{D\pipm}$ channel can be determined using
\begin{equation}
     \xpmdpi \equiv \xxi\xpmdk - \yxi\ypmdk,\quad \ypmdpi = \xxi\ypmdk + \yxi\xpmdk.
\end{equation}
This reduces the number of free parameters to $2\times(2M+1) + 10$.

In the quantum-correlated $\decay{\psiprpr}{\Dz\Dzb}$ system, the decay amplitude can be written as
\begin{equation}
    A = \frac{1}{\sqrt{2}}\left(\ampd^{(1)}\ampdb^{(2)} - \ampdb^{(1)}\ampd^{(2)}\right),
\end{equation}
where the superscripts (1) and (2) uniquely label the two $D$ mesons,
and $\ampd^{(i)}$ ($\ampdb^{(i)}$) is the decay amplitude of the corresponding $\Dz$ ($\Dzb$) decay.
The strong-phase parameters, $C_n, S_n$, are measured in this system by a double-tag (DT) method~\cite{PhysRevLett.56.2140},
in which one of the $D$ mesons is reconstructed as the signal decay,
and the other is reconstructed as a tag decay.
In addition to $\decay{D}{\KS h^{\prime+} h^{\prime-}}$ decays, $\decay{D}{\KL h^{\prime+} h^{\prime-}}$ decays are also included as signal
to provide further constraints on the strong-phase parameters.
The tag mode can be either a (quasi-)flavor-specific state $f$ like $\Kmp\pipm$, a \CP eigenstate like $\Kp\Km$, or another signal decay.
In addition, the single-tag (ST) method, where only one of the $D$ mesons is reconstructed as the tag decay, is adopted to determine ST yields for normalization as detailed in Sect.~\ref{sec:CPfit}.

Applying the weighting functions to flavor-tagged decay rates yields $2\times (2M+1)$ parameters,
\begin{equation}
    \begin{aligned}
    \Nobs_n   (\KSLhh | f) \propto [P_n^+ + (r_D^f)^2P_n^- - 2R_D^fr_D^fs_K(-1)^n(C_n\cos\delta_D^f - S_n\sin\delta_D^f)], \\
    \Nobsbar_n(\KSLhh | f) \propto [P_n^- + (r_D^f)^2P_n^+ - 2R_D^fr_D^fs_K(-1)^n(C_n\cos\delta_D^f + S_n\sin\delta_D^f)], 
    \end{aligned}
    \label{eq:flav_coef} 
\end{equation}
where $r_D$ and $\delta_D$ are the amplitude ratio and strong-phase difference, respectively, between the doubly-Cabibbo-suppressed (DCS) and Cabibbo-favored (CF) processes,
and $R_D$ is the coherence factor of the tag mode.
Depending on the signal decay, $s_K$ is $1$ for $\decay{D}{\KS h^{\prime+}h^{\prime-}}$ and $-1$ for $\decay{D}{\KL h^{\prime+} h^{\prime-}}$, due to the different \CP content of the neutral kaons.
The parameters $\Nobs_n$ ($\Nobsbar_n$) refer to the case where the tag mode is a DCS (CF) decay like $\decay{\Dz}{\Kp\pim} (\decay{\Dz}{\Km\pip})$.

The $2\times(2M+1)$ parameters from \CP tags are 
\begin{equation}
    \begin{aligned}
    \Nobs_n(\KSLhh | \CP) \propto \left[P_n^+ + P_n^- - 2s_K(2F_{+} - 1)(-1)^nC_n\right], \\
    \Nobsbar_n(\KSLhh | \CP) \propto \left[P_n^- + P_n^+ - 2s_K(2F_{+} - 1)(-1)^nC_n\right],
    \end{aligned}
    \label{eq:cp_coef}
\end{equation}
where $\Nobs_n$ ($\Nobsbar_n$) corresponds to the $w_n$($\overline{w}_n$)-weighted integrals.
The \CP-even fraction $F_{+}$ is 1 for \CP-even tags, 0 for \CP-odd tags, and ${F_+ = 0.9406\pm0.0032\stat\pm0.0021\syst}$ for the ${\pip\pim\piz}$ tag~\cite{BESIII:2024nnf}.

For the case where both signal and tag $D$ mesons are reconstructed by the $\decay{D}{\KSL h^{\prime+}h^{\prime-}}$ decays,
four sets of parameters, $\Nobs_{n_1n_2}^{\pm\pm}$, can be computed by the four possible weighting-function combinations ${\{w_{n_1}, \overline{w}_{n_1}\}\times \{w_{n_2}, \overline{w}_{n_2}\}}$.
As a result, there are a total of ${4\times(2M_{h'} + 1)\times(2M_{h''}+1)}$ parameters 
when the maximum Fourier order of the $\decay{D}{\KSL h^{\prime(\prime)+}h^{\prime(\prime)-}}$ decay is set to $M_{h^{\prime(\prime)}}$.
Letting $s_1$ and $s_2$ denote the signs of the first and second weights with the $+(-)$ sign itself corresponding to $w_n(\overline{w}_n)$,
and letting $\overline{s}_1$ and $\overline{s}_2$ denote the opposite signs, the parameters can be expressed as
\begin{align}
    \Nobs_{n_1n_2}^{s_1s_2}(\KS h^{\prime+}h^{\prime-} | \KSL h^{\prime\prime+}h^{\prime\prime-})  \propto & \left[P_{n_1}^{s_1}P_{n_2}^{{s_2}} + P_{n_1}^{\overline{s}_1}P_{n_2}^{\overline{s}_2} \right. \nonumber \\
    & \left.  - 2s_{K_1}s_{K_2}(-1)^{n_1}(-1)^{n_2}(C_{n_1}C_{n_2} + s_1s_2S_{n_1}S_{n_2})\right],
    \label{eq:dbldlz_coef}
\end{align}
where $s_{K_{1(2)}} = 1$ when the first (second) $D$ meson decays into the ${\KShh}$ final state, 
and $s_{K_{1(2)}} = -1$ when the first (second) $D$ meson decays into the ${\KL h^{\prime\prime+}h^{\prime\prime-}}$ final state.

In summary, the optimal Fourier method introduces an extra optimal weight in addition to the Fourier weights proposed in Ref.~\cite{Poluektov:2017zxp}.
Setting the maximum Fourier order to $M_{h}$ for $\decay{D}{\KSLhh}$ decays, applying the weighting functions to \lhcb data yields $2\times(2M_h+1)$ parameters for each $B$ charge, $B$ decay and $D$ decay.
The \CP observables and flavor parameters $P_n^{\pm}$ can be determined from a simultaneous fit to $\decay{\Bpm}{D\Kpm}$ and $\decay{\Bpm}{D\pipm}$ candidates when the input strong-phase parameters $C_n,S_n$ are known.
From \besiii data, ${2\times (2M_{h'}+1)}$ parameters of flavor tags, ${2\times (2M_{h'}+1)}$ parameters of \CP tags, 
and ${4\times(2M_{h'}+1)\times(2M_{h''}+1)}$ parameters of ${\KShh}$ tags are computed.
The flavor-tag parameters are used to determine $P_n^{\pm}$ in \besiii data, and the strong-phase parameters $C_n, S_n$, are obtained from fits to \CP-tag and ${\KShh}$-tag parameters.
The $C_n$ and $S_n$ parameters need to be shared between \besiii and \lhcb data, while separate $P_n^{\pm}$ parameters are obtained independently from flavor tags in \besiii and $\decay{\Bpm}{D\pipm}$ decays in \lhcb.
Therefore, a joint fit to the \besiii and \lhcb parameters allows for a simultaneous determination of the \CP observables and strong-phase parameters.

Pseudoexperiments are conducted to validate the approach, where it is found that the smallest statistical uncertainty on $\gamma$ is achieved with $M_{\pi} = 2$ and $M_{K} = 1$ when the dataset size is comparable to that analyzed in this paper. 
The increasing sensitivity with $M_h$ is limited by sizes of both \besiii and \lhcb datasets.
The average uncertainty of $\gamma$ from the optimal Fourier approach is also lower than that from the binned approach when both approaches are applied to the same ensemble.
The study is performed under ideal conditions where the generated amplitude models are the same as those used to compute the weighting functions.
As the underlying model of the data is not perfectly known, imperfect description of the models may not lead to the most optimal results.
Nevertheless, the choice of model does not bias the central values, and the orders $M_{\pi} = 2$ and $M_K = 1$ are chosen to determine the \CP observables $\xpmdk$, $\ypmdk$, $\xxi$ and $\yxi$.

\section{Detectors and simulation}
\label{sec:bes_detector}

The \besiii detector~\cite{BESIII:2009fln} records symmetric $e^+e^-$ collisions 
provided by the BEPCII storage ring~\cite{Yu:2016cof}
in the center-of-mass energy range from 1.84 to 4.95\gev with a peak luminosity of $1.1 \times 10^{33}\;\text{cm}^{-2}\text{s}^{-1}$ 
achieved at $\sqrt{s} = 3.773\gev$. 
Large data samples in this energy region have been collected by the \besiii experiment~\cite{BESIII:2020nme,Lu:2020,Zhang:2022bdc}.
The cylindrical core of the \besiii detector covers $93\%$ of the full solid angle and consists of a helium-based
 multilayer drift chamber~(MDC), a plastic scintillator time-of-flight
system~(TOF), and a CsI(Tl) electromagnetic calorimeter~(EMC),
which are all enclosed in a superconducting solenoidal magnet
providing a 1.0~T magnetic field, which was 0.9~T in 2012. 
The solenoid is surrounded by an octagonal flux-return yoke made of steel, interleaved with resistive-plate-counter muon-identification modules.

The charged-particle momentum resolution at $1\gevc$ is $0.5\%$, and the ${\rm d}E/{\rm d}x$ resolution is $6\%$ for electrons from Bhabha scattering. The EMC measures photon energies with a resolution of $2.5\%$ ($5\%$) at $1$\gev in the barrel (end-cap) region. The time resolution in the TOF barrel region is 68~ps, while that in the end-cap region is 110\ps.  The end-cap TOF system was upgraded in 2015 using multigap resistive-plate chamber technology, providing a time resolution of 60\ps~\cite{Guo:2017sjt}. This upgrade benefits 63\% of the data used in this analysis. More details can be found in Ref.~\cite{BESIII:2009fln}.

The simulation samples produced with a \geant-based software package~\cite{Agostinelli:2002hh}, which includes the geometric description of the \besiii detector and the detector response, are used to determine detection efficiencies and estimate backgrounds. The simulation models the beam energy spread and initial-state radiation (ISR) in $e^+e^-$ annihilations with the generator {\mbox{\textsc{kkmc}}\xspace}~\cite{Jadach:2000ir,Jadach:1999vf}.

Inclusive simulation samples are produced including $D\Db$ pairs corrected with quantum-correlation effects, non-$D\Db$ decays of the $\psi(3770)$, the ISR production of the $J/\psi$ and $\psi(3686)$ states, and continuum $\decay{e^+e^-}{u\uquarkbar,~d\dquarkbar,~s\squarkbar}$ processes incorporated in {\mbox{\textsc{kkmc}}\xspace}. All particle decays in the inclusive simulation samples are modeled with \evtgen~\cite{Lange:2001uf,Ping:2008zz} using branching fractions either taken from the Particle Data Group~\cite{PDG2024}, when available, or estimated with {\mbox{\textsc{Lundcharm}}\xspace}~\cite{PhysRevD.62.034003,Yang:2014vra}. Final-state radiation from charged final-state particles is incorporated using the \photos package~\cite{BARBERIO1991115}.
Simulated signal samples are produced for DT $\KSL h^{\prime+}h^{\prime-}$ \vs tag modes, where quantum-correlation effects are implemented. The $\decay{D}{\KS\pi^{+}\pi^{-}}$ decay is simulated with the amplitude model measured by the Belle experiment~\cite{BaBar:2018agf,BaBar:2018cka} and the $\decay{D}{\KL\pi^{+}\pi^{-}}$ decay is simulated with the $\KS\pi^{+}\pi^{-}$ model implemented with U-spin breaking parameters~\cite{BESIII:2022qvy}. The $\decay{D}{\KSL \Kp\Km}$ decay is simulated with the amplitude model developed by the \besiii experiment. Multibody tag-side decays are simulated with the aforementioned \evtgen according to the most recent models from experimental studies.

The \lhcb detector~\cite{LHCb-DP-2008-001,LHCb-DP-2014-002} is a single-arm forward
spectrometer covering the \mbox{pseudorapidity} range $2<\eta <5$,
designed for the study of particles containing \bquark or \cquark
quarks. The detector used to collect the data analyzed in this paper includes a high-precision tracking system
consisting of a silicon-strip vertex detector surrounding the $pp$
interaction region, a large-area silicon-strip detector located
upstream of a dipole magnet with a bending power of about
$4{\mathrm{\,T\,m}}$, and three stations of silicon-strip detectors and straw
drift tubes are placed downstream of the magnet.
The tracking system provides a measurement of the momentum of charged particles with
a relative uncertainty that varies from 0.5\% at low momentum to 1.0\% at 200\gevc.
The minimum distance from a track to a primary $pp$ collision vertex, the impact parameter, 
is measured with a resolution of $(15+29/\pt)\mum$,
where \pt is the component of the momentum transverse to the beam, in\,\gevc.
Different types of charged hadrons are distinguished using information
from two ring-imaging Cherenkov detectors. 
Photons, electrons and hadrons are identified by a calorimeter system consisting of
scintillating-pad and preshower detectors, an electromagnetic
and a hadronic calorimeter. Muons are identified by a
system composed of alternating layers of iron and multiwire
proportional chambers.
The online event selection is performed by a trigger, 
which consists of a hardware stage, based on information from the calorimeter and muon
systems, followed by a software stage, which applies a full event
reconstruction.
Triggered data further undergo a centralized, offline processing step
to deliver physics-analysis-ready data across the entire \lhcb physics program~\cite{Stripping}.

Simulation is required to correct for reconstruction and selection efficiencies in \lhcb data.
In the simulation, $pp$ collisions are generated using
\pythia~\cite{Sjostrand:2007gs,*Sjostrand:2006za} 
with a specific \lhcb configuration~\cite{LHCb-PROC-2010-056}.
Decays of unstable particles
are described by \evtgen~\cite{Lange:2001uf}, in which final-state
radiation is generated using \photos~\cite{davidson2015photos}.
The interaction of the generated particles with the detector and its response,
are implemented using the \geant
toolkit~\cite{Allison:2006ve, Agostinelli:2002hh} as described in
Ref.~\cite{LHCb-PROC-2011-006}. 
When generating simulation samples for background processes, the underlying $pp$ interaction is reused multiple times, with an independently generated background decay for each~\cite{LHCb-DP-2018-004}.
Some subdominant background processes are generated with {\mbox{\textsc{RapidSim}}\xspace}~\cite{Cowan:2016tnm} that mimics the \lhcb detector acceptance, reconstruction efficiency and dynamics of the decay.

\section{Event selection and signal extraction}
\label{sec:selection}

Quantum-correlated $D\Db$ decays collected by the BESIII detector are reconstructed by requiring one of the $D$ mesons to decay to a signal channel, ${\KSLhh}$, and the other to decay via a tag mode.
As described in Sect.~\ref{sec:analysis_strategy}, three categories of tag modes are used in the analysis.
The flavor and $C\!P$-tag modes are listed in Table~\ref{tab:usedtags}.
The decays where both $D$ mesons are reconstructed in a signal channel are selected according to the following self-conjugate DT modes: ${\KS\pip\pim}$ \vs $\{\KS\pip\pim,~\KS\pipm[\pimp],~\KS(\to\piz[\piz])\pip\pim,~[\KL]\pip\pim,~\KS\Kpm[\Kmp],~[\KL]\Kp\Km\}$ and ${\KS\Kp\Km}$ \vs $\{\KS\Kpm[\Kmp],~ [\KL]\Kp\Km,~[\KL]\pip\pim\}$,
where particles in square brackets are not reconstructed.

\begin{table}[tb]
\caption{Summary of tag modes selected against $\decay{D}{\KSL h^{\prime+}h^{\prime-}}$.}
\label{tab:usedtags}
\begin{center}
\begin{tabular}{ccccc}
\hline
Tag & $\KS h^{\prime+}h^{\prime-}$ & $\KL h^{\prime+}h^{\prime-}$  \\
\hline
Flavor   & $K\pi\pi^{0}$,  $K\pi$, $K\pi\pi\pi$, $Ke\nu_{e}$ & $K\pi\pi^{0}$,  $K\pi$, $K\pi\pi\pi$ \\
$C\!P$-even &  $KK$, $\pi\pi$, $\KS\pi^{0}\pi^{0}$, $\KL\pi^{0}$, $\pi\pi\pi^{0}$ & $KK$, $\pi\pi$, $\KS\pi^{0}\pi^{0}$, $\pi\pi\pi^{0}$\\
\multirow{3}{*}{$C\!P$-odd} & $\KS\pi^{0}$, $ \KS\eta(\to\gamma\gamma)$, $\KS\eta(\to\pi\pi\piz)$,  & $\KS\pi^{0}$, $ \KS\eta(\to\gamma\gamma)$, $\KS\eta(\to\pi\pi\piz)$,  \\ 
                            & $ \KS\eta^{\prime}(\to\pi\pi\gamma)$, $ \KS\eta^{\prime}(\to\pi\pi\eta)$,  & $ \KS\eta^{\prime}(\to\pi\pi\eta)$, $ \KS\eta^{\prime}(\to\pi\pi\gamma)$,  \\
                            & $\KS\omega(\to\pi\pi\piz)$, $\KL\pi^{0}\pi^{0}$ & $\KS\omega(\to\pi\pi\piz)$ \\
\hline
\end{tabular}
\end{center}
\end{table}

Final-state particles of charm-meson decays are reconstructed from candidates of charged tracks and neutral photons using selection criteria identical to those applied in the binned analysis of $\decay{D}{\KS \pi^+ \pi^-}$~\cite{BESIII:2025nsp}. 
In addition, the requirement of no extra $\pi^0$ candidates is removed for the $\KL \Kp \Km$ \vs $\KS \Kp\Km$ and $\KS \Kp \Km$ \vs $\KL \pi^+\pi^-$ final states, as higher statistics are obtained without a decrease in signal purity.
Meanwhile, to mitigate resolution effects in selected $D\Db$ decay samples, 
$D$, $\KS$ and $\KL$ candidates are kinematically constrained to their known masses~\cite{PDG2024}.

The charm-meson decays where all final-state particles are reconstructed are classified as fully reconstructed modes. 
Following Ref.~\cite{BESIII:2025nsp}, a mode-specific requirement on the energy difference $\Delta E \equiv E_{D}-\sqrt{s}/2$ is applied, 
and the beam-constrained mass ${M_{\rm BC} \equiv \sqrt{({\sqrt{s}/2})^2-|\boldsymbol{p}_{D}|^2c^2}}$ is utilized as the fit variable. 
Here, $\sqrt{s}/2$ is the beam energy, $E_{D}$ is the sum of final-state-particle energies and $\boldsymbol{p}_{D}$ represents the reconstructed momentum of the $D$ candidate both evaluated in the center-of-mass system. 
When there is a final-state particle such as a $\KL$ meson or neutrino that cannot be reconstructed within the BESIII detector, a partial reconstruction method is adopted. 
The squared missing mass is calculated by $M^2_{{\rm miss}}= E^2_{{\rm miss}}-|\boldsymbol{p}_{{\rm miss}}|^2c^2$, where $E_{{\rm miss}}=\sqrt{s}/2-E_{{\rm other}}$ and $\boldsymbol{p}_{{\rm miss}}=-\boldsymbol{p}_{\rm tag}-\boldsymbol{p}_{{\rm other}}$. 
Here, $\boldsymbol{p}_{\rm tag}$ represents the momentum of the fully reconstructed $D$ candidate, while $\boldsymbol{p}_{{\rm other}}$ and $E_{{\rm other}}$ are the total momentum and energy of the other reconstructed particles in the partially reconstructed modes, all evaluated in the center-of-mass system. 
In the $\decay{D}{K e \neue}$ mode, $U_{\rm miss} = E_{\rm miss} - |\boldsymbol{p}_{\rm miss}|c$ is employed as the fit variable as signal events form a peak around $U_{\rm miss} = 0 \gev$.

\begin{figure}[tb]
    \centering
    \includegraphics[width=0.49\linewidth]{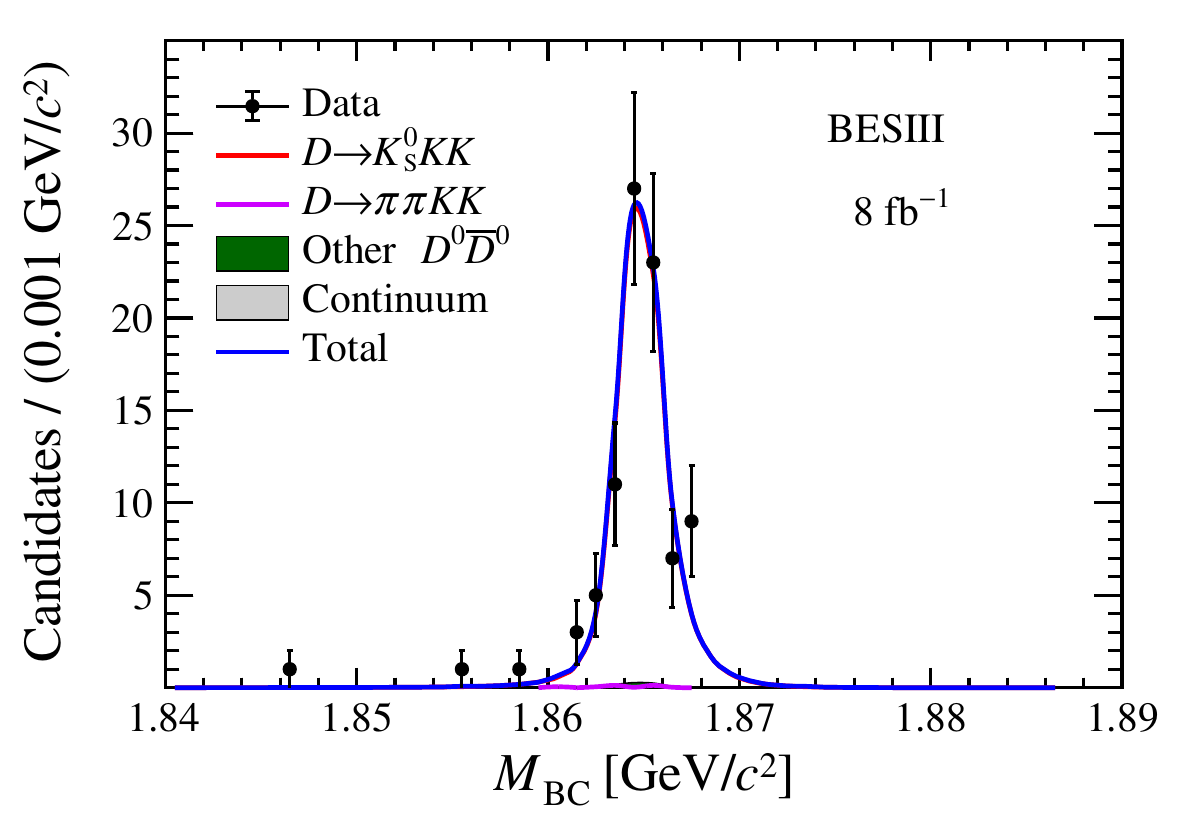}
    \includegraphics[width=0.49\linewidth]{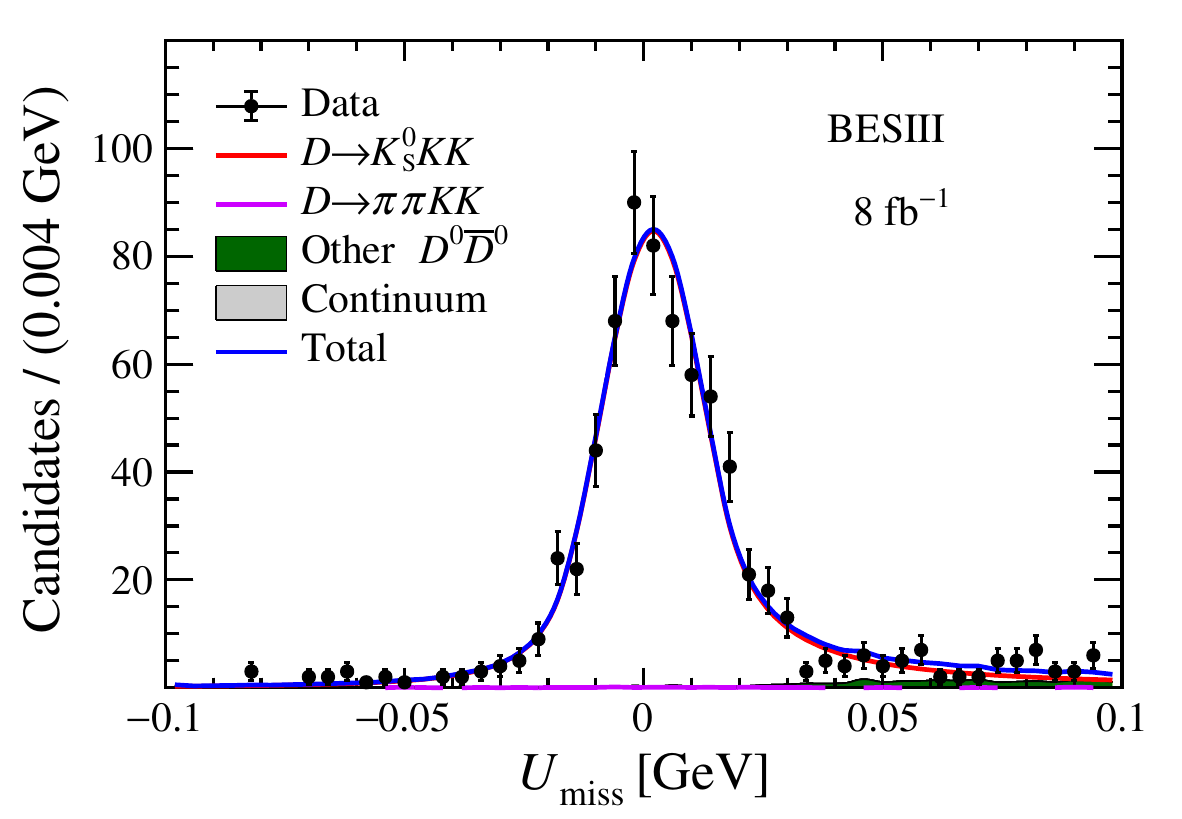}
    \includegraphics[width=0.49\linewidth]{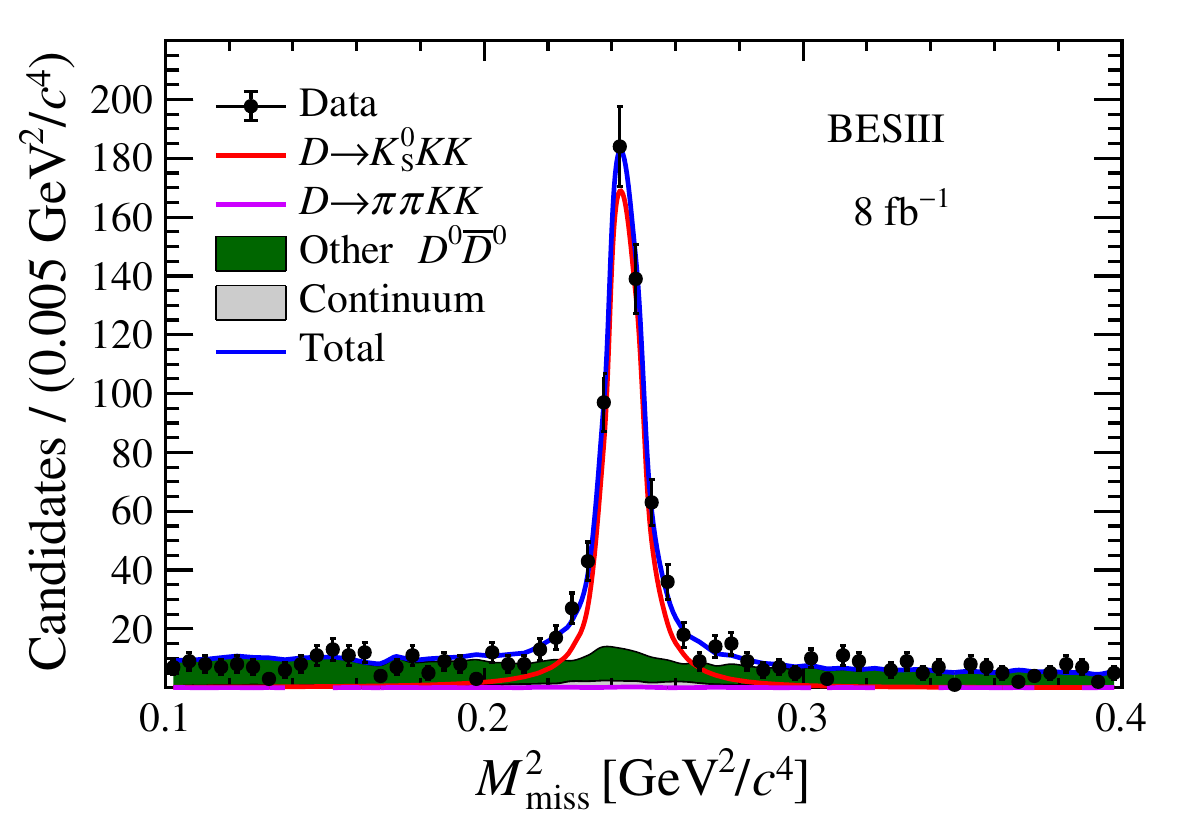}
    \includegraphics[width=0.49\linewidth]{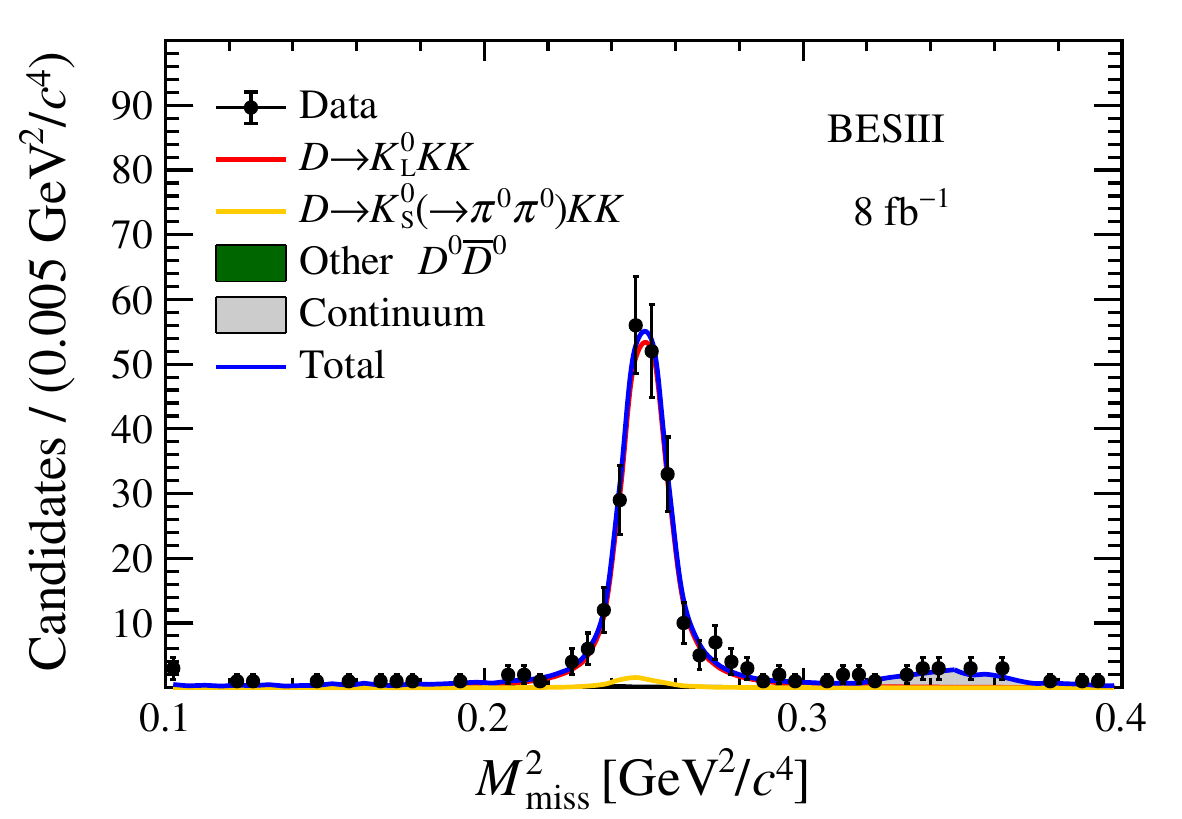}
    \caption{Example fit results for the $\decay{D}{\KSL \Kp\Km}$ channel at \besiii, where (top left) corresponds to the $\decay{D}{\KS KK}$ \vs $KK$ tag; (top right) the $\decay{D}{\KS KK}$ \vs $Ke[\nu_{e}]$ tag; (bottom left) the $\decay{D}{\KS [K]K}$ \vs $\decay{D}{\KS \pi\pi}$ tag; and (bottom right) the $\decay{D}{[\KL] KK}$ \vs $KK$ tag.}
    \label{fig:bes_k0kk_massfit_exam}
\end{figure}

Possible sources of background in $D$ decays are estimated through inclusive simulation samples.
Details on background studies, extraction of signal yields and efficiency determinations of the $\decay{D}{\KSL \pi^{+}\pi^{-}}$ signal channel can be found in Ref.~\cite{BESIII:2025nsp}.
For the channel $\decay{D}{\KSL \Kp\Km}$, the dominant peaking backgrounds are $\decay{D}{\pi^+\pi^- \Kp\Km}$ and $\decay{D}{\KS(\to\piz\piz)\Kp\Km}$ with a contamination rate of $0.5\%$ and $2.5\%$, respectively.
The nonpeaking backgrounds are mainly from other $D\Db$ decays and continuum processes. One-dimensional unbinned maximum-likelihood fits are performed to discriminate signal from background.
In the fits, double-sided Crystal Ball functions~\cite{Skwarnicki:1986xj} are used as signal shapes, with all of their parameters fixed from simulated signal samples. The signal models are then convolved with Gaussian functions whose mean values and widths vary freely in order to take into account the different resolution effects between data and simulation. 
The background shapes and contributions are directly derived from inclusive simulation samples except for those from the $\decay{\KS}{\pi^0\pi^0}$ decay in $\decay{D}{\KL \Kp\Km}$ \vs $C\!P$-tag modes. For those background components, the shapes are extracted from simulated signal samples and the background yields are estimated from data based on the signal yields of $\KS \Kp\Km$ \vs $C\!P$-eigenstate modes.
Example fit results for the $\decay{D}{\KSL \Kp\Km}$ channel are shown in Fig.~\ref{fig:bes_k0kk_massfit_exam}. Details on fits separated by individual tag modes along with their corresponding signal yields are given in Appendix~\ref{app:bes_k0kk_datafit_result}.

In the datasets collected by the \lhcb detector, decays of \decay{\KS}{\pip\pim} are reconstructed in two different categories.
The first involves \KS mesons that decay early enough for the
pions to be reconstructed in the vertex detector, and the
second involves \KS mesons that decay later such that track segments of the
pions cannot be formed in the vertex detector. These categories are
referred to as \emph{long} and \emph{downstream}, respectively. The
long category has better mass, momentum and vertex resolution than the
downstream category, while the downstream category has higher reconstruction efficiency and contributes approximately two thirds of the dataset. 
The $D$-meson candidates are then formed by combining a $\KS$ candidate with two tracks both with either the pion or kaon mass hypothesis. 
The $B$-meson candidate is subsequently built by combining a $D$ candidate and another pion or kaon track.
This constitutes eight categories in the $\decay{\Bpm}{D(\to\KShh)h^{\pm}}$ data.

\begin{figure}[!tb]
    \centering
    \includegraphics[width=0.48\linewidth]{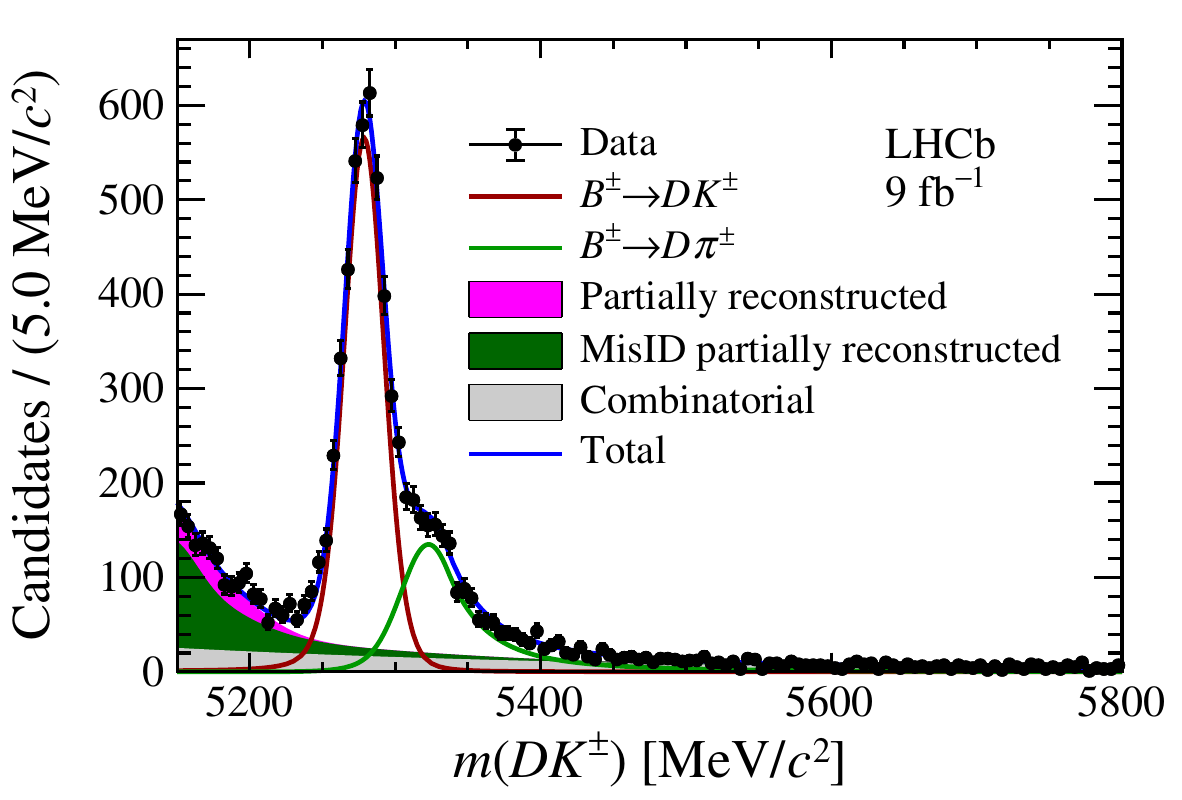}
    \includegraphics[width=0.48\linewidth]{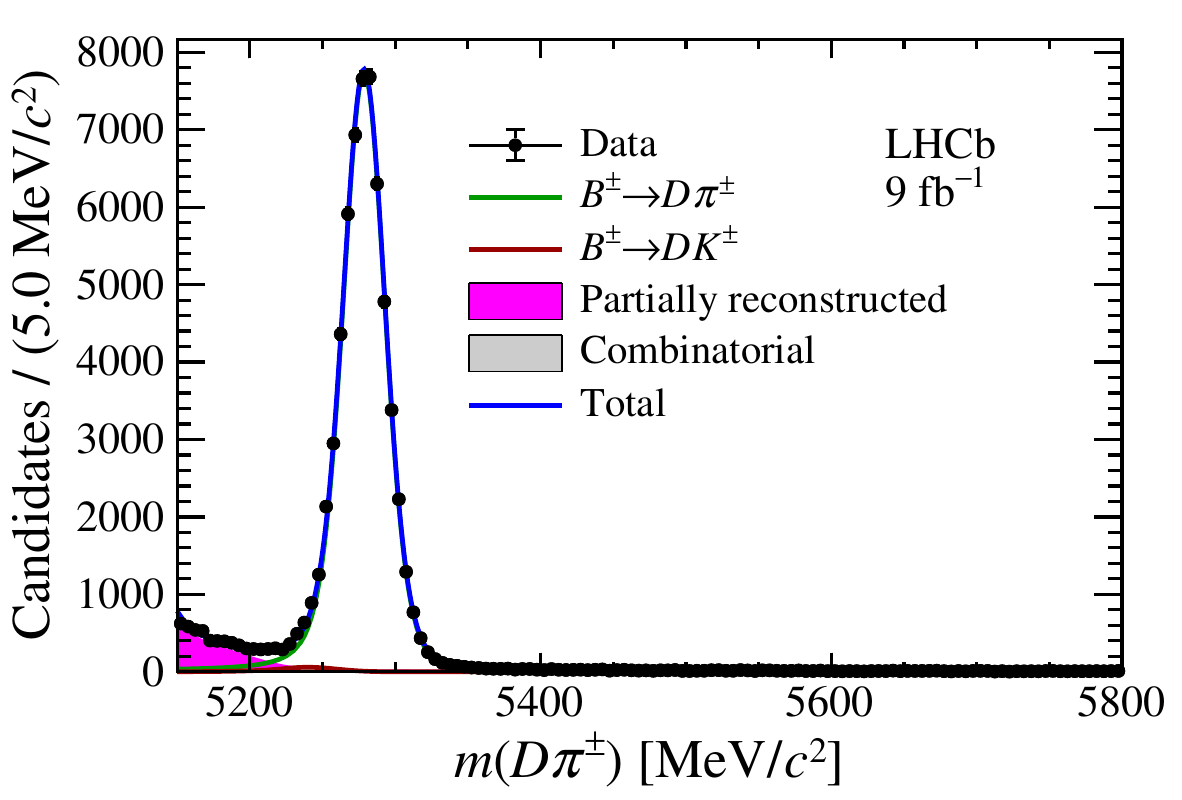}
    
    \includegraphics[width=0.48\linewidth]{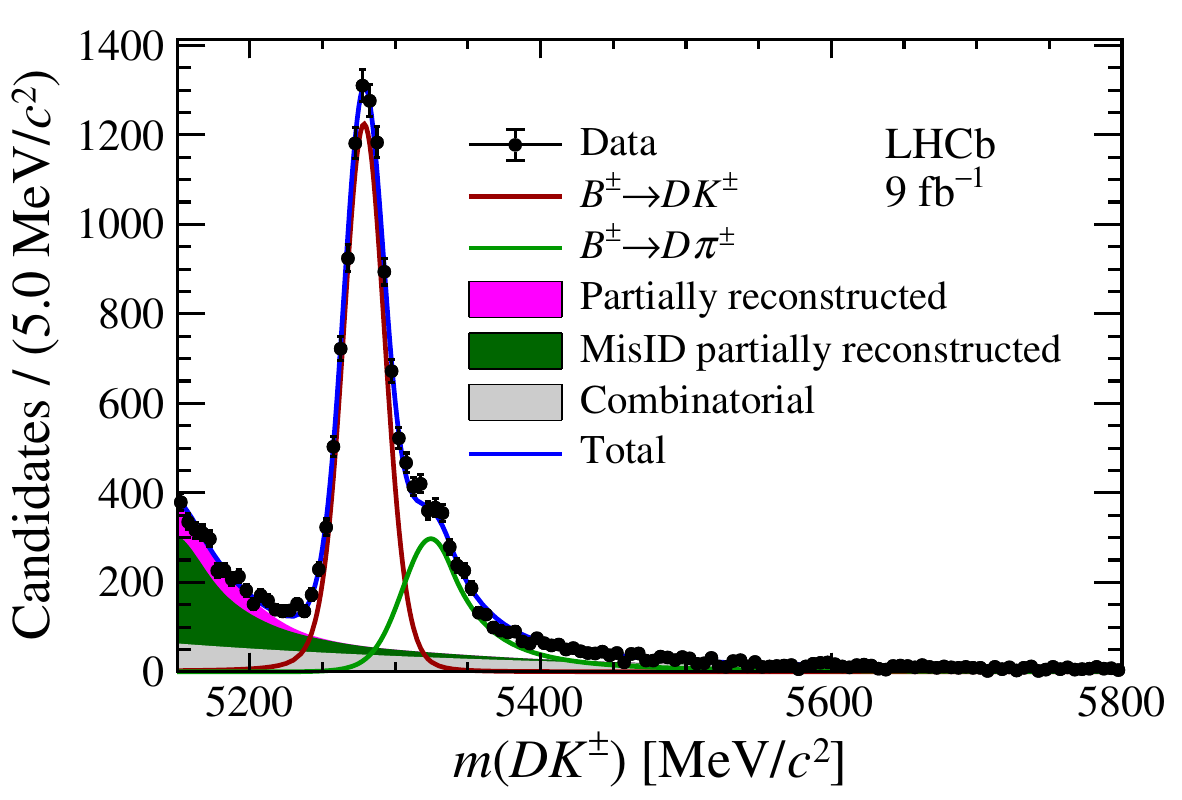}
    \includegraphics[width=0.48\linewidth]{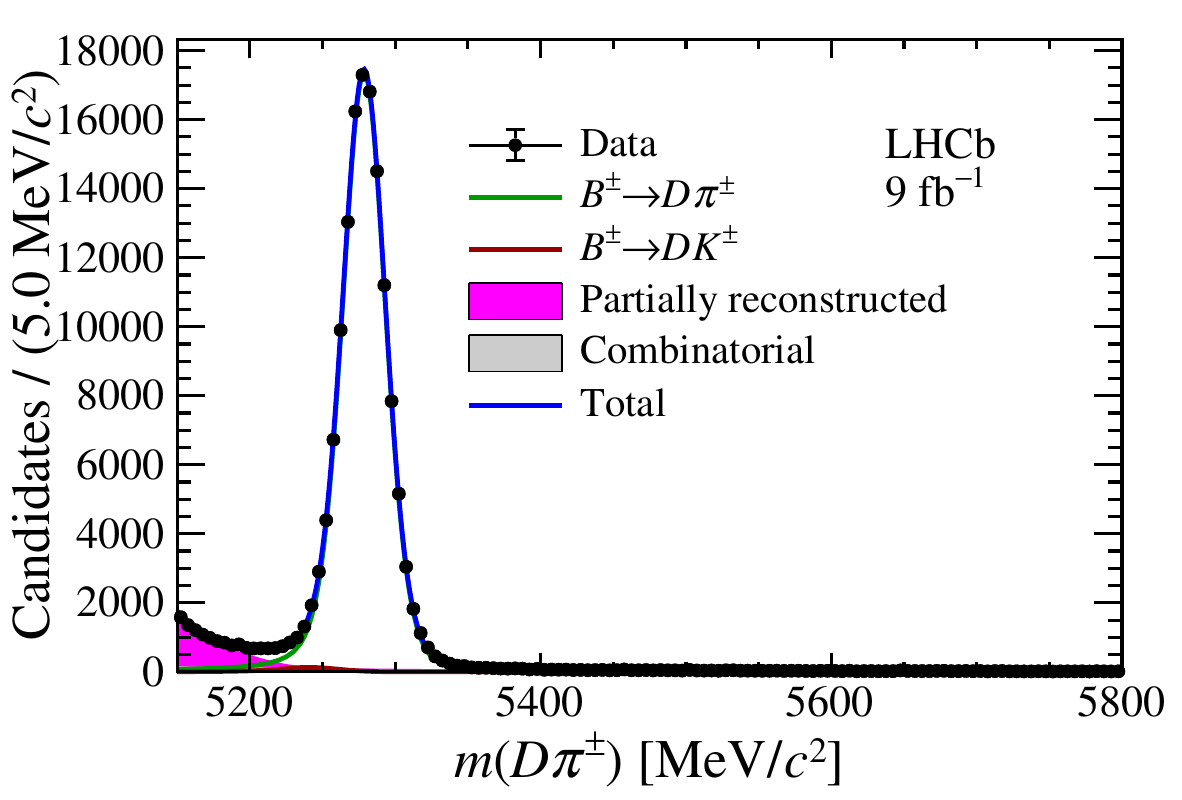}
    
    \includegraphics[width=0.48\linewidth]{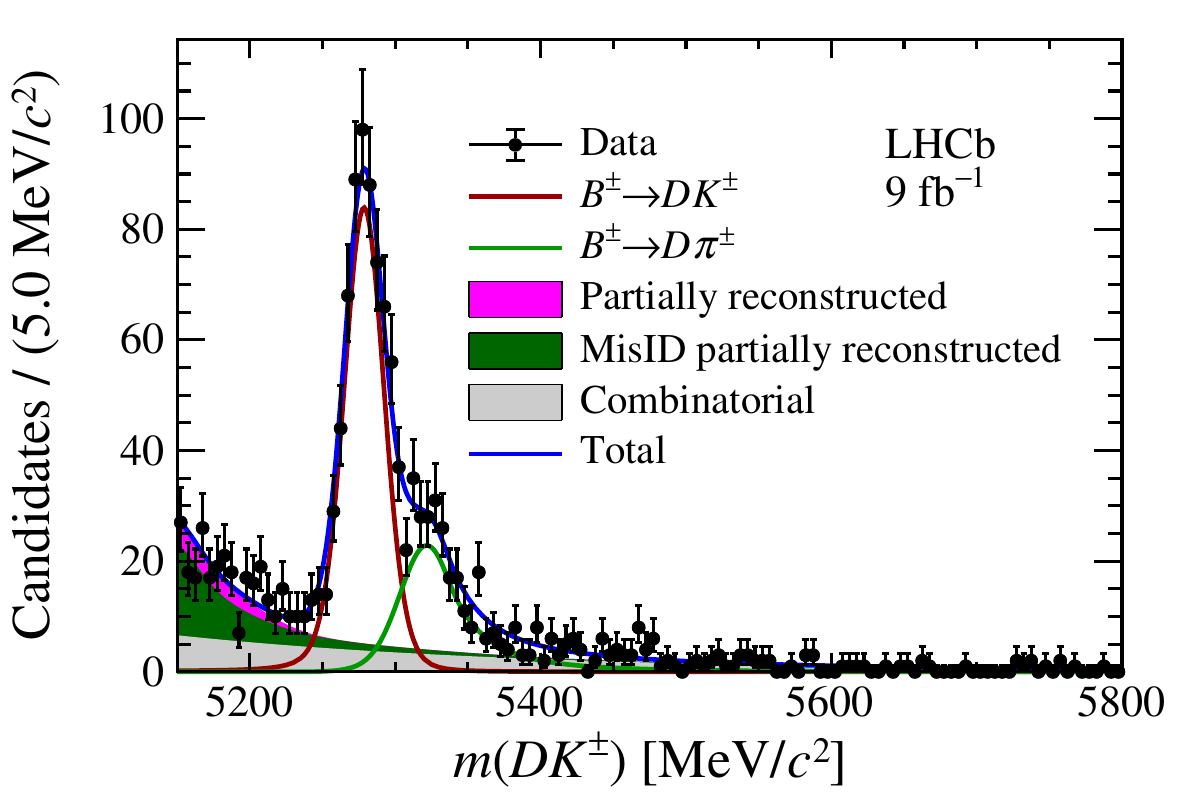}
    \includegraphics[width=0.48\linewidth]{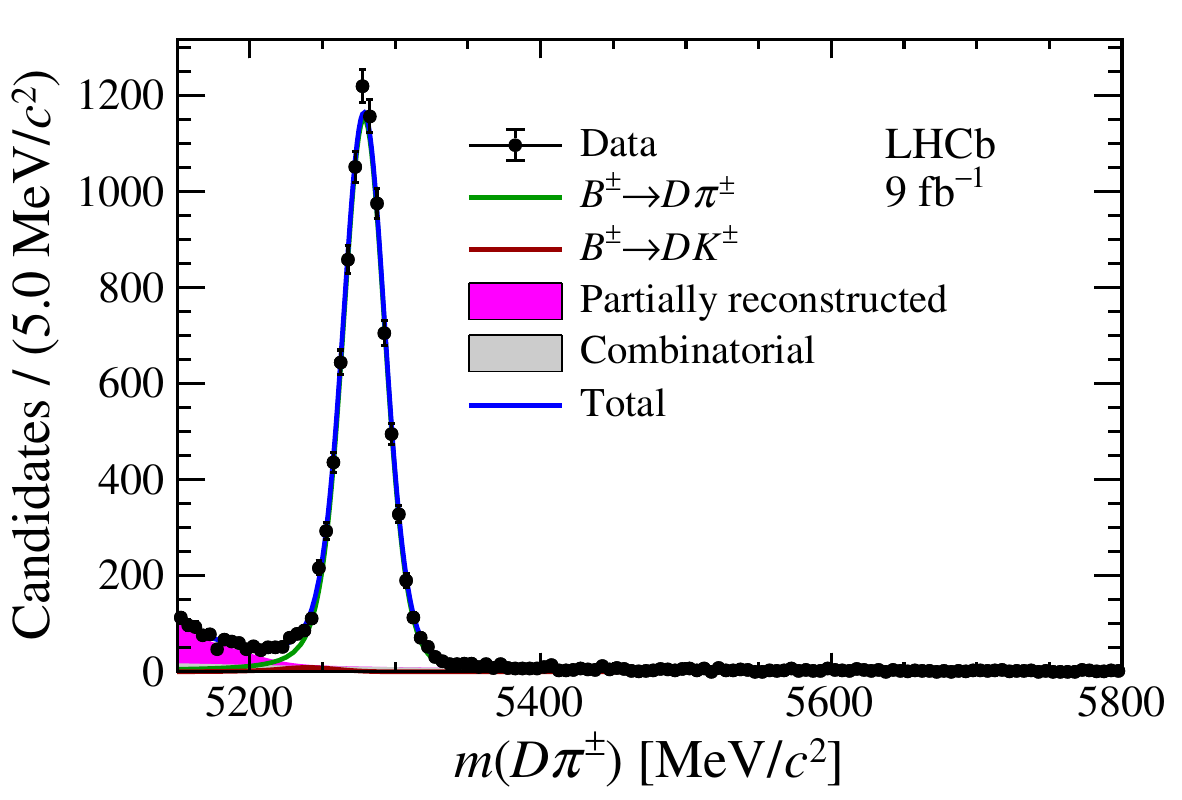}
    
    \includegraphics[width=0.48\linewidth]{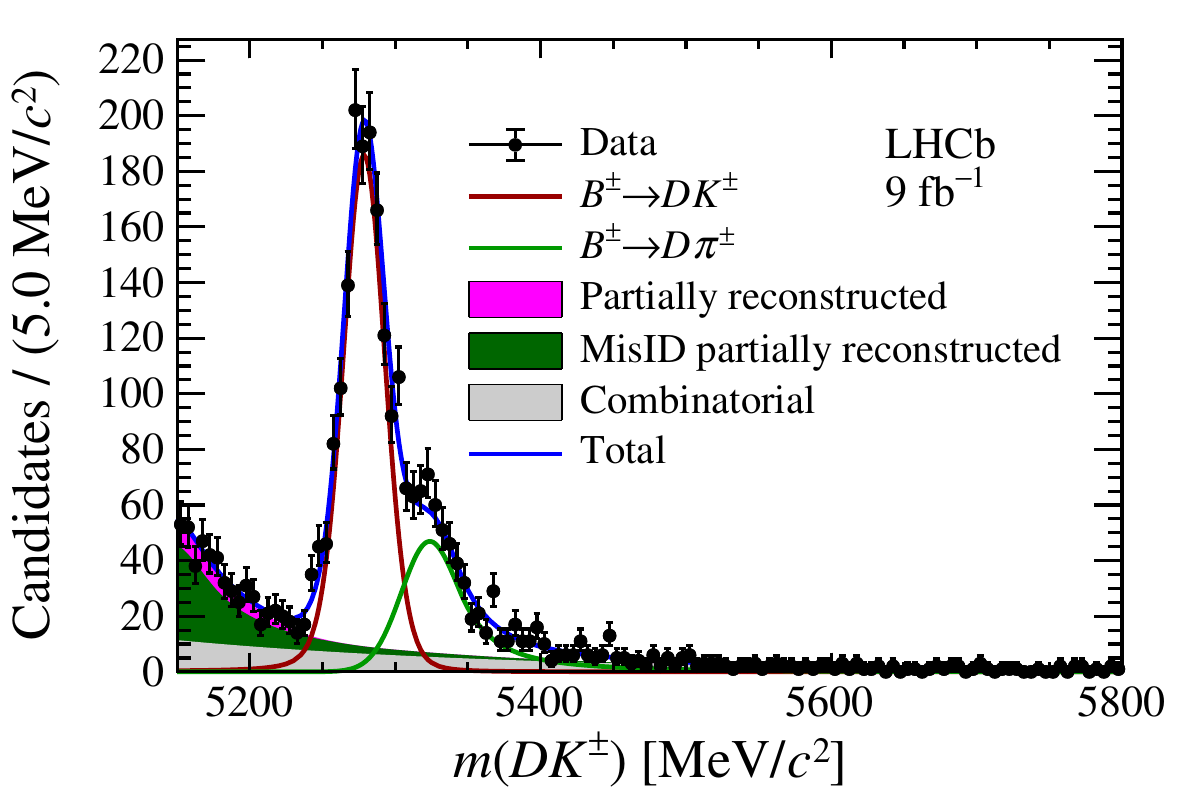}
    \includegraphics[width=0.48\linewidth]{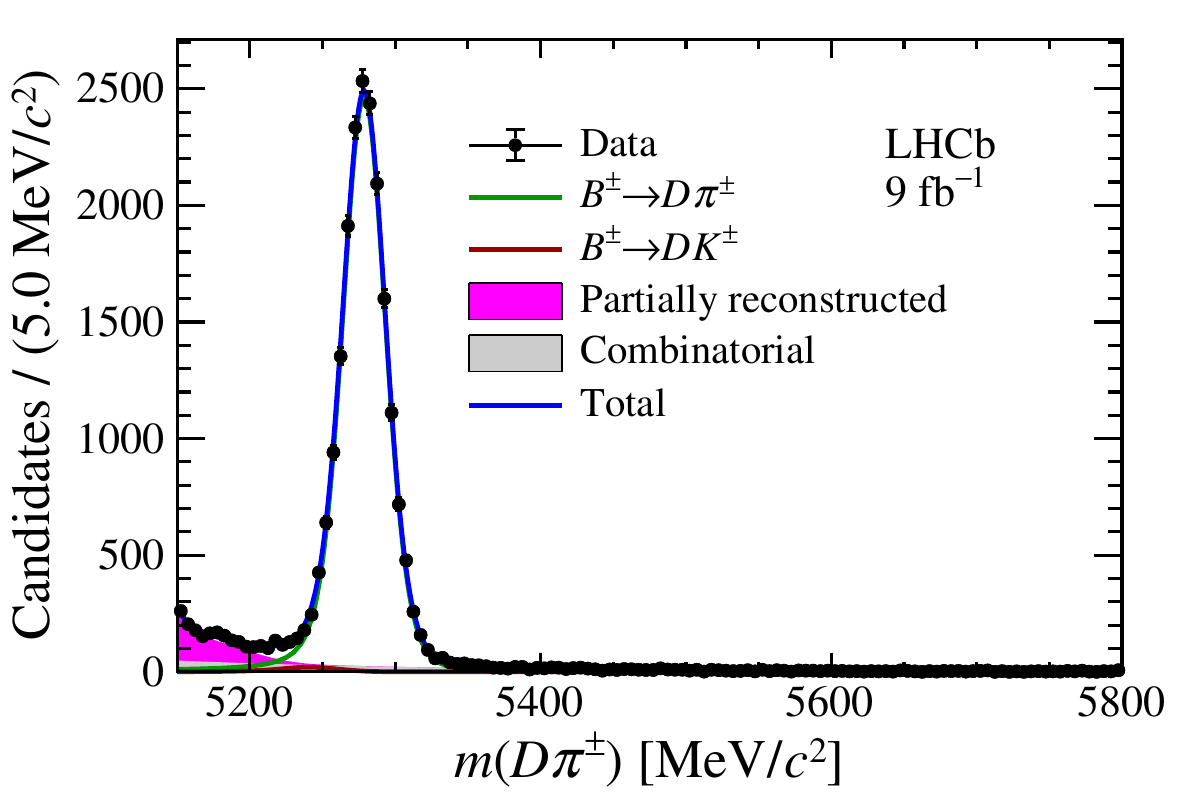}
    \caption{Mass distributions and fit results of the (left) $\decay{\Bpm}{D\Kpm}$ and (right)~$\decay{\Bpm}{D\pipm}$ data, with the $D$ meson decaying to either (top four) ${\KS\pip\pim}$ or (bottom four) ${\KS\Kp\Km}$ final states. The first and third rows correspond to the \emph{long} category, while the second and fourth rows correspond to the \emph{downstream} category.}
    \label{fig:lhcb_massfit}
\end{figure}

The selection criteria and fit model to the $Dh^{\pm}$ mass spectra are identical to those described in Ref.~\cite{LHCb-PAPER-2020-019}.
Signal and background yields in data are extracted by a two-stage fit.
The first stage is the same as the \emph{global fit} in Ref.~\cite{LHCb-PAPER-2020-019}, in which the parametrizations of signal and background components are determined.
The global fit includes: the signal $\decay{\Bpm}{Dh^{\pm}}$ decays; 
partially reconstructed $\decay{\Bpm}{\Dstar(\to D\gamma /\piz) h^{\pm}}$, $\decay{\Bz}{\Dstarmp(\to D\pimp)h^{\pm}}$, $\decay{\Bz}{Dh^{\pm}\pimp}$ and $\decay{\Bs}{D\Km\pip}$ decays, where a pion or photon is not reconstructed; the aforementioned background decays in $\decay{\Bpm}{D\pipm}$ with the companion pion misidentified as a kaon; and a combinatorial component in each category.
All shape parameters are subsequently fixed in the second stage, in which another fit is performed in a tightened mass range, from 5150 to 5800\mevcc, to reject most of the partially reconstructed background events.
The mass distributions with fit results also included are shown in Fig.~\ref{fig:lhcb_massfit}.

\section{Observables from data}
\label{sec:coefficient}

The observables are obtained from weighted sums over data events, 
in which background contributions are subtracted and efficiency effects are corrected.
The signal contribution in \lhcb data is isolated by the \sPlot technique~\cite{Pivk:2004ty}
from which per-event weights, denoted as $f_i$, are obtained based on the mass fit results.
However, \sPlot is not applicable to \besiii data due to the known correlation between the fit variables and Dalitz plot coordinates.
The $D$-decay observables are computed from both \besiii data and simulated background samples,
and the background observables are subtracted from the data observables.
Efficiency corrections are applied to both datasets by applying inverse efficiency weights derived from relative efficiency distributions.

\begin{figure}[tb]
    \centering
    \includegraphics[width=0.49\linewidth]{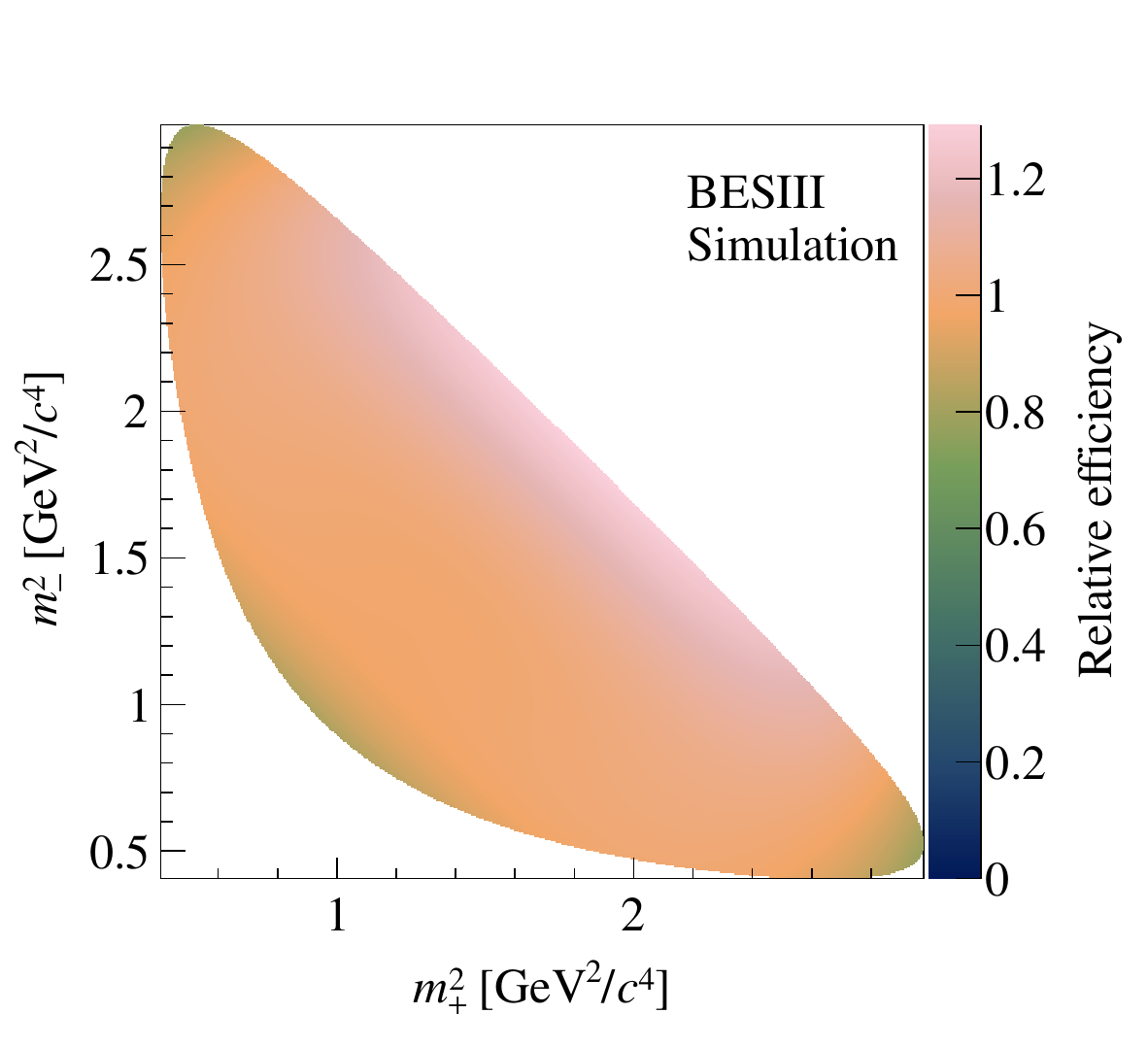}
    \includegraphics[width=0.49\linewidth]{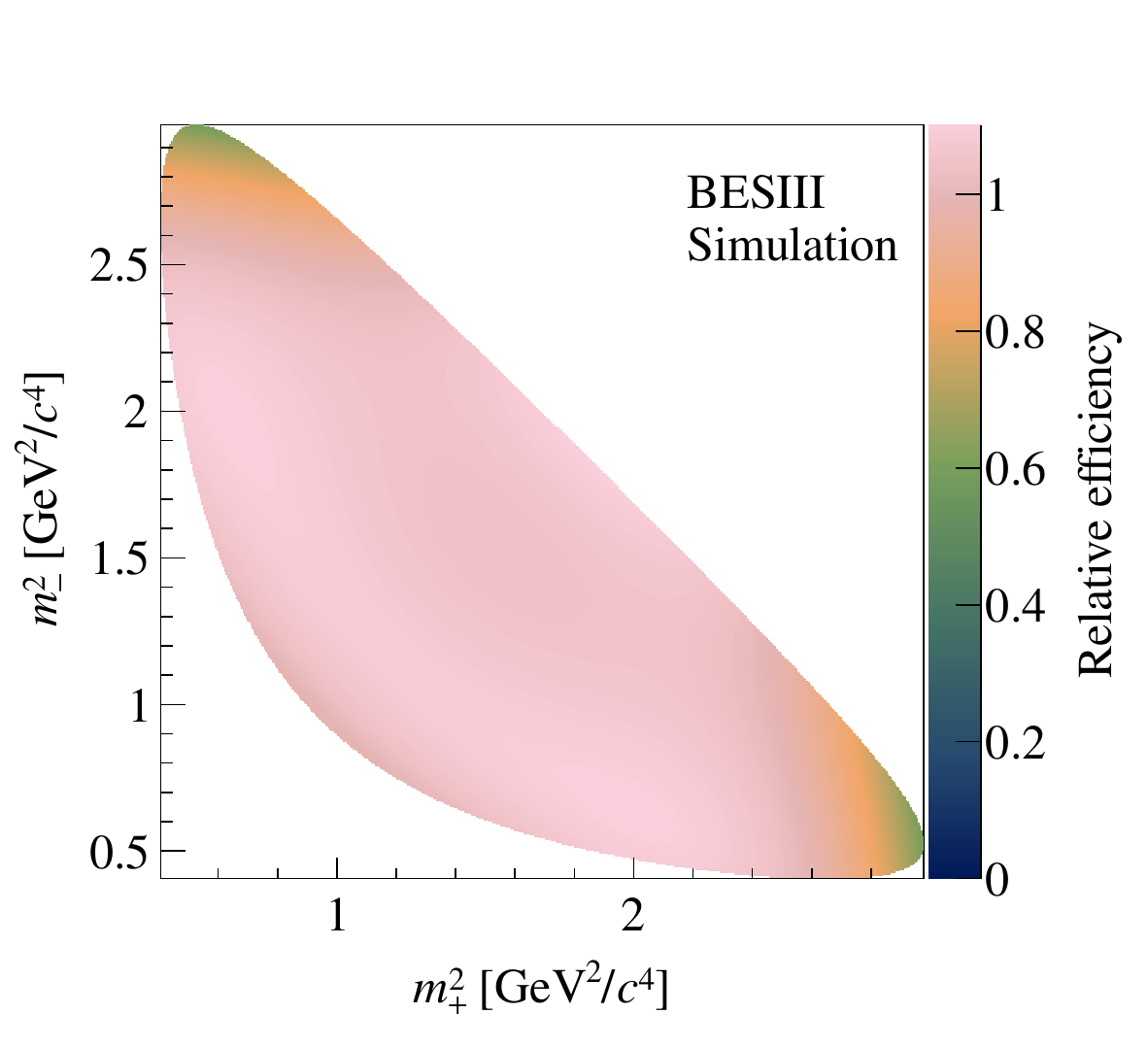}
    
    \includegraphics[width=0.49\linewidth]{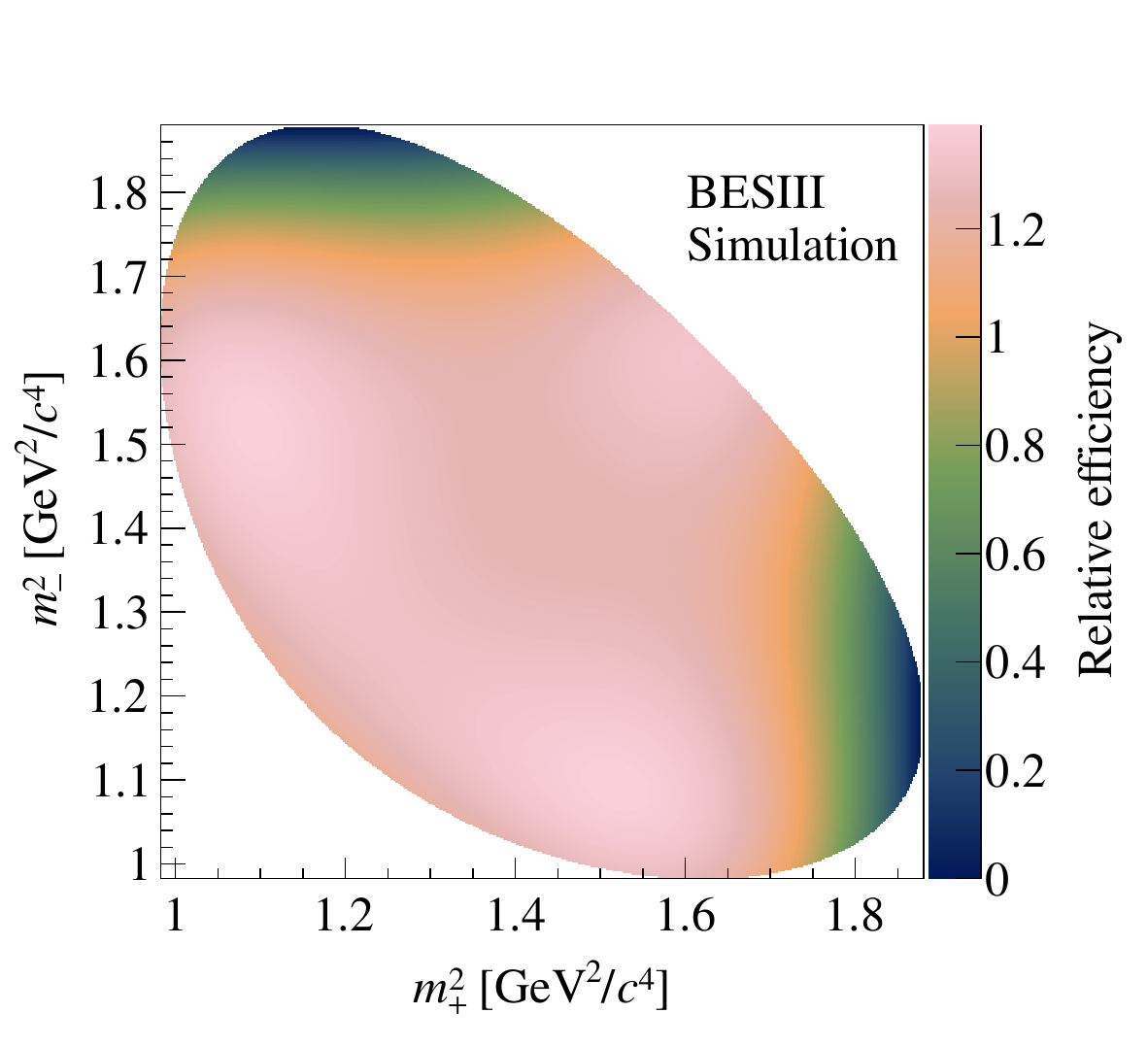}
    \includegraphics[width=0.49\linewidth]{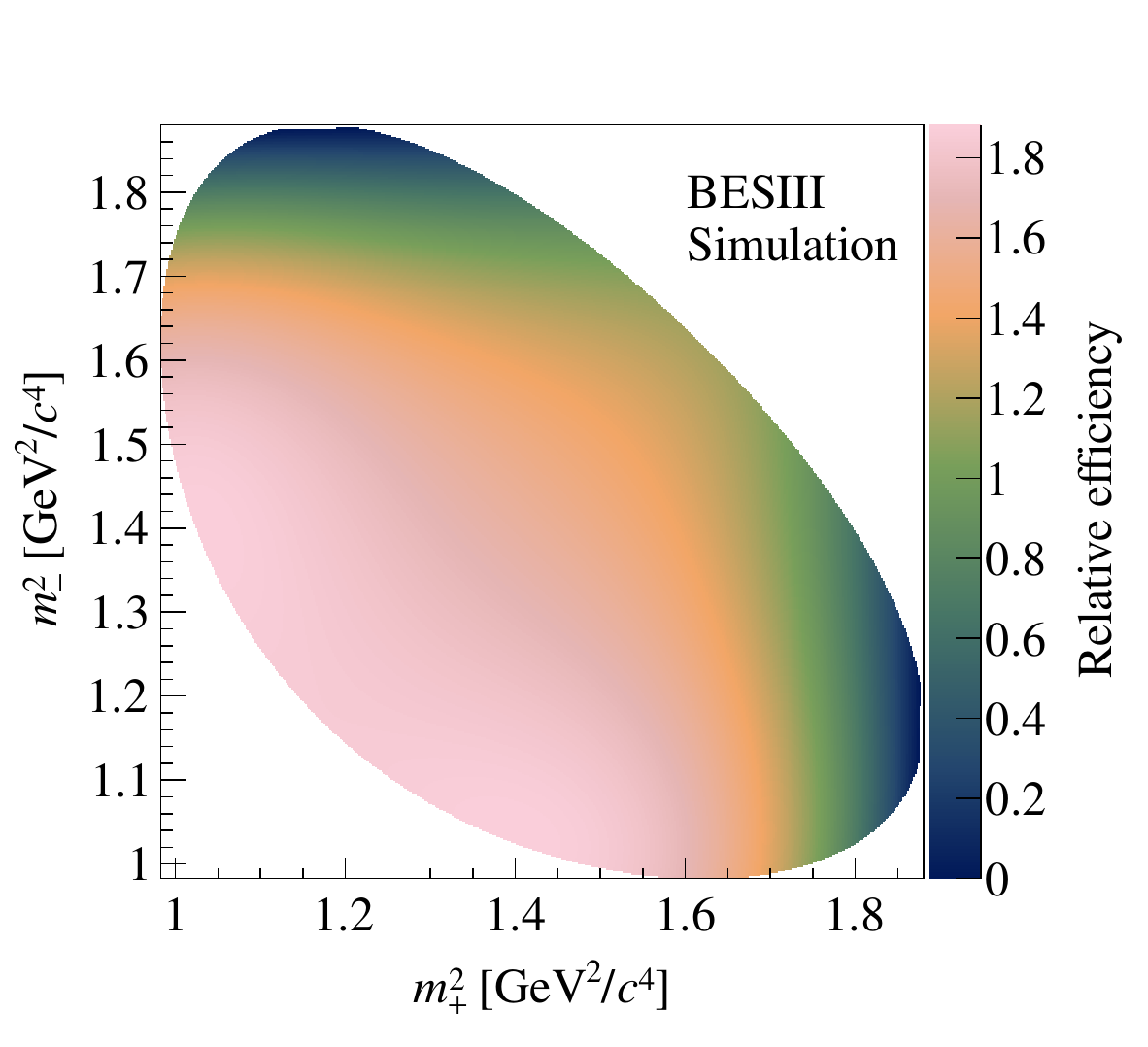}
    \caption{Example efficiency distributions from \besiii simulation. For the $K\pi$ tag mode, the plots correspond to the (top left)~$\decay{D}{\KS \pip\pim}$, (top right)~$\decay{D}{\KL \pip\pim}$, (bottom left)~$\decay{D}{\KS \Kp\Km}$ and (bottom right)~$\decay{D}{\KL \Kp\Km}$ signal modes. 
    The average efficiencies over the Dalitz plot are scaled to unity.}
    \label{fig:bes_eff}
\end{figure}

\begin{figure}[tb]
    \centering
    \includegraphics[width=0.49\linewidth]{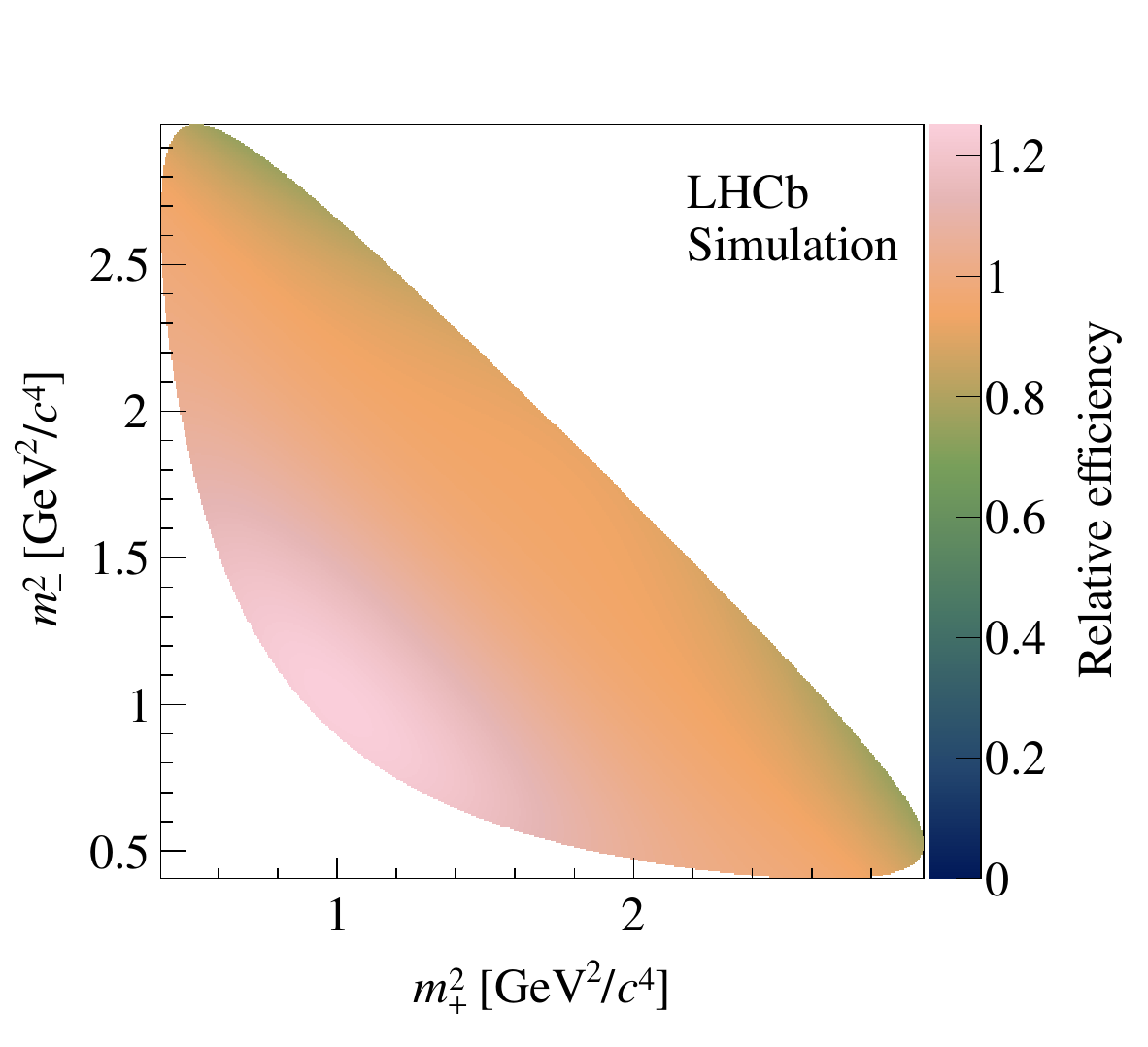}
    \includegraphics[width=0.49\linewidth]{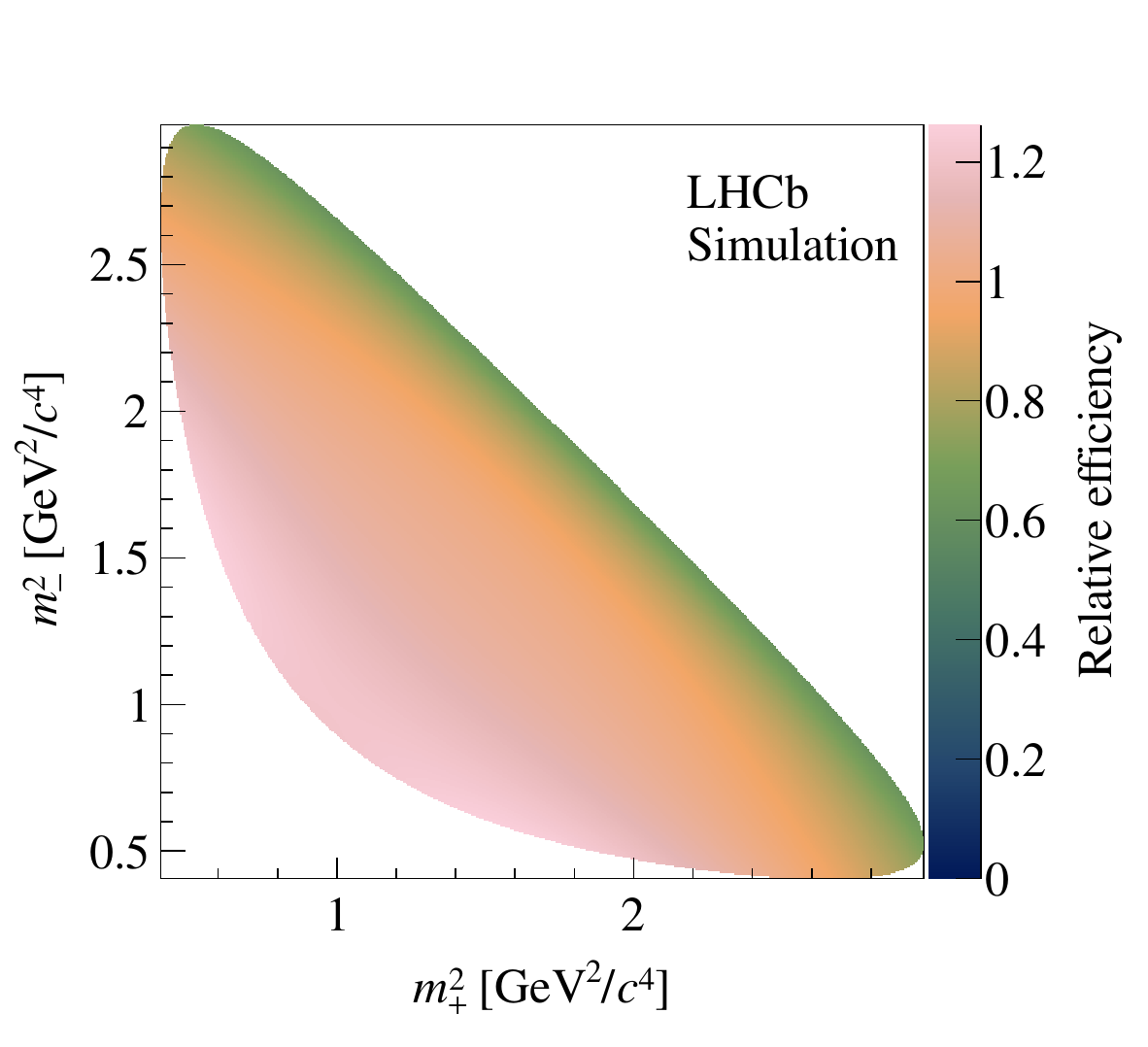}
     
    \includegraphics[width=0.49\linewidth]{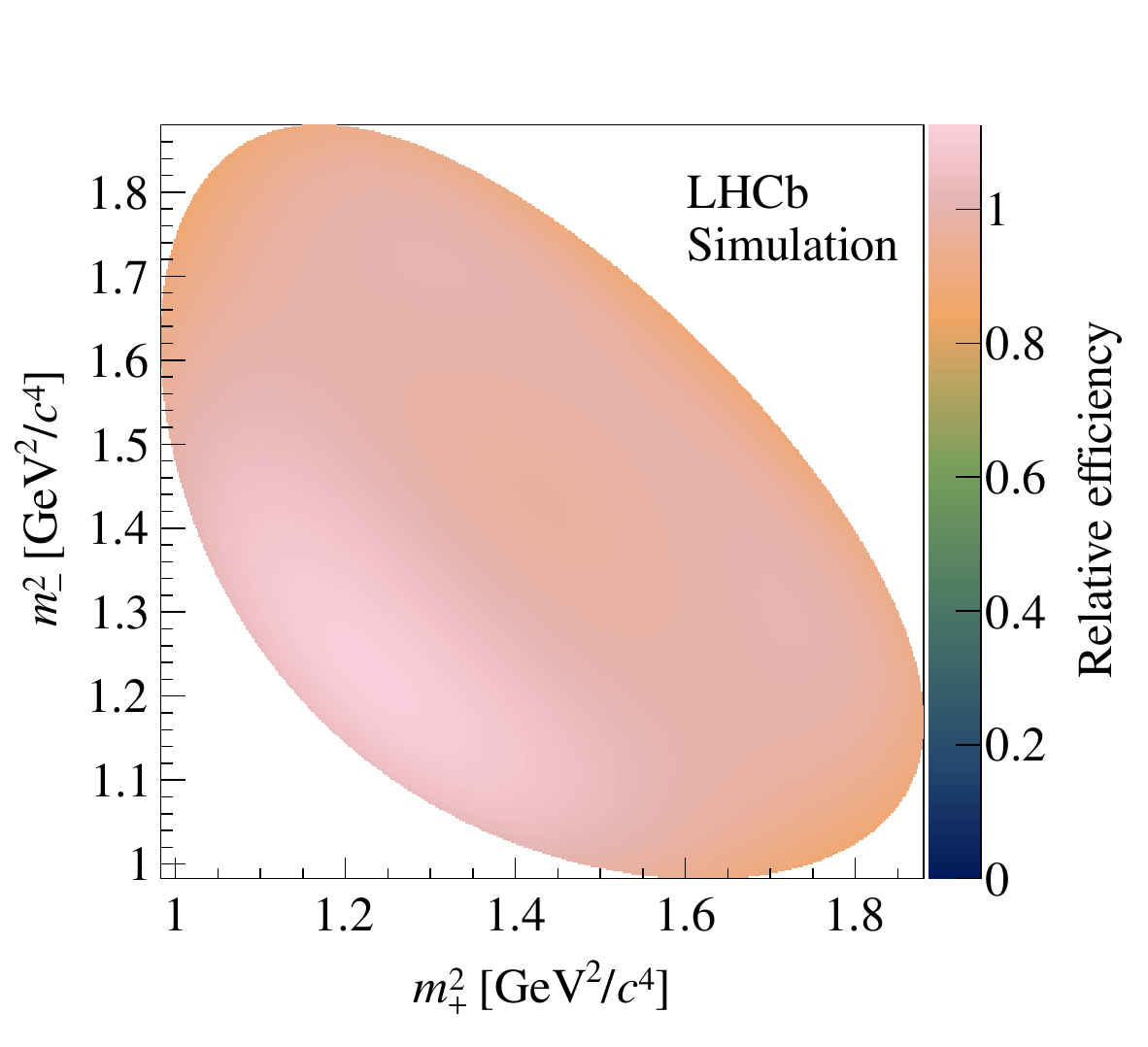}
    \includegraphics[width=0.49\linewidth]{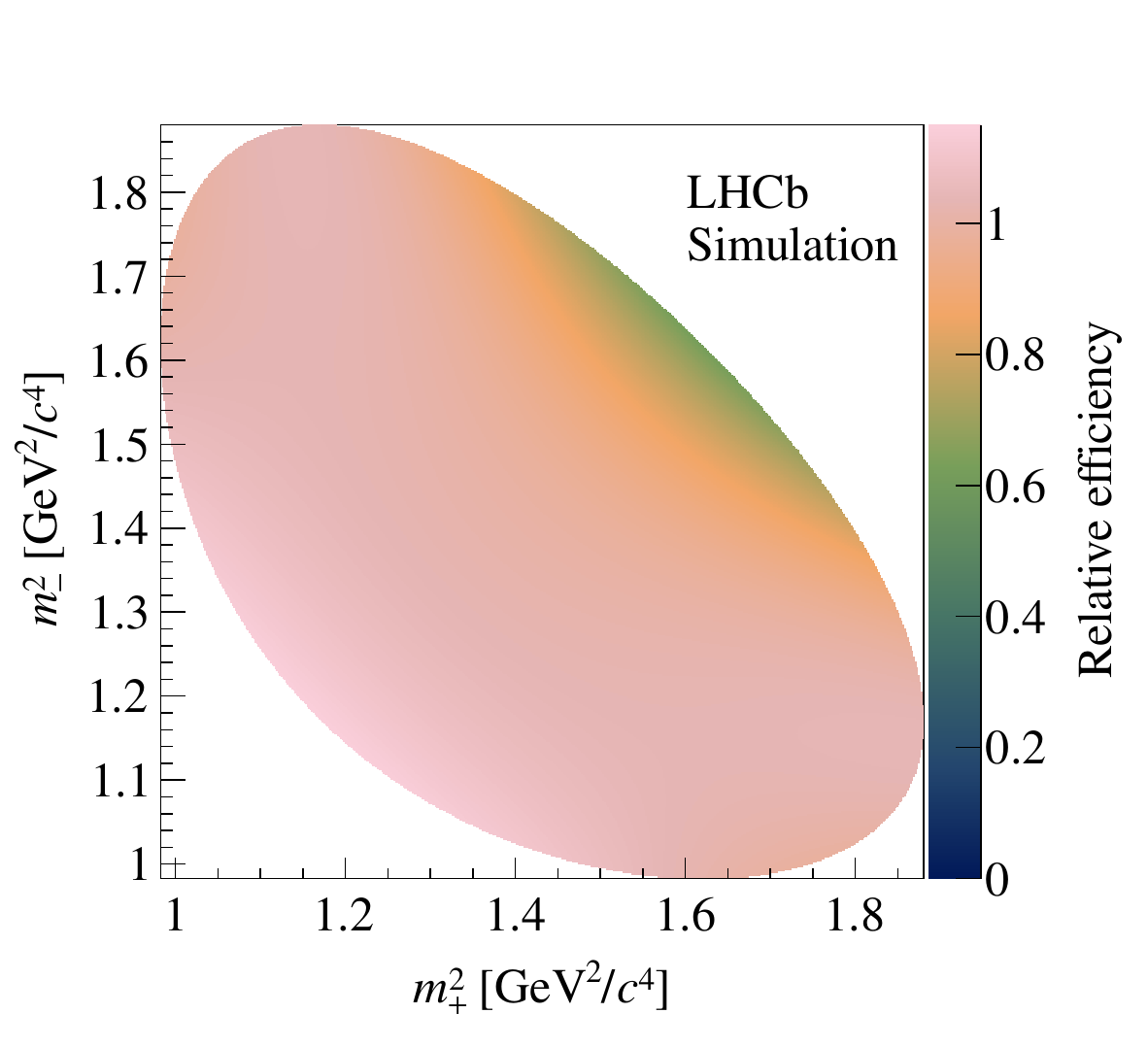}
    \caption{Efficiency distributions of $\decay{\Bpm}{D\pipm}$ decays with (top) $\decay{D}{\KS\pip\pim}$ and (bottom) $\decay{D}{\KS\Kp\Km}$ decays for the (left)~\emph{long} and (right)~\emph{downstream} categories of the $\KS$ candidate.
    The average efficiencies over the Dalitz plot are scaled to unity.}
    \label{fig:lhcb_eff}
\end{figure}

Relative efficiency distributions are described as functions of the Dalitz-plot coordinates and obtained from simulated signal samples.
The $\decay{D}{\KShh}$ decays in both \besiii and \lhcb simulation samples are generated with a uniform phase-space distribution; thus, the Dalitz-plot distributions post selection are proportional to the efficiencies across the Dalitz plot, 
which include the effects of detector acceptance, reconstruction and selection.
Efficiency distributions are obtained by fitting each of the Dalitz-plot distributions of simulated signal samples with a two-dimensional polynomial function, with terms up to fourth order.
In addition, extra per-event efficiency weights are introduced to correct the difference between data and simulation samples for $D$ decays in \besiii. The main discrepancies exist in the tracking efficiency, the particle-identification (PID) efficiency of $K\pi$ separation, and the reconstruction efficiency of $\KS$ candidates. They are investigated using a control sample of $D\Db$ hadronic events~\cite{BESIII:2025nsp}. Subsequently, corrections are applied to the signal decay modes: $\KS\pi^{+}\pi^{-}$, $\KS\pipm[\pimp]$, $\KS(\to\piz[\piz])\pip\pim$, $[\KL]\pi^{+}\pi^{-}$, $\KS \Kp\Km$, $\KS\Kpm[\Kmp]$, and $[\KL] \Kp\Km$. For tag modes, these differences largely cancel out in the ratio of ST to DT yields for fully reconstructed tag modes, and are found to be negligible for the partially reconstructed tag modes, such as $Ke\nu_{e}$, $\KL\pi^{0}$, and $\KL\pi^{0}\pi^{0}$.
Figure~\ref{fig:bes_eff} shows example efficiency distributions from \besiii simulation.
Figure~\ref{fig:lhcb_eff} shows the efficiency distributions from $\decay{\Bpm}{D\pipm}$ simulation at \lhcb, which are also used in the $\decay{\Bpm}{D\Kpm}$ decays,
as the Dalitz-plot distributions of simulated samples are found to be consistent between $\decay{\Bpm}{D\pipm}$ and $\decay{\Bpm}{D\Kpm}$ decays.

\begin{figure}[!tb]
    \centering
    \includegraphics[width=0.49\linewidth]{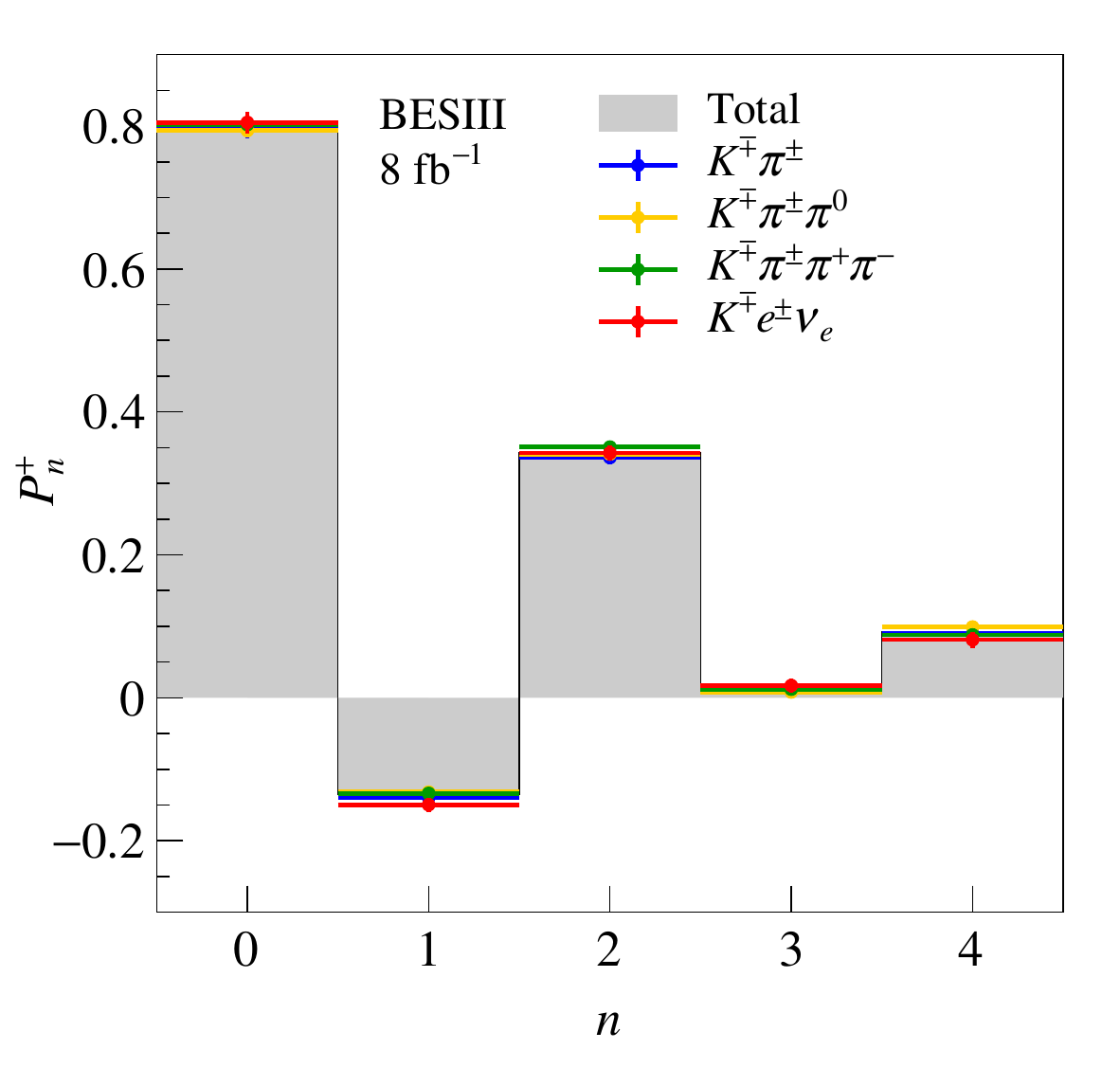}
    \includegraphics[width=0.49\linewidth]{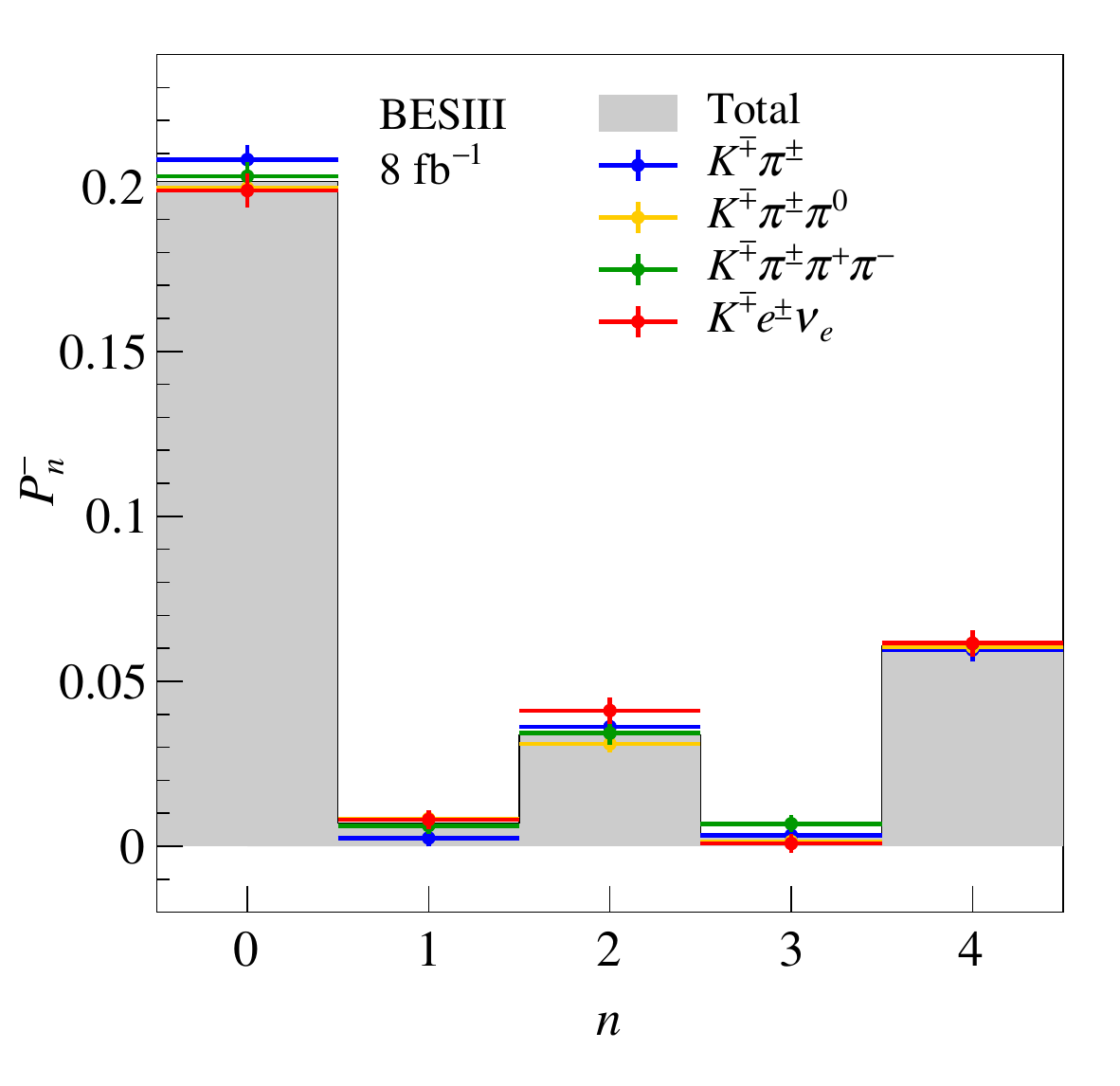}
    \caption{Flavored parameters (left) $P_n^{+}$ and (right) $P_n^{-}$ of the $\decay{\Dz}{\KS\pip\pim}$ decay from different tags. The total observables are obtained from combining those of all tags. The observables are normalized to ${P_0^{+} + P_0^{-} = 1}$.}
    \label{fig:ksflav_coef}
\end{figure}

\begin{figure}[!tb]
    \centering
    \includegraphics[width=0.49\linewidth]{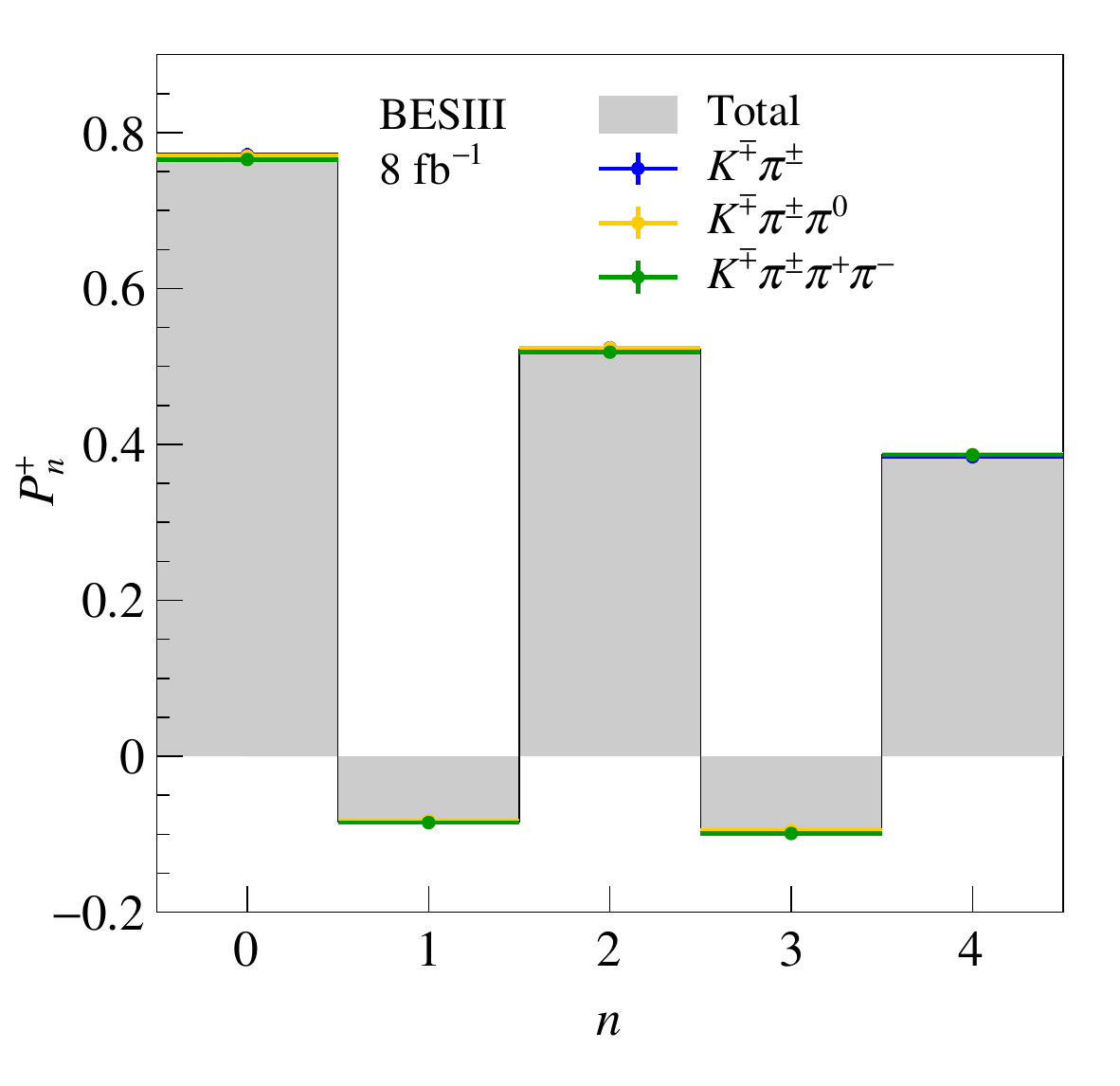}
    \includegraphics[width=0.49\linewidth]{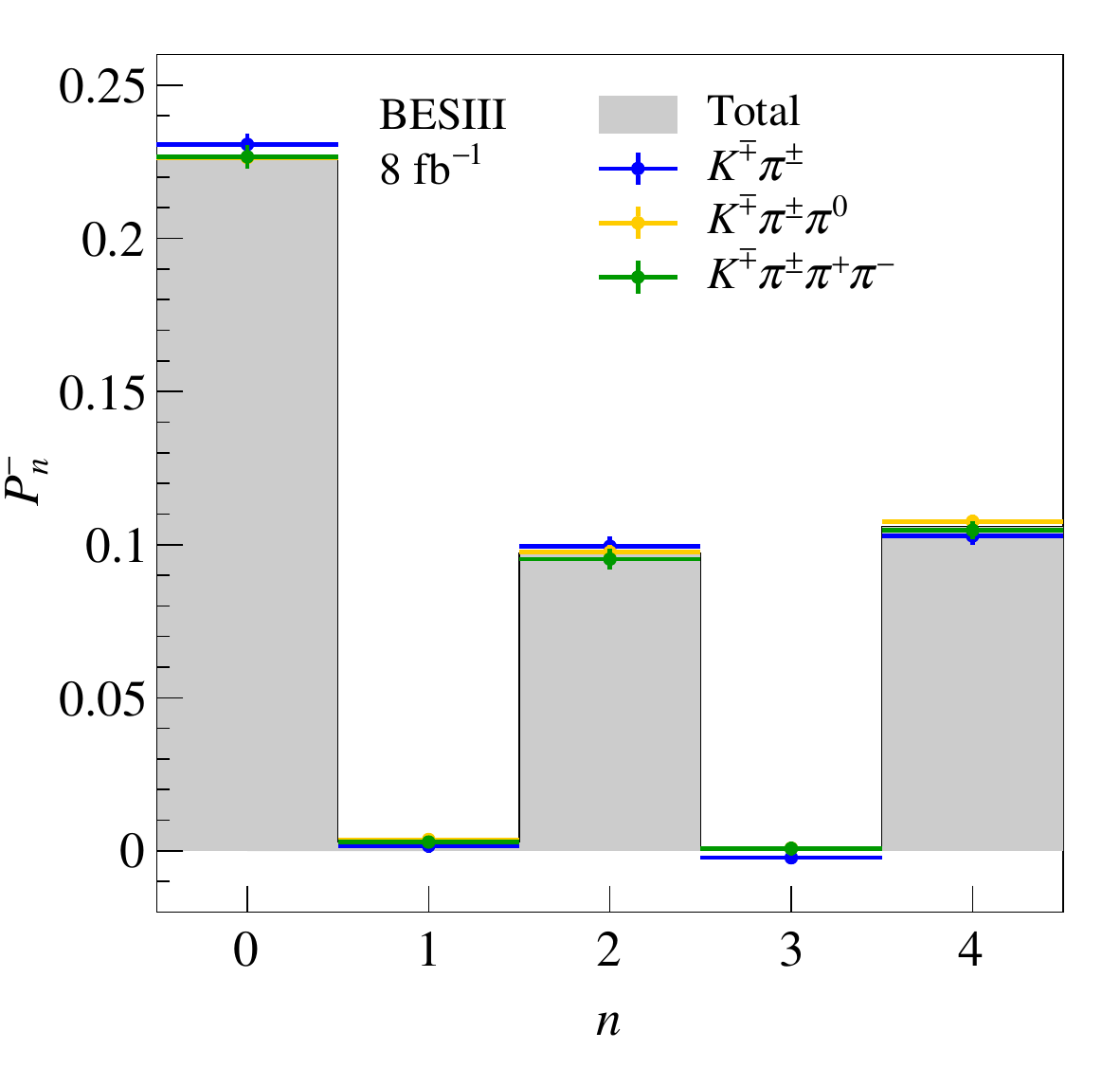}
    \caption{Flavored parameters (left) $P_n^{+}$ and (right) $P_n^{-}$ of the $\decay{\Dz}{\KL\pip\pim}$ decay from different tags. The total observables are obtained from combining those of all tags. The observables are normalized to ${P_0^{+} + P_0^{-} = 1}$.}
    \label{fig:klflav_coef}
\end{figure}

The weighted signal yields from flavor tags are computed by
\begin{equation}
    \Nobs_{n} = \sum_{i\in \Dz} w_{n,i}/\epsilon_i, \quad \Nobsbar_{\!n} = \sum_{i \in \Dzb} w_{n,i}/\epsilon_i \,,
    \label{eq:D_flav_coef_calc}
\end{equation}
where the weights are obtained from the functions $w_n$ defined in Eq.~\ref{eq:wp_weights}, efficiency distributions $\epsilon$ and the Dalitz-plot coordinates of candidate $i$.
The computation of $\Nobs_n$ ($\Nobsbar_{\!n}$) is performed by summing over $\Dz$ ($\Dzb$)-dominated $\KS h^+ h^-$ data, where the tag side decays into DCS (CF) $\Dz$ final states, \eg $\Kp\pim$ ($\Km\pip$).
These observables from the flavor tags are used to compute the flavored parameters, $P_n^{\pm}$.
The DCS contributions in these channels are corrected according to Eq.~\ref{eq:flav_coef},
where the strong-phase parameters $C_n, S_n$ are fixed to model predictions and the hadronic parameters $r_D, \delta_D$ and $R_D$ are fixed to values that are listed in Table~\ref{tab:flav_strong_phase}.
The corrected parameters $P_n^{\pm}$ from different tags, as shown in Figs.~\ref{fig:ksflav_coef} and \ref{fig:klflav_coef} for $\decay{D}{\KS\pip\pim}$ and $\decay{D}{\KL\pip\pim}$ decays, respectively,
are found to be consistent.

\begin{table}[tb]
    \centering
    \caption{Parameters of the flavor tags that are used to correct flavor observables, taken from Ref.~\cite{Ablikim2021}.}
    \label{tab:flav_strong_phase}
    \begin{tabular}{c|c r @{$\;$} c @{$\;$} l  r @{$\;$} c @{$\;$} l }
        \hline
        Tag & $r_D~(\times 10^{-2})$ & \multicolumn{3}{c}{$\;\delta_D~({\degrees})$} & \multicolumn{3}{c}{$R_D$} \\
        \hline
        $K\pi$       & $5.87 \pm 0.02$ & $190.2$ &$^{+}_{-}$ &$^{3.6}_{4.8}$ &  & $1$ & \\
        $K\pi\pi^0$  & $4.41 \pm 0.11$ & $196$ &$\pm$ &11 & $0.79$ & $\pm$ & $0.04$ \\
        $K\pi\pi\pi$ & $5.50 \pm 0.07$ & $161$ &$^{+}_{-}$ &$^{28}_{18}$ & $0.44$ & $^{+}_{-}$ &$^{0.10}_{0.09}$\\
        \hline
    \end{tabular}
\end{table}

Similarly, weighted yields from \CP tags are computed as
\begin{equation}
    \Nobs_{n} = \sum_{i} w_{n,i}/\epsilon_i, \quad \Nobsbar_{\!n} = \sum_{i} \overline{w}_{n,i}/\epsilon_i\,, \label{eq:D_CP_coef_calc}
\end{equation}
and four combinations of weighted yields of self-conjugated tags are computed by combining per-event weights from the phase space of both sides 
\begin{equation}
    \Nobs_{n_1n_2}^{s_1s_2} = \sum_i \left.\left(\left\{w_{n_1,i_1}, \overline{w}_{n_1, i_1}\right\}/\epsilon_{i_1}\right)
    \cdot
    \left(\left\{w_{n_2, i_2}, \overline{w}_{n_2, i_2}\right\}/\epsilon_{i_2}\right)\right.,
\end{equation}
where the first (second) set of weights is based on the Dalitz-plot coordinates of the first~(second) $D$ meson.

By summing over the same data samples, the observables are naturally correlated with each other.
The covariance matrices between two weighted sums, $\sum_i w_1^i$ and $\sum_i w_2^i$, are estimated by Poisson bootstrapping with $V = \sum_i w_1^iw_2^i$.
Therefore, the covariance between observables of Eq.~\ref{eq:D_CP_coef_calc}, \eg $\Nobs_{n_1}$ and $\Nobs_{n_2}$, are
\begin{equation}
    {\rm cov}(\Nobs_{n_1}, \Nobs_{n_2}) = \sum_i w_{n_1, i}w_{n_2, i}/\left(\epsilon_i\right)^2,
    \label{eq:coef_cov}
\end{equation}
and the same computation is applied to other combinations.
For the $\KSL\pi^{+}\pi^{-}$ signal samples, covariances between $C\!P$-tag observables are computed by Eq.~\ref{eq:coef_cov}. 
Meanwhile, the correlations between self-conjugated tag observables are derived from pseudoexperiments with a much larger sample size, which are generated with the amplitude model as the correlations from data are found to bias the results under current statistics.
The covariance estimates in both \CP and self-conjugate tag observables with $\KSL\Kp\Km$ are also computed from an ensemble due to low statistics.
The uncertainties of those observables are calculated from data as a special case of Eq.~\ref{eq:coef_cov}, in which the two weights are identical, reducing the expression to the sum of squared weights.

The weighted yields from \lhcb data are computed by
\begin{equation}
   \smash{\optbar{\Nobs}}_{\!n}^{\pm} = \sum_{i\in \Bpm\to D h^\pm} \optbar{w}_{\hspace*{-0.2em}n, i}f_i/\epsilon_i, 
   \label{eq:lhcb_coef}
\end{equation}
which includes the extra signal weight, $f_i$.
The covariance between \lhcb observables are analogously computed by Eq.~\ref{eq:coef_cov}, \eg between $\Nobs_{n_1}^{-}$ and $\Nobsbar_{\!n_2}^{-}$, as
\begin{equation}
    {\rm cov}(\Nobs_{n_1}^{-}, \Nobsbar_{\!n_2}^{-}) = \sum_{i\in\Bm\to Dh^{-}} w_{n_1, i}\overline{w}_{n_2, i}\left(f_i\right)^2/\left(\epsilon_i\right)^2.
    \label{eq:lhcb_coef_cov}
\end{equation}

\begin{figure}[tb]
    \begin{center}
        \includegraphics[width=0.49\textwidth]{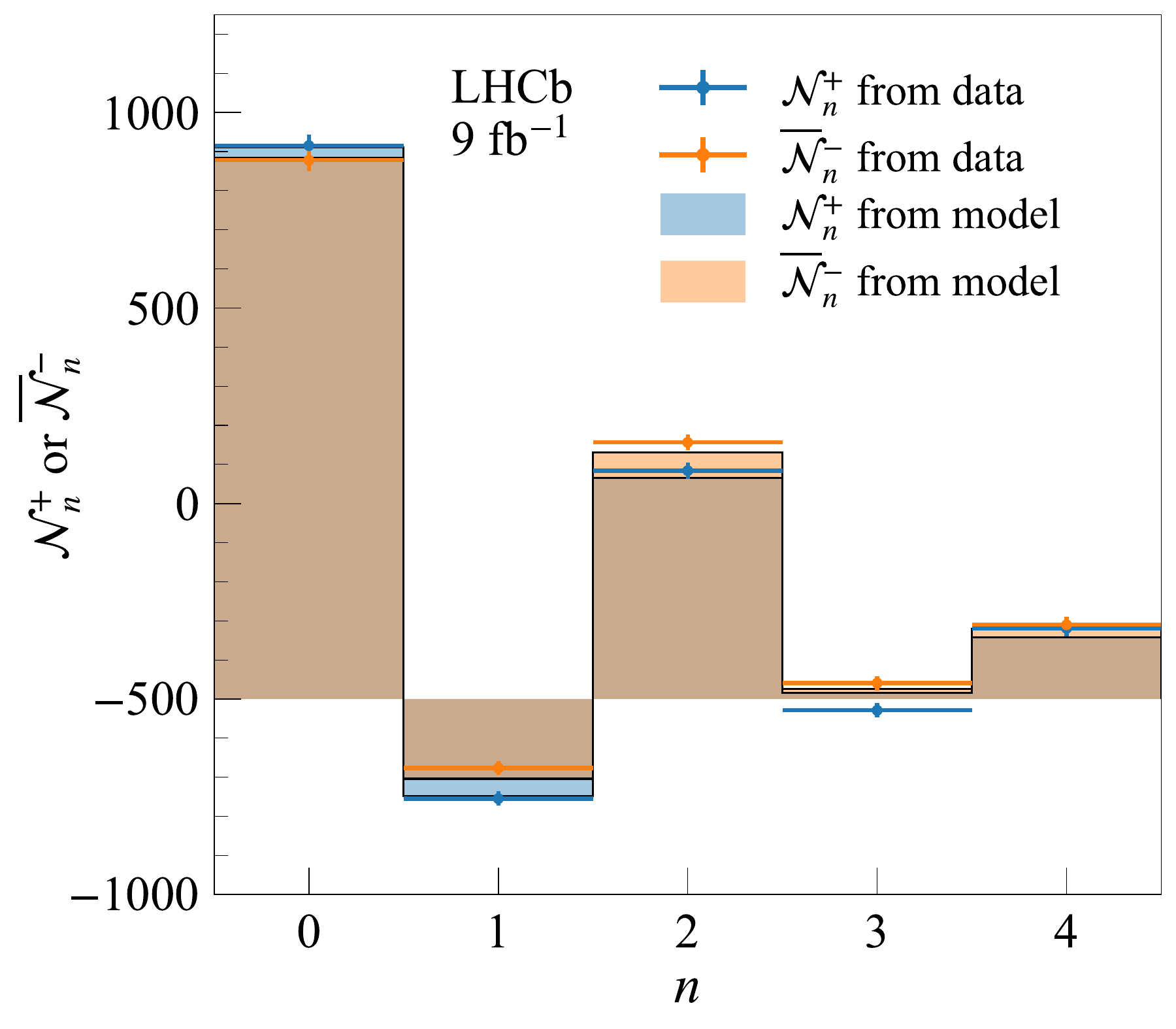}
        \includegraphics[width=0.49\textwidth]{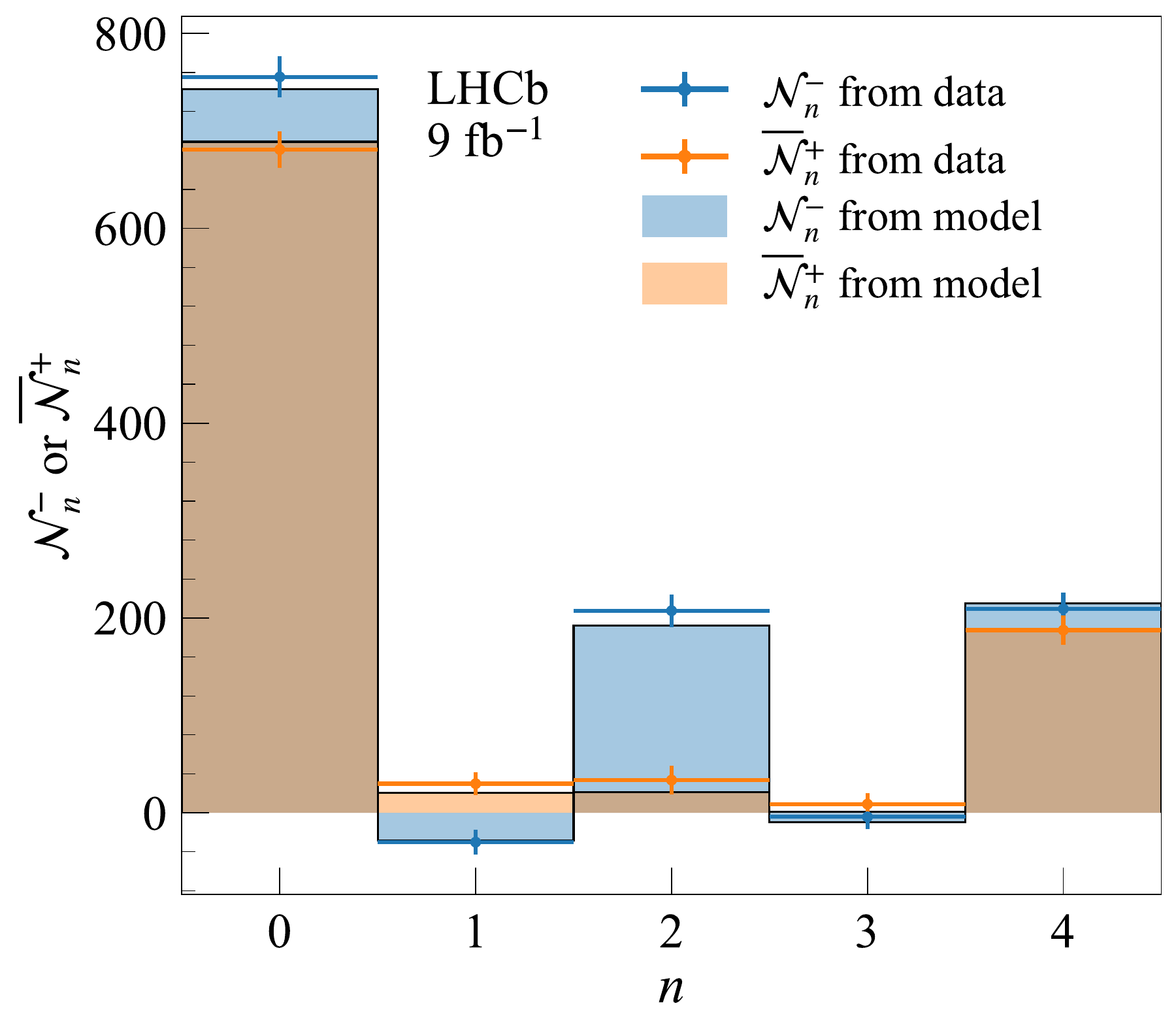}
        
        \includegraphics[width=0.49\textwidth]{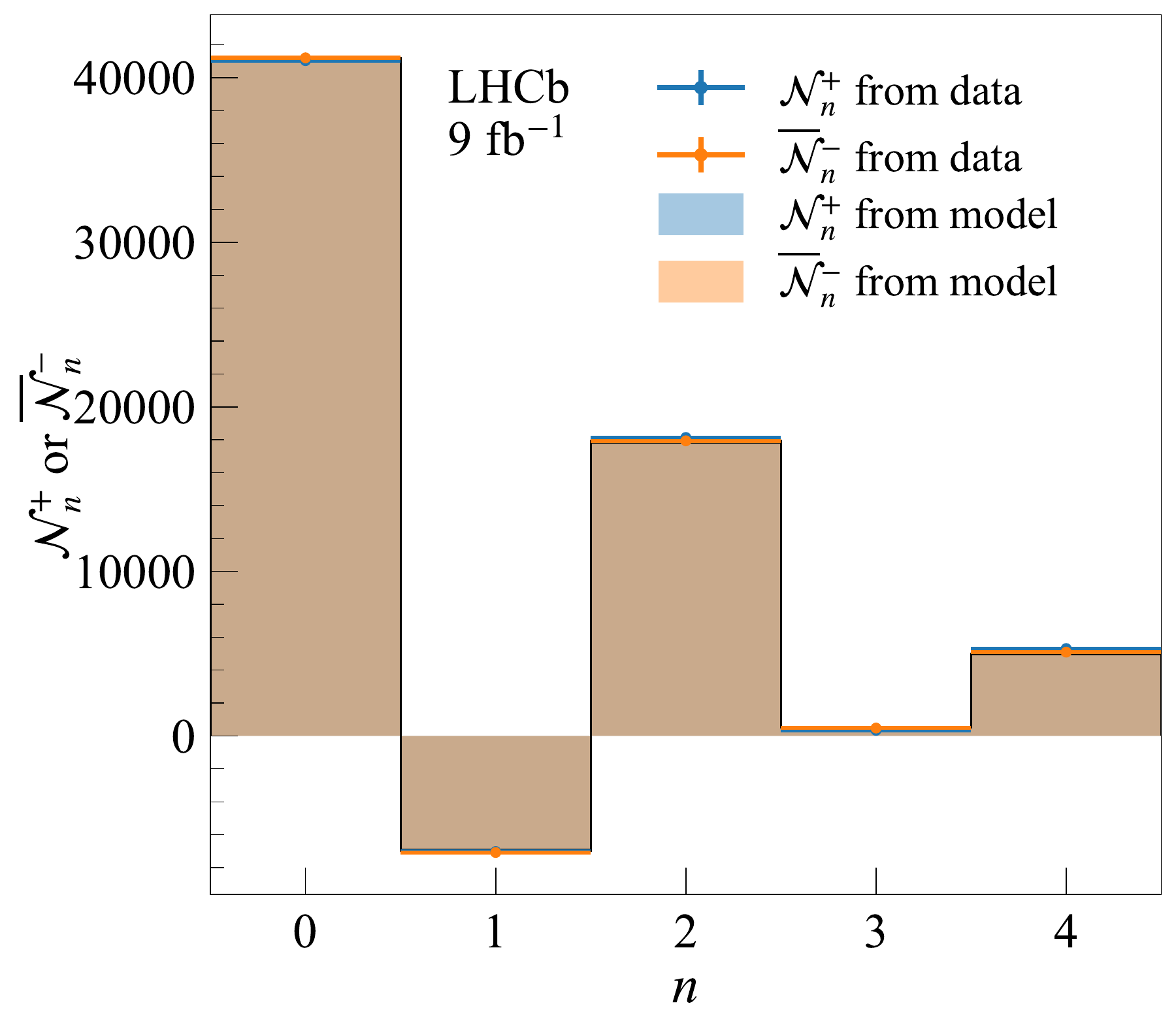}
        \includegraphics[width=0.49\textwidth]{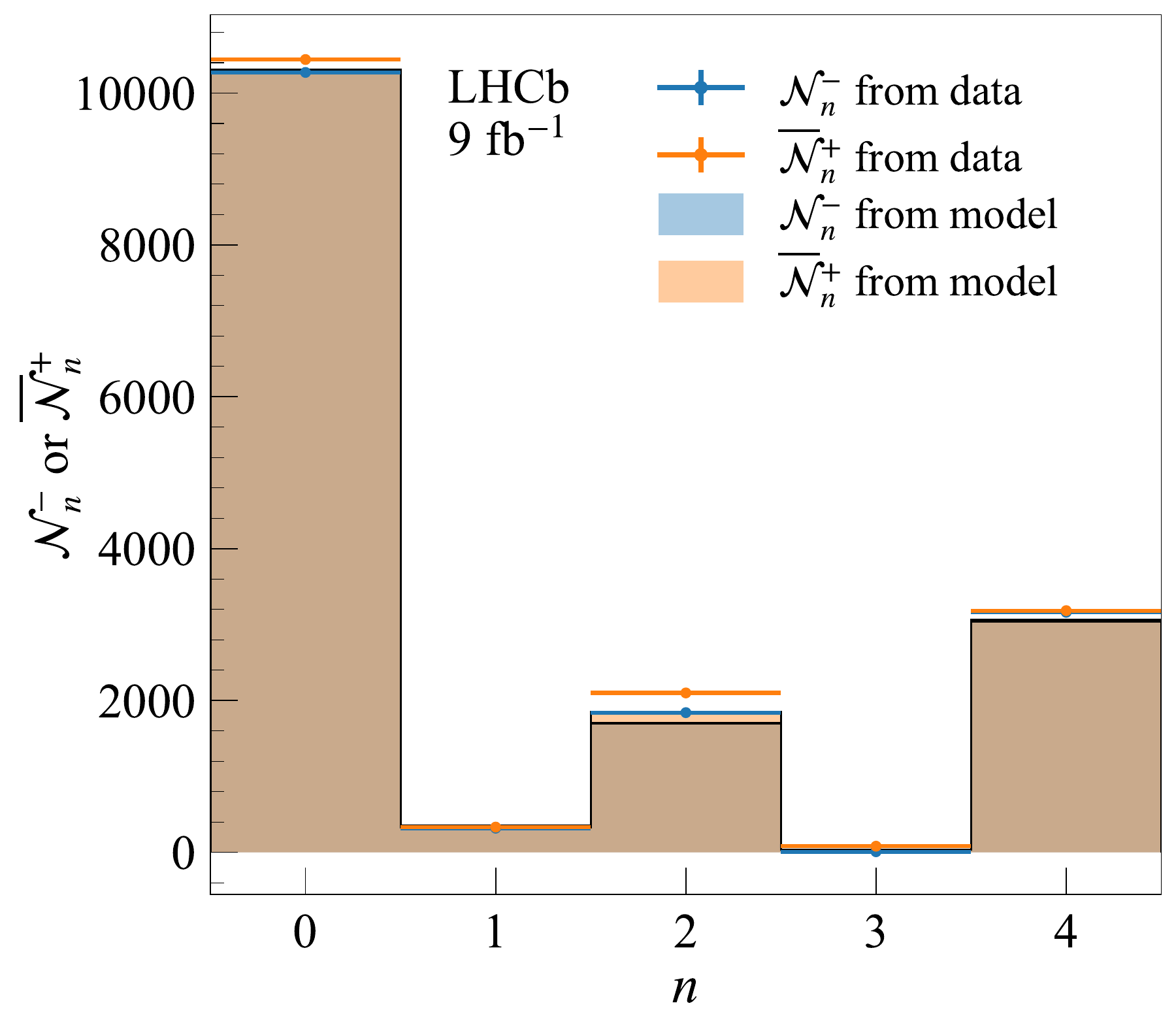}
    \end{center}
    \caption{Observables from the (top) $\decay{\Bpm}{D\Kpm}$ and (bottom) $\decay{\Bpm}{D\pipm}$ data sample with $\decay{D}{\KS\pip\pim}$ decays in the \emph{downstream} category.
    The \CP-conjugated observable pairs, (left)~$\Nobs_{n}^{+}$ or $\Nobsbar_{\!n}^{-}$ and (right) $\Nobs_{n}^{-}$ or $\Nobsbar_{\!n}^{+}$, are plotted together to visualize \CP asymmetries.}
    \label{fig:lhcb_coef}
\end{figure}

Figure~\ref{fig:lhcb_coef} shows the weighted yields obtained from the $\decay\Bpm{D h^{\pm}}$, $\decay{D}{\KS\pip\pim}$ data in the \emph{downstream} category. The model predictions are computed from the amplitude model measured in Refs.~\cite{BaBar:2018agf,BaBar:2018cka,BaBar:2010nhz} and physics parameters measured in Ref.~\cite{LHCb-PAPER-2020-019}.
Sizable \CP-violation effects are seen in the observables from the $\decay{\Bpm}{D\Kpm}$ data, 
while much smaller \CP asymmetries are present in the $\decay{\Bpm}{D\pipm}$ channel due to the small value of $\rdpi$.

\section{Fit to observables from data}
\label{sec:CPfit}

The parameters of interest are extracted by the least-squares approach,
where a $\chi^2$ test statistic is constructed with the observables from data.
This step is referred to as the \emph{\CP fit}.
In Sect.~\ref{sec:CPjointfit}, a joint fit to both BESIII and LHCb observables is performed to determine the \CP observables,
while in Sect.~\ref{sec:CPstrongphasefit}, a fit using only BESIII observables is carried out to determine the strong-phase parameters independently.

\subsection{Joint fit to determine $\boldsymbol{\CP}$ observables}
\label{sec:CPjointfit}

For each \CP tag in \besiii data, the ${2\times(2M_{h'}+1)}$ expected observables are computed with Eq.~\ref{eq:cp_coef} as
\begin{equation}
    \begin{aligned}
    \mean{\Nobs_n} = h_{\CP}\left[P_n^{+} + P_n^{-} - 2s_K(2F_{+} - 1)(-1)^nC_n\right], \\
    \mean{\Nobsbar_{\!n}} = h_{\CP}\left[P_n^{-} + P_n^{+} - 2s_K(2F_{+} - 1)(-1)^nC_n\right], 
    \end{aligned}
\end{equation}
where ${h_{\CP} = N_{\CP\pm}^{\rm ST}/2N_{f}^{\rm ST}}$ is the normalization factor, 
with $N_{\CP}^{\rm ST}$ and $N_f^{\rm ST}$ defined as efficiency-corrected ST yields of the \CP tag and flavor tags, respectively.
The ST yields and efficiencies are fixed to those measured in Ref.~\cite{BESIII:2025nsp}.
The flavor parameters $P_n^{\pm}$ are fixed to those computed from flavor tags, 
and the strong-phase parameters $C_n$ are allowed to vary.
The $\chi^2$ test statistic from \CP tags is constructed as
\begin{equation}
    \chi^2_{\CP\text{-tag}} = \left[\Nobs_n - \mean{\Nobs_n}, \Nobsbar_{\!n} - \mean{\Nobsbar_{\!n}}\right]^{\rm T}V_{\CP}^{-1}\left[\Nobs_n - \mean{\Nobs_n}, \Nobsbar_{\!n} - \mean{\Nobsbar_{\!n}}\right].
\end{equation}
Here, the observed weighted yields $\Nobs_n$ are computed from the weighted sums over the relevant data samples, in which the background observables, computed from the weighted sums over simulated background samples, are subtracted.
The covariance matrices $V_{\CP}$ are computed from the weighted sums as described in Sect.~\ref{sec:coefficient}.

The ${4\times(2M_{h'} + 1)\times (2M_{h''}+1)}$ expected observables of the self-conjugate tags are computed from Eq.~\ref{eq:dbldlz_coef} as
\begin{equation}
    \mean{\Nobs_{n_1n_2}^{s_1s_2}} = h_{DD}\left[P_{n_1}^{s_1}P_{n_2}^{{s_2}} + P_{n_1}^{\overline{s}_1}P_{n_2}^{\overline{s}_2}
     - 2s_{K_1}s_{K_2}(-1)^{n_1}(-1)^{n_2}(C_{n_1}C_{n_2} + s_1s_2S_{n_1}S_{n_2})\right],
\end{equation}
where $h_{DD} = \kappa N_{DD} /(2N_{f,1}^{\rm ST}N_{f,2}^{\rm ST})$ is the normalization factor.
The factor $\kappa$ accounts for the different reconstruction methods between the two $D$ mesons,
with ${\kappa = \left. 2\BF(\decay{\KS}{\piz\piz})\middle/\BF(\decay{\KS}{\pip\pim})\right.}$ for the $\KS(\to\piz[\piz])\pip\pim$ \vs $\KS\pip\pim$ channel,
${\kappa = 2}$ for the channels where the two $D$ mesons decay to different final states,
and ${\kappa = 1}$ for the channels where the two $D$ mesons decay to the same final state.
The branching fractions $\BF(\decay{\KS}{\piz\piz})$ and $\BF(\decay{\KS}{\pip\pim})$ are fixed to their known values~\cite{PDG2024}.
The number of $D\Db$ pairs are measured to be ${N_{DD} = (28.66\pm 0.25)\times 10^6}$~\cite{BESIII:2024lbn},
and $N_{f, 1(2)}^{\rm ST}$ correspond to the efficiency-corrected ST yield of flavor tags of the first (second) $D$-meson decays. The $\chi^2$ test statistic of the self-conjugate tags is constructed in the same way as the \CP tags,
where 4 sets of observables, $\Nobs_{n_1n_2}^{\pm\pm}$, are included.

An additional weak constraint from the amplitude model is included to help determine the strong-phase parameters, following the same approach as in Ref.~\cite{BESIII:2025nsp}.
Differences in strong-phase parameters between $\decay{\Dz}{\KLhh}$ and $\decay{\Dz}{\KShh}$ decays, $\Delta C_n, \Delta S_n$, are constrained to those from the amplitude models.
The amplitude models of the $\decay{\Dz}{\KShh}$ decays are measured by the \belle and \babar experiments~\cite{BaBar:2018agf,BaBar:2018cka,BaBar:2010nhz},
while the amplitude models of the $\decay{\Dz}{\KLhh}$ decays are derived from the corresponding ${\KShh}$ models and the assumption of U-spin symmetry.
Further U-spin breaking effects are taken into account in $\decay{\Dz}{\KL\pip\pim}$ decays from the \besiii measurement~\cite{BESIII:2020khq}, 
and estimated with the approach described in Ref.~\cite{CLEO:2010iul} for $\decay{\Dz}{\KL\Kp\Km}$ decays as no measurement is available at the moment.
Uncertainties of $\Delta C_n, \Delta S_n$, denoted as $\delta\Delta C_n, \delta\Delta S_n$, are estimated from pseudoexperiments,
in which the amplitude model is altered by sampling input parameters within their measured uncertainties in Refs.~\cite{BaBar:2018agf,BaBar:2018cka,BaBar:2010nhz},
or by using alternative amplitude models that are measured independently~\cite{Belle:2010xyn,BESIII:2022qvy}.
The $\chi^2$ test statistic of this constraint, \eg, for $\Delta C_n$, is constructed as
\begin{equation}
    \chi^2_{\Delta C_n} = \sum_{n = 0}^{2M_{h}+1}\left(\frac{C_n^{\prime} - C_n - \mean{\Delta C_n}}{\delta\Delta C_n}\right)^2,
\end{equation}
where $C_n$ and $C_n^{\prime}$ are the measured strong-phase parameters of the $\decay{\Dz}{\KShh}$ and $\decay{\Dz}{\KLhh}$ decays, respectively.
Furthermore, the expected values of $\mean{\Delta C_n}$ are computed from the amplitude models,
and $M_h$ denotes the highest Fourier order in the fit with ${h = \pi, K}$.

The $\chi^2$ test statistic constructed from \besiii data is
\begin{equation}
    \chi^2_{C} 
    = \chi^2_{\KSLhh~\text{\textit{vs.}}~\CP\text{-tag}} + \chi^2_{\KShh~\vs~\KSL h^{\prime\prime+}h^{\prime\prime-}} + \chi^2_{\Delta(C,S)_n},
\label{eq:beschi2}
\end{equation}
in which $\chi^2_{\KSLhh~\text{\textit{vs.}}~\CP\text{-tag}}$ contains 12 categories: 4 signal $D$ decays and 3 tag modes summarized in Table~\ref{tab:usedtags}. The $\chi^2_{\KShh~\vs~\KSL h^{\prime\prime+}h^{\prime\prime-}}$ term contains 9 self-conjugate tags as explained in Sect.~\ref{sec:selection}.

For \lhcb data, the $\chi^2$ test statistic can be constructed by the same approach, 
using ${2\times(2M_{h'}+1)}$ expected signal observables $\mean{\smash{\optbar{\Nobs}}^{\pm}}$ computed from Eq.~\ref{eq:Bfou} in each category,
the weighted signal yields $\smash{\optbar{\Nobs}}^{\pm}$ from Eq.~\ref{eq:lhcb_coef}, and their covariance matrices $V_{B^{\pm}}$ from Eq.~\ref{eq:lhcb_coef_cov},
\begin{equation}
    \chi^2_{B} = \left[\Nobs_n^{\pm} - \mean{\Nobs_n^{\pm}}, \Nobsbar_{\!n}^{\pm} - \mean{\Nobsbar_{\!n}^{\pm}}\right]^{\rm T}V_{\Bpm}^{-1}\left[\Nobs_n^{\pm} - \mean{\Nobs_n^{\pm}}, \Nobsbar_{\!n}^{\pm} - \mean{\Nobsbar_{\!n}^{\pm}}\right].
\end{equation}
This construction represents 16 categories of \lhcb data.
The normalization factors of each category, $h_{\Bpm}$, ${2\times(2M_h+1)}$ flavor parameters of \lhcb data in each $D$ decay and $\KS$ reconstruction type, $P_n^{\pm}$, and \CP observables of $\decay{\Bpm}{D\Kpm}$ decays, $\xpmdk$ and $\ypmdk$, are free parameters in the fit.
The $\CP$ observables of $\decay{\Bpm}{D\pipm}$ decays, as explained in Sect.~\ref{sec:analysis_strategy}, are described by four $\decay{\Bpm}{D\Kpm}$ parameters and two $\decay{\Bpm}{D\pipm}$ parameters, which are also free parameters.
Finally, the strong-phase parameters are shared between $\chi^2_C$ and $\chi^2_B$, so that the $\CP$ observables can be extracted from a joint fit to both \besiii and \lhcb data. Comparing with those fitted with only the \besiii data at the same orders $M_{\pi} = 2$ and $M_K = 1$, strong-phase parameters from the joint fit have slightly lower uncertainties in $S_n$, indicating extra sensitivity provided by \lhcb data as presented in Appendix~\ref{app:strong_phase_alt}. 
The joint fit is found to have a good quality, with the minimized $\chi^2$ corresponding to a $p$-value of 1\%.

\begin{figure}
    \centering
    \includegraphics[width=0.9\linewidth]{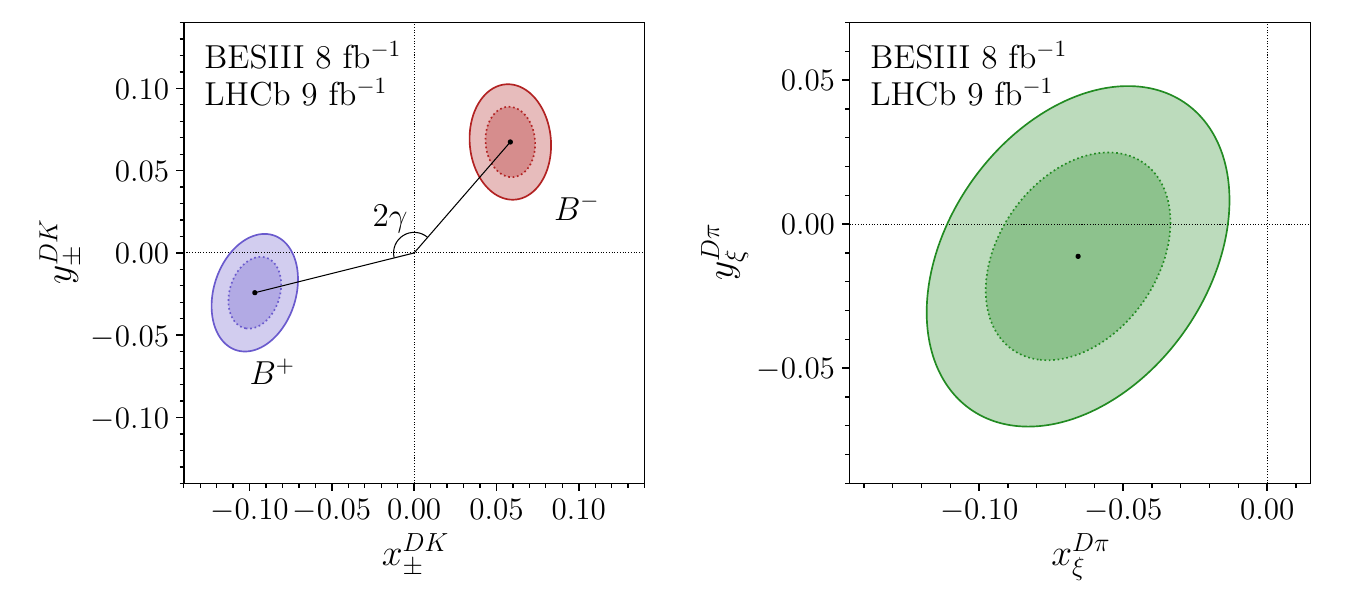}
    \caption{Results of \CP observables from the joint fit. 
    The contours correspond to 68\% and 95\% confidence levels of observable pairs, (left) $(\xpdk, \ypdk)$ in blue and $(\xmdk, \ymdk)$ in red, and (right) $(\xxi, \yxi)$.}
    \label{fig:cp_obs_res}
\end{figure}

The \CP observables from the joint fit are shown in Fig.~\ref{fig:cp_obs_res},
including the two-dimensional likelihood contours for the pairs of observables ${(\xmdk, \ymdk)}$, ${(\xpdk, \ypdk)}$ and ${(\xxi, \yxi)}$.
The contours are determined from the statistical uncertainties and correlations from the joint fit,
and those from systematic uncertainty studies as detailed in Sect.~\ref{sec:systematic}.
The likelihood distributions are assumed to be Gaussian-like and were validated by the binned measurement~\cite{LHCb-PAPER-2020-019}.
The lengths of the two vectors from the origin to $(\xpmdk, \ypmdk)$ are the individual $r_{\Bpm}^{DK}$ values from $\decay{\Bpm}{D\Kpm}$ decays.
The nonzero opening angle between these two vectors, which equals $2\gamma$, shows a clear sign of \CP violation.

Several cross-checks have been performed to validate the fit procedure.
The fits to data are performed with higher Fourier orders, ${(M_{\pi}, M_{K}) = (2, 2)}$, $(3, 1)$ and $(3, 2)$,
and the results of \CP observables and strong-phase parameters are found to be consistent.
Fits are also performed using only the \lhcb data with both the binned and unbinned approaches, in which the unbinned strong-phase parameters are fixed to those measured in Fig.~\ref{fig:strong_phase_res_k0hh},
and the binned strong-phase parameters are fixed to those measured in Refs.~\cite{BESIII:2025nsp,BESIII:2020hpo}.
Correlation between the resultant \CP observables was studied with pseudoexperiments and was found to be around 0.8 for each of the observables.
Using the same pseudoexperiments their consistency in six-dimensional parameter space was found to correspond to a $p$-value of 39\%.

Searches for intrinsic bias are conducted by generating and fitting an ensemble of pseudoexperiments.
The yields and mass distributions are based on the fits to data in Sect.~\ref{sec:selection},
and the distributions across the phase space are generated from the baseline amplitude models~\cite{BaBar:2018agf,BaBar:2018cka,BaBar:2010nhz}.
Standardized-residual distributions of the \CP observables are found to be consistent with the standard normal distribution.

\subsection{Determination of strong-phase parameters}
\label{sec:CPstrongphasefit}

The strong-phase parameters $C_n, S_n$ serve as crucial inputs for future measurements of $\gamma$ with the novel approach.
To avoid potential correlations between future $\gamma$ measurements and this measurement, results of these parameters determined by a fit to \besiii data alone with the $\KSLpp$ and $\KSLkk$ channels combined are also reported in this analysis.
The Fourier orders are chosen to be one order higher than in the baseline with $M_{\pi} = 3$ and $M_K = 2$, so that the truncation choice does not limit future higher-precision measurements.
The strong-phase parameters are obtained with the $\chi^2$ test statistic as written in Eq.~\ref{eq:beschi2} and listed in Table~\ref{tab:strong_phase_res_comb_final}, where the first uncertainties are statistical and the second systematic. The correlations between each source of uncertainty are included in the Supplemental Material~\cite{supplemental}.
Figure~\ref{fig:strong_phase_res_k0hh} shows good agreement between the model predictions and fitted results of the strong-phase parameters, where the statistical and systematic uncertainties have been combined.

\begin{figure}[tb]
    \centering
    \includegraphics[width=\linewidth]{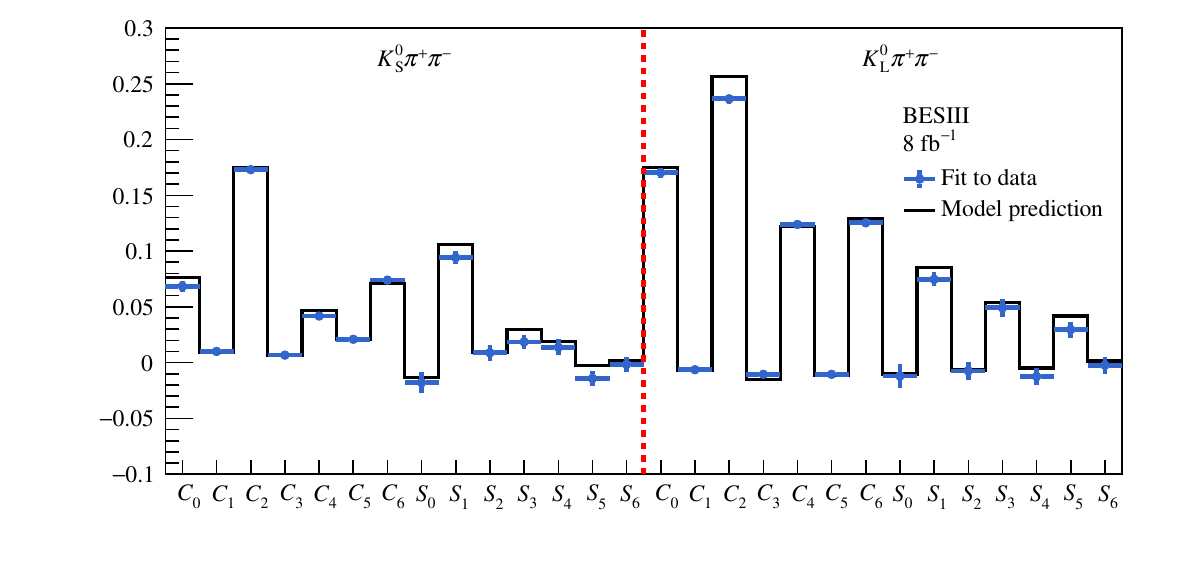}
    \includegraphics[width=\linewidth]{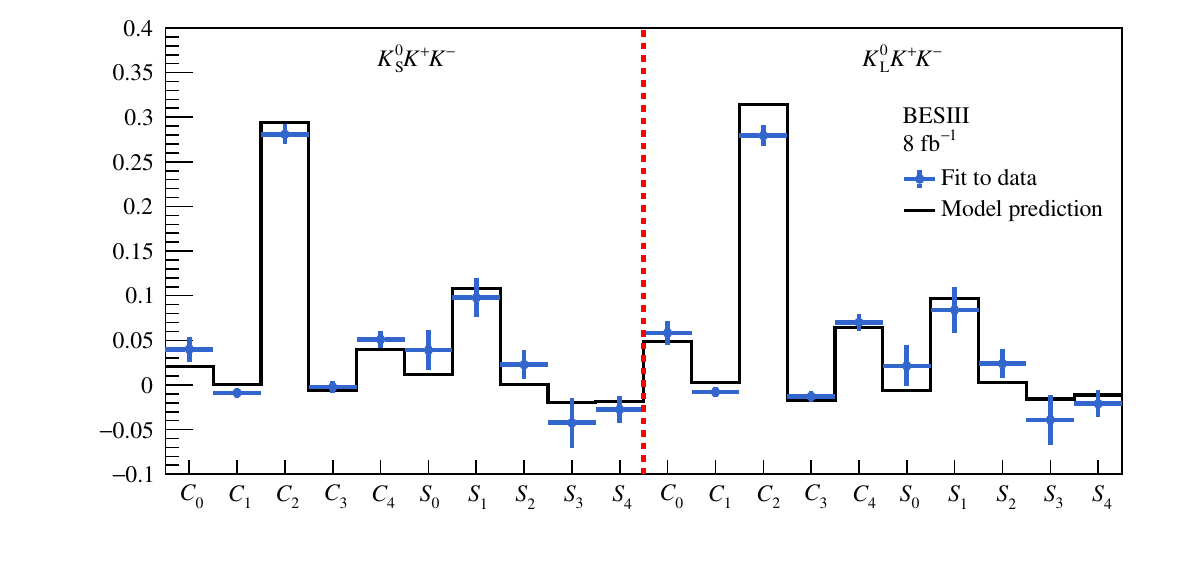}
    \caption{Strong-phase parameters of the (top) $\decay{D}{\KSL\pip\pim}$ and (bottom) $\decay{D}{\KSL\Kp\Km}$ decays from \besiii data.}
    \label{fig:strong_phase_res_k0hh}
\end{figure}

\begin{table}[tb]
    \caption{Results of the strong-phase parameters $C_n, S_n$ from \besiii-only data under Fourier orders $M_{\pi}=3$, $M_{K} = 2$.}
    \label{tab:strong_phase_res_comb_final}
    \begin{center}
        \begin{tabular}{c|c|rr}
            \hline
            $n$ & Weight & \multicolumn{1}{c}{$C_n$} & \multicolumn{1}{c}{$S_n$} \\
            \hline
            \multicolumn{4}{c}{$\KSpp$} \\
            \hline
             0 & $w^{\rm opt}$            & $0.0684 \pm 0.0038 \pm 0.0029$ & $-0.0179 \pm 0.0087 \pm 0.0037$ \\
             1 & $w^{\rm opt}\sin{\phi}$  & $0.0100 \pm 0.0023 \pm 0.0007$ & $0.0941 \pm 0.0051 \pm 0.0021$ \\
             2 & $w^{\rm opt}\cos{\phi}$  & $0.1730 \pm 0.0028 \pm 0.0014$ & $0.0089 \pm 0.0068 \pm 0.0021$ \\
             3 & $w^{\rm opt}\sin{2\phi}$ & $0.0067 \pm 0.0021 \pm 0.0007$ & $0.0186 \pm 0.0064 \pm 0.0013$ \\
             4 & $w^{\rm opt}\cos{2\phi}$ & $0.0418 \pm 0.0031 \pm 0.0012$ & $0.0138 \pm 0.0068 \pm 0.0021$ \\
             5 & $w^{\rm opt}\sin{3\phi}$ & $0.0209 \pm 0.0024 \pm 0.0007$ & $-0.0141 \pm 0.0067 \pm 0.0015$ \\
             6 & $w^{\rm opt}\cos{3\phi}$ & $0.0740 \pm 0.0030 \pm 0.0009$ & $-0.0016 \pm 0.0065 \pm 0.0015$ \\
             \hline
             \multicolumn{4}{c}{$\KLpp$} \\
             \hline
             0 & $w^{\rm opt}$            & $0.1703 \pm 0.0038 \pm 0.0015 $ & $-0.0120 \pm 0.0100 \pm 0.0027$ \\
             1 & $w^{\rm opt}\sin{\phi}$  & $-0.0063 \pm 0.0015 \pm 0.0006$ & $0.0747 \pm 0.0062 \pm 0.0016$ \\
             2 & $w^{\rm opt}\cos{\phi}$  & $0.2365 \pm 0.0034 \pm 0.0017 $ & $-0.0074 \pm 0.0079 \pm 0.0019$ \\
             3 & $w^{\rm opt}\sin{2\phi}$ & $-0.0105 \pm 0.0020 \pm 0.0007 $ & $0.0491 \pm 0.0080 \pm 0.0011$ \\
             4 & $w^{\rm opt}\cos{2\phi}$ & $0.1238 \pm 0.0030 \pm 0.0011$ & $-0.0124 \pm 0.0074 \pm 0.0015$ \\
             5 & $w^{\rm opt}\sin{3\phi}$ & $-0.0105 \pm 0.0022 \pm 0.0007$ & $0.0297 \pm 0.0070 \pm 0.0014$ \\
             6 & $w^{\rm opt}\cos{3\phi}$ & $0.1254 \pm 0.0031 \pm 0.0013 $ & $-0.0023 \pm 0.0077 \pm 0.0009$ \\
            \hline
            \multicolumn{4}{c}{$\KSkk$} \\
            \hline
            0 & $w^{\rm opt}$            & $0.0401 \pm 0.0111 \pm 0.0088$ & $0.0391 \pm 0.0251 \pm 0.0058$ \\
            1 & $w^{\rm opt}\sin{\phi}$  & $-0.0089 \pm 0.0055 \pm 0.0017$ & $0.0978 \pm 0.0203 \pm 0.0073$ \\
            2 & $w^{\rm opt}\cos{\phi}$  & $0.2809 \pm 0.0097 \pm 0.0059$ & $0.0227 \pm 0.0159 \pm 0.0028$ \\
            3 & $w^{\rm opt}\sin{2\phi}$ & $-0.0024 \pm 0.0063 \pm 0.0018$ & $-0.0424 \pm 0.0297 \pm 0.0060$ \\
            4 & $w^{\rm opt}\cos{2\phi}$ & $0.0509 \pm 0.0080 \pm 0.0054$ & $-0.0275 \pm 0.0163 \pm 0.0027$ \\
            \hline
            \multicolumn{4}{c}{$\KLkk$} \\
            \hline
            0 & $w^{\rm opt}$            & $0.0583 \pm 0.0112 \pm 0.0082$ & $0.0213 \pm 0.0254 \pm 0.0058$ \\
            1 & $w^{\rm opt}\sin{\phi}$  & $-0.0079 \pm 0.0055 \pm 0.0016$ & $0.0840 \pm 0.0242 \pm 0.0077$ \\
            2 & $w^{\rm opt}\cos{\phi}$  & $0.2794 \pm 0.0110 \pm 0.0054$ & $0.0238 \pm 0.0163 \pm 0.0028$ \\
            3 & $w^{\rm opt}\sin{2\phi}$ & $-0.0128 \pm 0.0064 \pm 0.0014$ & $-0.0391 \pm 0.0276 \pm 0.0056$ \\
            4 & $w^{\rm opt}\cos{2\phi}$ & $0.0701 \pm 0.0082 \pm 0.0050$ & $-0.0211 \pm 0.0152 \pm 0.0024$ \\
            \hline
        \end{tabular}
    \end{center}
\end{table}

Based on a pseudoexperiment study described in Sect.~\ref{sec:CPjointfit}, the average standardized residuals of three strong-phase parameters of $\decay{D}{\KSLpp}$ decays and half of all strong-phase parameters of $\decay{D}{\KSLkk}$ decays are found to deviate from zero.
Their biases are less than 30\% of the statistical uncertainties and are due to low statistics in data,
so are not used to correct central values of the strong-phase parameters, but assigned as a source of systematic uncertainty as described in Sect.~\ref{sec:systematic}.
The statistical uncertainties of five strong-phase parameters of $\decay{D}{\KSLkk}$ decays are found to be underestimated under the current sample size,
thus corrections of statistical uncertainties are applied to the fit results.

\section{Systematic uncertainties}
\label{sec:systematic}

The systematic uncertainties affecting this measurement are considered from two groups: those associated with the \besiii inputs and those related to \lhcb.
The uncertainties of the strong-phase parameters are summarized in Tables~\ref{tab:kspp_systematic_coef_besiii}--\ref{tab:klkk_systematic_coef_besiii},
while those for the \CP observables are listed in Table~\ref{tab:all_systematic_cpv}.

The dominant sources of systematic uncertainties from \besiii are identified as follows: biases estimated from pseudoexperiments; statistical fluctuations in the flavor-tag observables and ST yields; uncertainties associated with charm mixing and the hadronic parameters used for flavor tags; limited statistics of the uniform phase-space simulation samples for efficiency profiles, and the finite statistics of background samples for modeling the background shapes and yields. 

Most of these uncertainties can be addressed through dedicated pseudoexperiments. The sets of values for the flavor and charm-mixing parameters, ST yields, efficiency profiles, signal efficiencies, background models and yields are generated independently according to Gaussian distributions with means and standard deviations set to their baseline values and uncertainties. The \CP fit is then repeated for each varied set, and the standard deviation of the resulting pull distribution for each strong-phase parameter is assigned as the relative systematic uncertainty.

Since the DCS contributions in flavor tags and some of the correlation matrices depend on the amplitude models, the imperfection of the models can bias the strong-phase parameters. To assess this, alternative amplitude models are employed, and the differences in the extracted strong-phase parameters are assigned as the corresponding systematic uncertainties.

Another source of uncertainty originates from the tracking efficiency of pions from $\KS$ decays. This effect is due to the relatively low momentum and displaced topology of these secondary pions, making their tracking efficiencies less well constrained than those of prompt tracks. Although the overall uncertainties related to PID and tracking are found to be negligible after correcting the efficiency with the method described in Sect.~\ref{sec:coefficient}, residual mismodeling of the pions from $\KS$ decays can still affect the signal efficiencies. To quantify this, the signal efficiencies and efficiency profiles are varied according to the tracking-efficiency corrections derived from $\pi^{\pm}$ control samples. The resulting shifts in the refitted strong-phase parameters are taken as systematic uncertainties. 

Possible mismodeling of the detector resolution between data and simulation is taken into consideration. The efficiency profiles are derived separately using truth-level and reconstruction-level information of reconstructed events in the simulation samples. The differences in the extracted strong-phase parameters with these two kinds of efficiency profiles are treated as related systematic uncertainties.

For each source of systematic uncertainty from \besiii, its impact on the strong-phase parameters is studied by analyzing data or pseudoexperiments at $M_{\pi}=3$ and $M_K=2$. All sources of systematic uncertainties can be found in Tables~\ref{tab:kspp_systematic_coef_besiii}--\ref{tab:klkk_systematic_coef_besiii}.

\begin{table}[tb]
    \caption{Summary of all uncertainties on the strong-phase parameters of $\decay{\Dz}{\KS \pip\pim}$ decays from \besiii data. All uncertainties are quoted in units of $10^{-4}$.}
    \label{tab:kspp_systematic_coef_besiii}
    \begin{center}
    {\footnotesize
    \setlength\tabcolsep{4pt}
    \centering
    \begin{tabular}{l|rrrrrrrrrrrrrr}
        \hline
        Source & $C_{0}$ & $C_{1}$ & $C_{2}$ & $C_{3}$ & $C_{4}$ & $C_{5}$ & $C_{6}$ & $S_{0}$ & $S_{1}$ & $S_{2}$ & $S_{3}$ & $S_{4}$ & $S_{5}$ & $S_{6}$ \\
        \hline
        Statistical uncertainty & 38 & 23 & 28 & 21 & 31 & 24 & 30 & 87 & 51 & 68 & 64 & 68 & 67 & 65 \\
        \hline
        Fitter bias & 1 & $<$1 & 4 & $<$1 & 2 & 1 & 1 & $<$1 & 2 & 9 & 3 & 6 & 1 & 3 \\
        Flavor parameters & 1 & 3 & 8 & 5 & 9 & 5 & 8 & 2 & 10 & 1 & 8 & 1 & 7 & 6 \\
        Input parameters & 6 & 1 & 2 & $<$1 & 3 & $<$1 & $<$1 & 20 & 4 & 7 & 1 & 3 & 2 & 1 \\
        DCS corrections & 6 & 1 & $<$1 & $<$1 & 4 & 1 & $<$1 & 2 & 2 & 2 & 4 & 1 & $<$1 & 1 \\
        Correlation matrices & $<$1 & 1 & $<$1 & 1 & $<$1 & $<$1 & $<$1 & 3 & 1 & 3 & $<$1 & 2 & $<$1 & 2 \\
        ST yields & 22 & 4 & 8 & 1 & 2 & 2 & 1 & 12 & $<$1 & 5 & 4 & 5 & 6 & 6 \\
        Efficiency profiles & 3 & 3 & 3 & 2 & 3 & 1 & 2 & 3 & 1 & 2 & 2 & 2 & 2 & 2 \\
        DT efficiencies & 3 & 1 & 1 & $<$1 & $<$1 & $<$1 & $<$1 & 1 & $<$1 & 1 & $<$1 & 1 & $<$1 & 1 \\
        Background shapes & 3 & 1 & 3 & 2 & 3 & 2 & 2 & 11 & 6 & 7 & 7 & 6 & 6 & 6 \\
        Background contributions & 6 & 1 & 3 & $<$1 & 1 & 1 & 2 & 1 & 1 & 1 & 1 & 1 & 1 & 1 \\
        Tracking of $\pipm$ in $\KS$ & 6 & $<$1 & $<$1 & 1 & $<$1 & 1 & 7 & 5 & $<$1 & 2 & 1 & 2 & 3 & 3 \\
        Resolution differences & 3 & 4 & 5 & 2 & 3 & 3 & 2 & 14 & 8 & 6 & 5 & 15 & 10 & 10 \\
        \hline
        Total systematic uncertainty & 29 & 7 & 14 & 7 & 12 & 7 & 9 & 37 & 21 & 21 & 13 & 21 & 15 & 15 \\
        \hline   
    \end{tabular}
    }
    \end{center}
\end{table}

\begin{table}[tb]
    \caption{Summary of all uncertainties on the strong-phase parameters of $\decay{\Dz}{\KL \pip\pim}$ decays from \besiii data. All uncertainties are quoted in units of $10^{-4}$.}
    \label{tab:klpp_systematic_coef_besiii}
    \begin{center}
    {\footnotesize
    \setlength\tabcolsep{4pt}
    \centering
    \begin{tabular}{l|rrrrrrrrrrrrrr}
        \hline
        Source & $C_{0}$ & $C_{1}$ & $C_{2}$ & $C_{3}$ & $C_{4}$ & $C_{5}$ & $C_{6}$ & $S_{0}$ & $S_{1}$ & $S_{2}$ & $S_{3}$ & $S_{4}$ & $S_{5}$ & $S_{6}$ \\
        \hline
        Statistical uncertainty & 38 & 15 & 34 & 20 & 30 & 22 & 31 & 100 & 62 & 79 & 80 & 74 & 70 & 77 \\
        \hline
        Fitter bias & 4 & $<$1 & 1 & $<$1 & 2 & 1 & 3 & 3 & 11 & 2 & 4 & $<$1 & 2 & $<$1 \\
        Flavor parameters & 7 & 2 & 7 & 3 & 5 & 3 & 5 & 20 & 7 & 13 & 7 & 10 & 6 & 7 \\
        Input parameters & 3 & $<$1 & 1 & $<$1 & 2 & $<$1 & $<$1 & 3 & $<$1 & 4 & 1 & 3 & 1 & 2 \\
        DCS corrections & 4 & $<$1 & 2 & $<$1 & 4 & $<$1 & 2 & 1 & 1 & 1 & $<$1 & $<$1 & 1 & $<$1 \\
        Correlation matrices & 1 & $<$1 & $<$1 & $<$1 & 1 & $<$1 & 1 & 3 & 1 & 2 & 4 & 2 & $<$1 & 1 \\
        ST yields & 9 & 1 & 6 & 2 & 4 & 2 & 2 & 9 & $<$1 & 4 & 1 & 8 & 5 & 4 \\
        Efficiency profiles & 3 & 1 & 3 & 2 & 2 & 2 & 2 & 4 & 1 & 3 & 1 & 2 & 2 & 2 \\
        DT efficiencies & 4 & $<$1 & 3 & $<$1 & 2 & $<$1 & 1 & 1 & $<$1 & 1 & $<$1 & 1 & $<$1 & 1 \\
        Background shapes & 4 & 2 & 6 & 2 & 4 & 2 & 5 & 11 & 7 & 8 & 8 & 7 & 7 & 6 \\
        Background contributions & 8 & 1 & 5 & 1 & 4 & 1 & 2 & 2 & 2 & 1 & 2 & 1 & 1 & 1 \\
        Tracking of $\pipm$ in $\KS$ & $<$1 & $<$1 & $<$1 & $<$1 & $<$1 & $<$1 & $<$1 & 5 & $<$1 & 3 & 1 & 1 & 1 & 1 \\
        Resolution differences & 5 & 5 & 12 & 6 & 4 & 5 & 9 & 8 & 4 & 7 & 3 & 8 & 8 & 4 \\
        \hline
        Total systematic uncertainty & 15 & 6 & 17 & 7 & 11 & 7 & 13 & 27 & 16 & 19 & 11 & 15 & 14 & 9 \\
        \hline   
    \end{tabular}
    }
    \end{center}
\end{table}

\begin{table}[tb]
    \caption{Summary of all uncertainties on the strong-phase parameters of $\decay{\Dz}{\KS\Kp\Km}$ decays from \besiii data. All uncertainties are quoted in units of $10^{-4}$.}
    \label{tab:kskk_systematic_coef_besiii}
    \begin{center}
    {\footnotesize
    \setlength\tabcolsep{4pt}
    \centering
    \begin{tabular}{l|rrrrrrrrrr}
        \hline
        Source & $C_{0}$ & $C_{1}$ & $C_{2}$ & $C_{3}$ & $C_{4}$ & $S_{0}$ & $S_{1}$ & $S_{2}$ & $S_{3}$ & $S_{4}$ \\
        \hline
        Statistical uncertainty & 111 & 55 & 97 & 63 & 80 & 251 & 203 & 159 & 297 & 163 \\
        \hline
        Fitter bias & 14 & 7 & 1 & 8 & 4 & 18 & 31 & 7 & 33 & 7 \\
        Flavor parameters & 76 & 14 & 45 & 14 & 50 & 21 & 37 & 11 & 33 & 11 \\
        Input parameters & 4 & 1 & 4 & 1 & 3 & 8 & 6 & 4 & 8 & 4 \\
        DCS corrections & 4 & 1 & 8 & $<$1 & 3 & 3 & 6 & $<$1 & 1 & $<$1 \\
        Correlation matrices & $<$1 & $<$1 & $<$1 & $<$1 & $<$1 & $<$1 & 1 & 1 & 2 & $<$1 \\
        ST yields & 11 & 1 & 1 & 1 & 1 & 1 & 1 & 1 & 1 & 1 \\
        Efficiency profiles & 7 & 2 & 8 & 3 & 7 & 16 & 10 & 5 & 11 & 6 \\
        DT efficiencies & 5 & 1 & 1 & 1 & 1 & 3 & 1 & 1 & 2 & 1 \\
        Background shapes & 10 & 4 & 8 & 4 & 6 & 28 & 32 & 16 & 30 & 20 \\
        Background contributions & 34 & 4 & 31 & 4 & 16 & 13 & 22 & 11 & 16 & 9 \\
        Tracking of $\pipm$ in $\KS$ & 3 & $<$1 & 1 & $<$1 & $<$1 & $<$1 & 2 & $<$1 & 1 & 1 \\
        Resolution differences & 12 & 1 & 16 & 1 & 7 & 36 & 36 & 14 & $<$1 & 7 \\
        \hline
        Total systematic uncertainty & 88 & 17 & 59 & 18 & 54 & 58 & 73 & 28 & 60 & 27 \\
        \hline   
    \end{tabular}
    }
    \end{center}
\end{table}

\begin{table}[tb]
    \caption{Summary of all uncertainties on the strong-phase parameters of $\decay{\Dz}{\KL\Kp\Km}$ decays from \besiii data. All uncertainties are quoted in units of $10^{-4}$.}
    \label{tab:klkk_systematic_coef_besiii}
    \begin{center}
    {\footnotesize
    \setlength\tabcolsep{4pt}
    \centering
    \begin{tabular}{l|rrrrrrrrrr}
        \hline
        Source & $C_{0}$ & $C_{1}$ & $C_{2}$ & $C_{3}$ & $C_{4}$ & $S_{0}$ & $S_{1}$ & $S_{2}$ & $S_{3}$ & $S_{4}$ \\
        \hline
        Statistical uncertainty & 112 & 55 & 110 & 64 & 82 & 254 & 242 & 163 & 276 & 152 \\
        \hline
        Fitter bias & 21 & 8 & 10 & 7 & 5 & 18 & 34 & 7 & 33 & 5 \\
        Flavor parameters & 70 & 13 & 39 & 10 & 46 & 21 & 41 & 11 & 31 & 10 \\
        Input parameters & 3 & 1 & 2 & 2 & 2 & 8 & 7 & 4 & 7 & 4 \\
        DCS corrections & 11 & $<$1 & 1 & $<$1 & 5 & 3 & 6 & $<$1 & 1 & $<$1 \\
        Correlation matrices & 1 & 1 & 1 & 1 & 1 & 4 & 4 & 1 & 3 & 2 \\
        ST yields & 7 & 1 & 1 & 1 & 2 & 1 & 1 & 1 & 1 & 9 \\
        Efficiency profiles & 11 & 2 & 7 & 3 & 8 & 16 & 10 & 5 & 11 & 5 \\
        DT efficiencies & 6 & 1 & 1 & 1 & 1 & 3 & 2 & 1 & 2 & 1 \\
        Background shapes & 9 & 3 & 11 & 4 & 6 & 28 & 31 & 16 & 27 & 17 \\
        Background contributions & 31 & 4 & 32 & 5 & 15 & 13 & 27 & 11 & 15 & 8 \\
        Tracking of $\pipm$ in $\KS$ & 4 & $<$1 & 2 & 1 & 1 & $<$1 & 2 & $<$1 & 1 & 1 \\
        Resolution differences & 9 & 1 & 10 & 1 & 2 & 36 & 34 & 14 & $<$1 & 7 \\
        \hline
        Total systematic uncertainty & 82 & 16 & 54 & 14 & 50 & 58 & 77 & 28 & 56 & 24 \\

        \hline   
    \end{tabular}
    }
    \end{center}
\end{table}

In the analysis of \lhcb data, the major difference between this measurement and Ref.~\cite{LHCb-PAPER-2020-019} is the approach to extract observables. 
In this measurement, background subtraction is performed with the \sPlot technique, which assumes that the $B$-candidate mass and Dalitz-plot coordinates are independent variables.
Potential correlations between these variables may introduce bias on the \CP observables,
and are evaluated by inspecting the Dalitz-plot-dependent $B$-candidate mass distributions in simulated samples.
Pseudoexperiments generated with such correlations are fitted with the baseline model.
Mean biases of the \CP observables are taken as systematic uncertainties.

Efficiency profiles from the simulated signal are needed in this measurement to correct $\decay{\Bpm}{D h^{\pm}}$ observables,
thus imperfect modeling in simulation is considered as a source of systematic uncertainty.
Correction profiles are obtained by comparing Dalitz-plot distributions between simulated signal and $\decay{\Bpm}{D\pipm}$ signal from data.
This is possible as the $\decay{\Bpm}{D\pipm}$ decays have clean signal peaks in data and minimal \CP violation.
New efficiency profiles are obtained by multiplying the baseline profiles with the correction profiles, and are used to generate pseudoexperiments.
Weighted signal yields from the ensemble are computed with the baseline efficiency profiles, and mean biases of the \CP observables are taken as systematic uncertainties.
The sample size of simulated signal can affect the precision of the efficiency profiles, so its impact is assessed by sampling efficiency parameters from the fits to simulated samples and computing a number of new profiles.
The subsequent computation of observables and fits to data indicate that the standard deviations
of the \CP observables are two orders of magnitude smaller than other systematic uncertainties.
Therefore, no systematic uncertainty due to the simulation sample size is assigned.

The effect of detector resolution is taken into account by smearing the Dalitz-plot distributions in pseudoexperiments.
The smearing is based on the $m_{\pm}^2$ resolution as determined in simulation and scaled to account for data-simulation differences.
The baseline fit model is then applied to the ensemble, and mean biases of the \CP observables are taken as systematic uncertainties.
This source is equivalent to the bin-migration effect described in Ref.~\cite{LHCb-PAPER-2020-019}.

Systematic uncertainties related to the global fit are estimated by including extra effects when generating pseudoexperiments, and fitting the ensemble with the baseline model.
Several background decays are not considered in the baseline fit due to their low yields,
which include the semileptonic $\decay{B}{D\mu\neum}$ and $\decay{B}{D(\to\KS\pi l\neul)h}$ ($l=\mu, e$) decays,
the baryonic $\decay{\Lb}{Dp\pi^-}$ and $\decay{\Lb}{\Lc(\to p\KS\pip\pim) h^-}$ decays,
the charmless $\decay{\Bpm}{\KS h^+h^- h^{\prime\pm}}$ decays,
and the $\decay{\Bpm}{D(\to\Kpm\pimp)\KS\pipm}$ decays where the final-state tracks are wrongly assigned.
Their yields relative to signal are estimated from either \lhcb simulation or fast simulation~\cite{Cowan:2016tnm}, as well as the $B$-candidate mass and Dalitz-plot distributions.
Alternative parameter sets describing the $B$-candidate mass are obtained by resampling data and simulated samples,
and propagated to the fit in the tighter mass region and the \CP fit.
The relative yields of partially reconstructed background and PID efficiencies for misidentified background are fixed in the global fit.
They are varied within their uncertainties in the mass fits, and standard deviations of the \CP observables from the \CP fit are assigned as systematic uncertainties.
The partially reconstructed backgrounds are treated as a whole with the same Dalitz-plot distributions and uniform $Dh$ mass distributions across the phase space, 
which ignores underlying physics effects like \CP violation in $\decay{B}{D^{(*)}K^{(*)}}$ decays.
These effects are taken into account when generating pseudoexperiments, where the parameters in Ref.~\cite{LHCb-CONF-2018-002} are used.
Global fit-related systematic uncertainties are evaluated by adapting the same approach as in Ref.~\cite{LHCb-PAPER-2020-019}.

The amount of \CP violation in $\KS$ decay and interactions between detector materials and neutral kaons are ignored in the measurement.
These effects are included to generate a pseudosignal following the approach described in Ref.~\cite{Bjorn:2019kov}.
The baseline fit model is used to fit the pseudosignal, and biases on the \CP observables are assigned as systematic uncertainties.
The impact of ignoring $D$ mixing in \lhcb data is assessed by the same approach,
where the signal model including $D$ mixing is described in Ref.~\cite{Bondar:2010qs}.

From pseudoexperiments, increasing bias on the \CP observables is found when the Fourier orders increase.
This is due to sizeable correlations present within the covariance matrix determined from weighted sums of data in Eq.~\ref{eq:lhcb_coef_cov}.
Nevertheless, biases found in pseudoexperiments under baseline Fourier orders are small and assigned as a source of systematic uncertainty.

All uncertainties of \CP observables are summarized in Table~\ref{tab:all_systematic_cpv}, 
and the systematic uncertainties are found to be one order of magnitude smaller than the statistical ones.
For the sources that have been studied in both this measurement and Ref.~\cite{LHCb-PAPER-2020-019},
their impact is found to be at the same level.
The extra sources due to the new method are not dominant,
making the total systematic uncertainties comparable to those in Ref.~\cite{LHCb-PAPER-2020-019}.
The overall systematic uncertainties on the \CP observables from \besiii are at the same order as those from \lhcb sources, which are much smaller than the statistical uncertainties.

\begin{table}[tb]
    \caption{Summary of all uncertainties on the \CP observables from \besiii and \lhcb data.
    All uncertainties are quoted in units of $10^{-2}$.}
    \label{tab:all_systematic_cpv}
    {\footnotesize
    \renewcommand{\arraystretch}{1.05}
    \begin{center}
        \begin{tabular}{l|rrrrrr}
            \hline \\[-1em]
            Source & $\sigma(\xmdk)$ & $\sigma(\ymdk)$ & $\sigma(\xpdk)$ & $\sigma(\ypdk)$ & $\sigma(\xxi)$ & $\sigma(\yxi)$ \\ \\[-1em]
            \hline
            Statistical uncertainty                        & $0.98$ & $1.39$ & $1.02$ & $1.40$ & $2.09$ & $2.37$ \\
            \hline
            Correlation between $m(Dh)$ and $m_{\pm}^2$    & $0.03$ & $0.05$ & $<0.01$ & $0.01$ & $0.08$ & $0.05$ \\
            Data/simulation disagreement                   & $0.02$ & $0.10$ & $0.06$ & $0.10$ & $0.04$ & $0.03$ \\
            Resolution                                     & $0.05$ & $0.11$ & $0.15$ & $0.10$ & $0.10$ & $0.11$ \\ 
            Small backgrounds                              & $0.15$ & $0.11$ & $0.13$ & $0.18$ & $0.13$ & $0.08$ \\
            Mass-shape parameters                          & $0.07$ & $0.07$ & $0.05$ & $0.09$ & $0.15$ & $0.10$ \\ 
            Fixed-yield ratios                             & $0.02$ & $0.02$ & $0.03$ & $0.02$ & $0.02$ & $0.01$ \\ 
            PID efficiencies                               & $0.02$ & $0.03$ & $0.03$ & $0.03$ & $0.02$ & $0.01$ \\ 
            Low-mass physics effects                       & $0.01$ & $0.06$ & $0.08$ & $0.13$ & $0.03$ & $0.05$ \\ 
            \CP violation in $\KS$ decays                  & $0.01$ & $0.09$ & $0.03$ & $0.07$ & $0.18$ & $0.06$ \\ 
            $D$ mixing                                     & $0.03$ & $0.03$ & $0.04$ & $0.04$ & $0.01$ & $0.02$ \\ 
            Fit bias                                    & $0.03$ & $0.03$ & $0.06$ & $0.09$ & $0.07$ & $0.02$ \\
            \hline
            Systematic uncertainty from \lhcb        & $0.18$ & $0.23$ & $0.24$ & $0.31$ & $0.30$ & $0.20$ \\
            \hline
            Systematic uncertainty from \besiii      & $0.07$ & $0.09$ & $0.08$ & $0.06$ & $0.01$ & $0.02$ \\
            \hline
            Total systematic uncertainty                   & $0.20$ & $0.25$ & $0.26$ & $0.31$ & $0.30$ & $0.20$ \\
            \hline
        \end{tabular}
    \end{center}
    }
\end{table}

\section{Results of \boldmath{$\gamma$}}
\label{sec:result}

From the joint analysis of \besiii and \lhcb data, the \CP observables, determined from $\decay{\Bpm}{D(\to\KShh)h^{\pm}}$ decays, are
\begin{align*}
    \xmdk &= (~~\,5.84\pm0.98\pm0.20) \times 10^{-2}, \\
    \ymdk &= (~~\,6.74\pm1.39\pm0.25) \times 10^{-2}, \\
    \xpdk &= (   -9.68\pm1.02\pm0.26) \times 10^{-2}, \\
    \ypdk &= (   -2.42\pm1.40\pm0.31) \times 10^{-2}, \\
    \xxi  &= (   -6.57\pm2.09\pm0.30) \times 10^{-2}, \\
    \yxi  &= (   -1.12\pm2.37\pm0.20) \times 10^{-2},
\end{align*}
where the first uncertainties are statistical and the second systematic.
Statistical and systematic uncertainties from \besiii and \lhcb are also analyzed separately to study their individual impact.
The values and correlations of each source of uncertainty are summarized in Tables~\ref{tab:corr_cpv_lhcb_stat}--\ref{tab:corr_cpv_bes_syst} of Appendix~\ref{app:correlations}.

The measured \CP observables can be reinterpreted by a frequentist treatment as described in Refs.~\cite{LHCb-PAPER-2016-032,LHCb-PAPER-2021-033}, which is denoted as the \textsc{Plugin} method.
The solution in the region ${0 < \gamma < 180{\degrees}}$ is chosen, which is
\begin{align*}
    \gamma& = (71.3\pm 5.0){\degrees}, \\
      \rdk& = 0.0949^{+0.0086}_{-0.0085}, \\
      \ddk& = (121.6^{+5.6}_{-5.9}){\degrees}, \\
     \rdpi& = 0.0064^{+0.0021}_{-0.0019}, \\
     \ddpi& = (311^{+17}_{-20}){\degrees}.
\end{align*}

\begin{figure}[tb]
    \centering
    \includegraphics[width=0.7\linewidth]{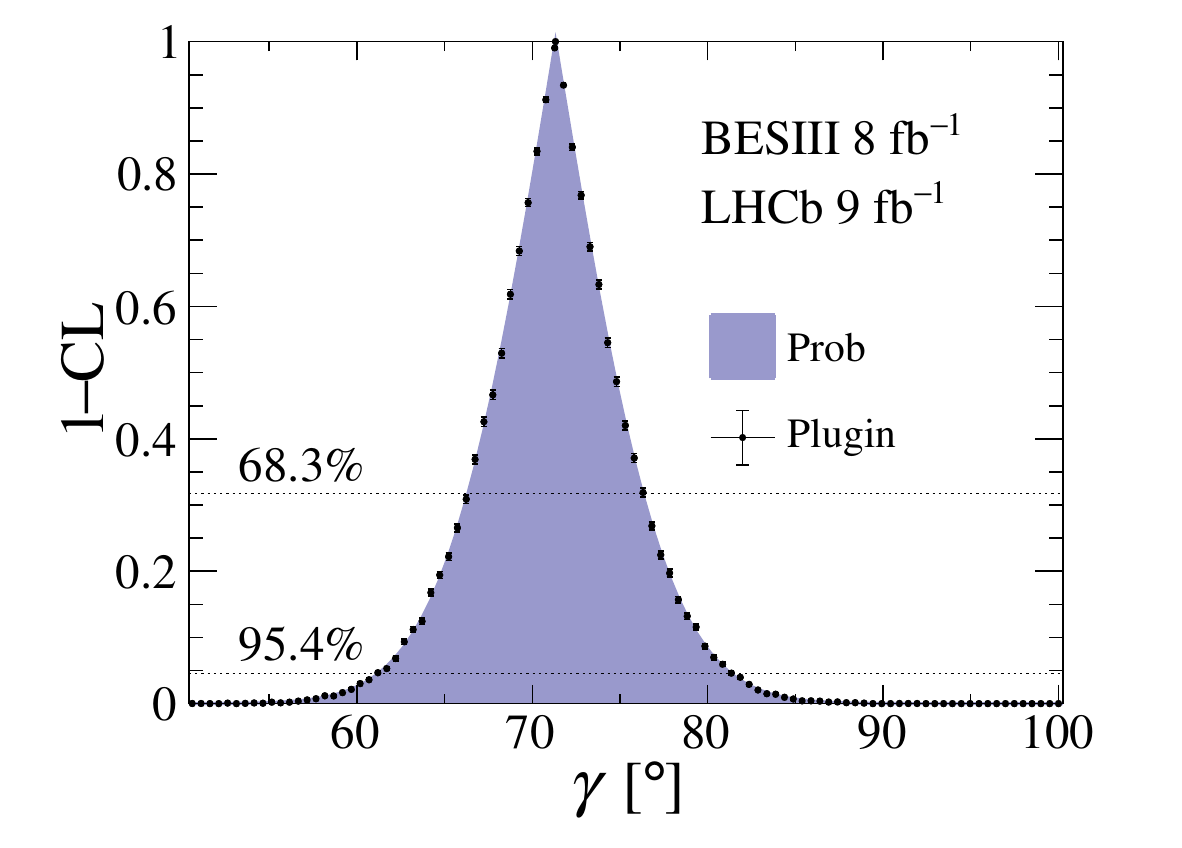}
    \caption{One-dimensional scan of the confidence level of the CKM angle $\gamma$ from the \textsc{Plugin} and \textsc{Prob} methods.}
    \label{fig:gamma_1dcl}
\end{figure}

\begin{figure}[tb]
    \centering
    \includegraphics[width=0.49\linewidth]{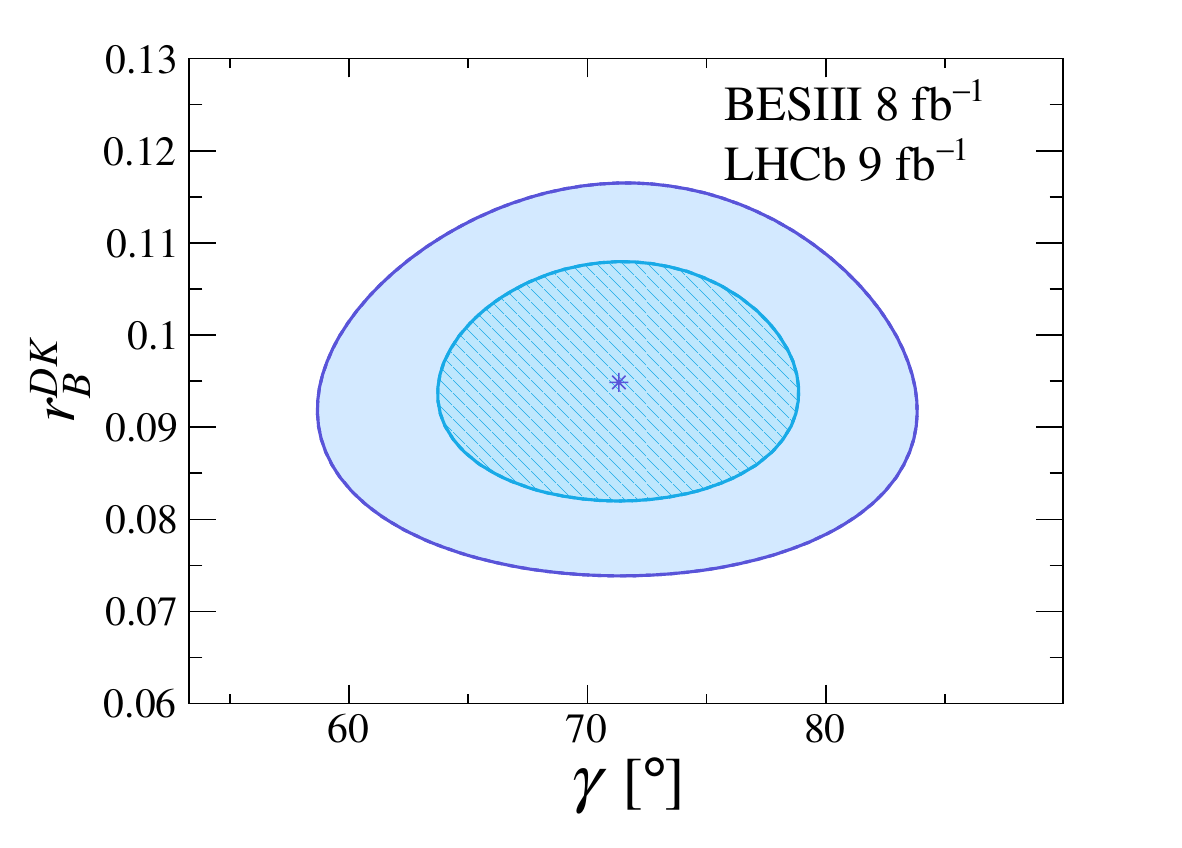}
    \includegraphics[width=0.49\linewidth]{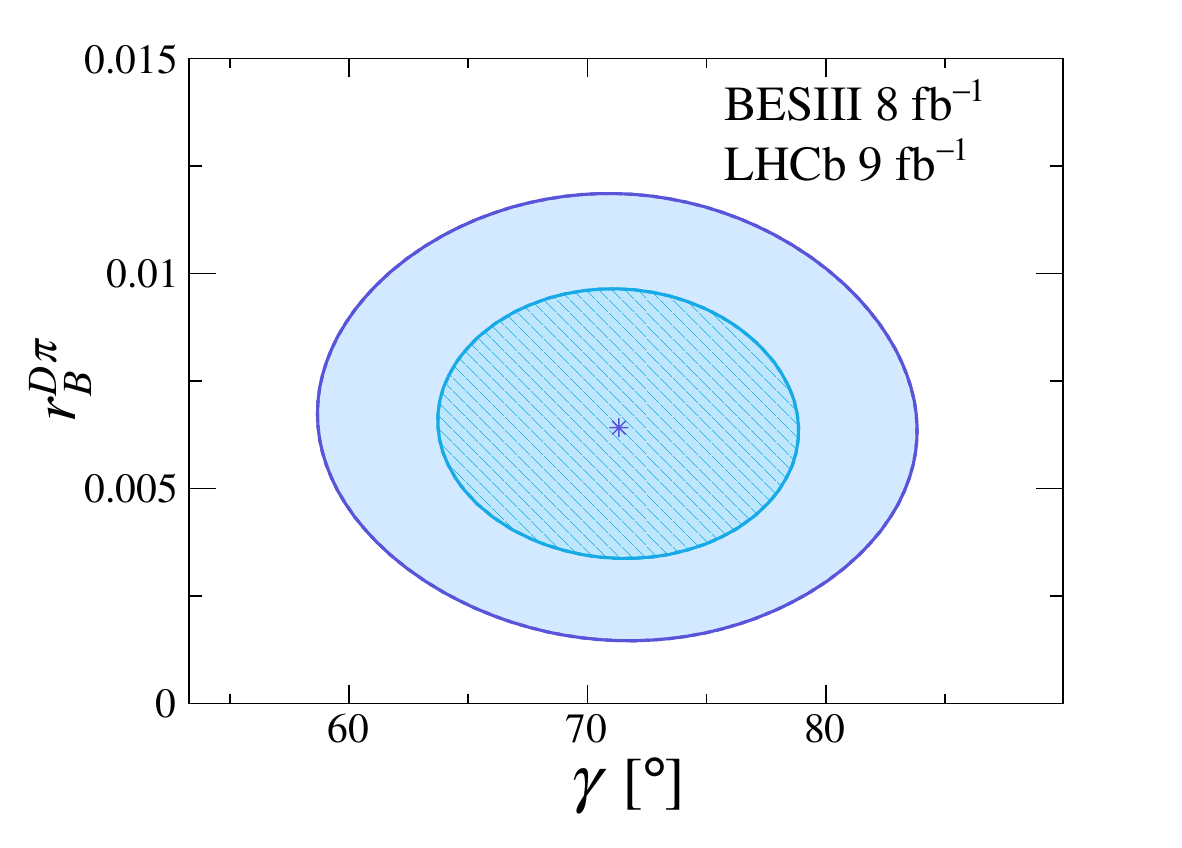}
    
    \includegraphics[width=0.49\linewidth]{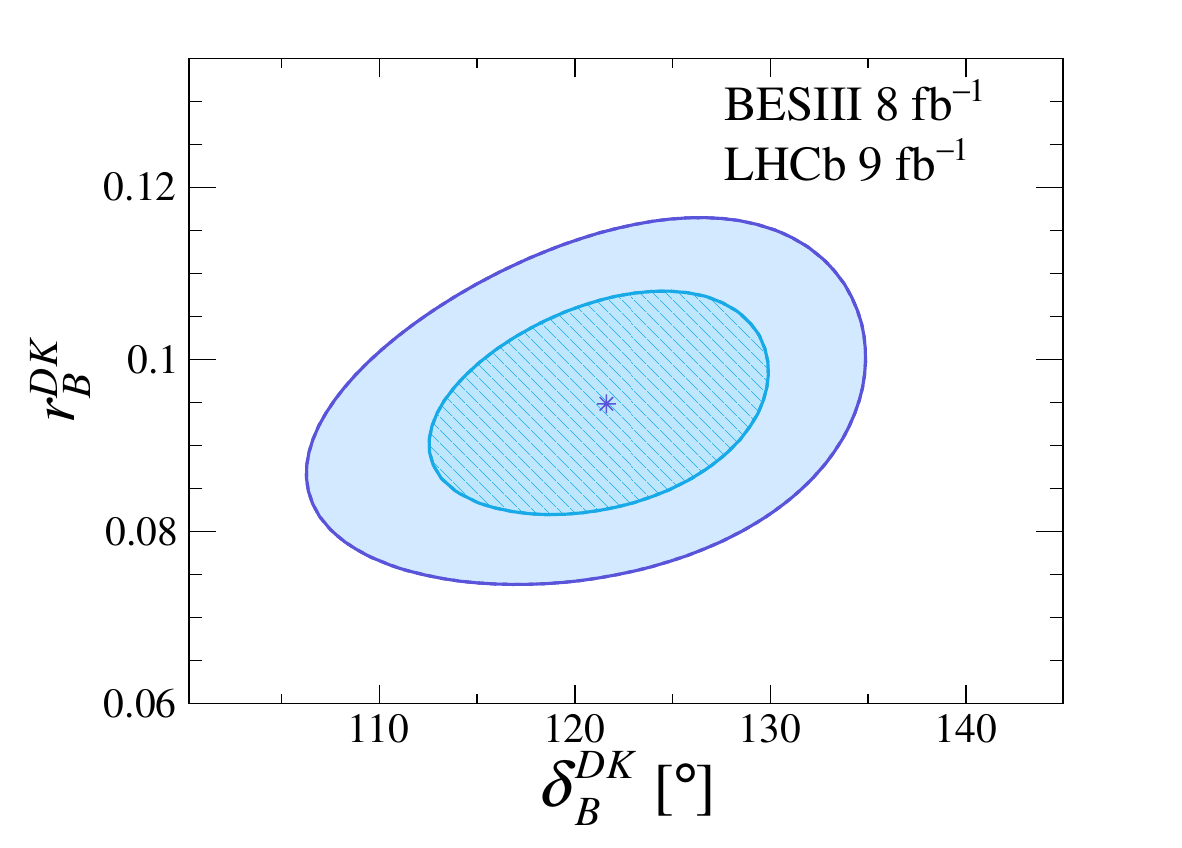}
    \includegraphics[width=0.49\linewidth]{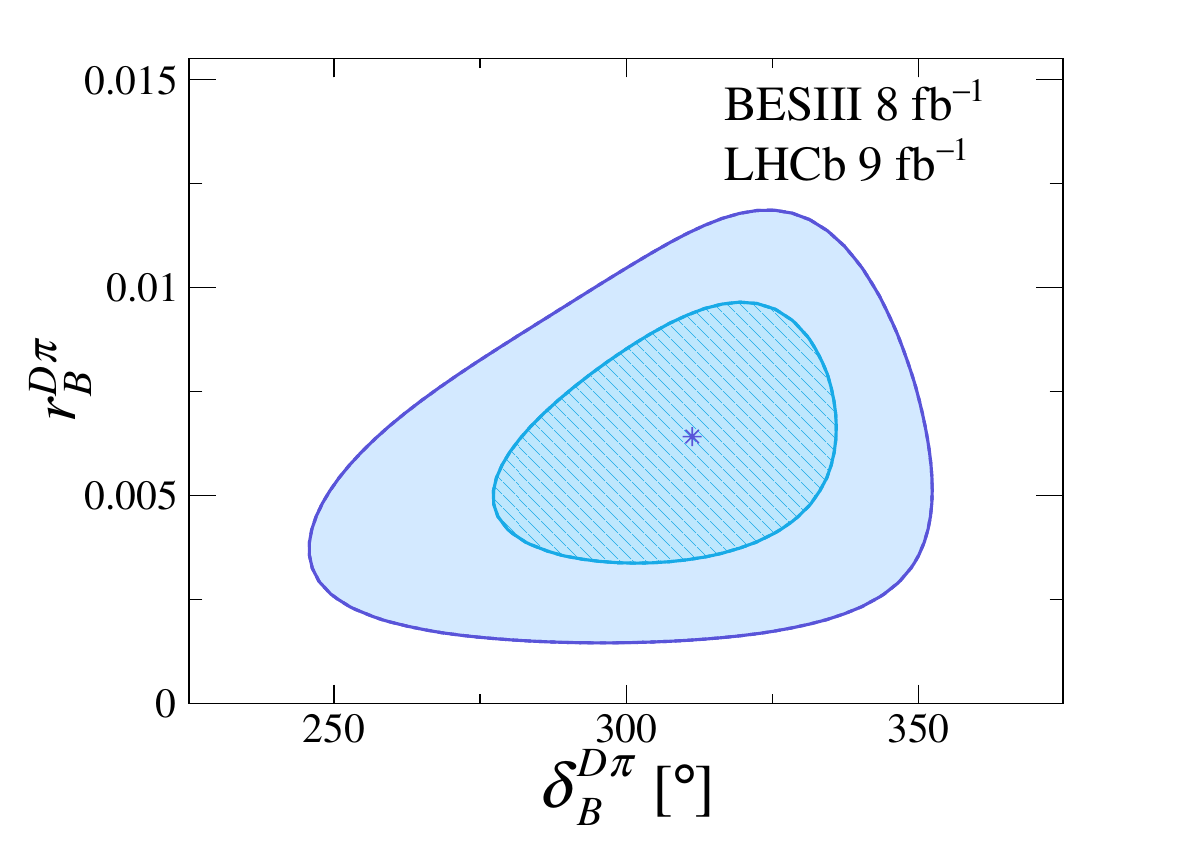}
    \caption{Two-dimensional confidence levels of the indicated parameter pairs from the \textsc{Prob} method. The contours correspond to 68\% and 95\% confidence levels of the observable pairs.}
    \label{fig:2dcl}
\end{figure}

These are consistent with the world-average values~\cite{CKMfitter2005,HFLAV23} and the previous measurement using the binned phase-space approach~\cite{LHCb-PAPER-2020-019} when considering statistical correlations estimated by pseudoexperiments.
The uncertainty of $\gamma$ is smaller than the previous measurement, marking the most precise single measurement of $\gamma$ to date.
A one-dimensional confidence-level scan of $\gamma$ is shown in Fig.~\ref{fig:gamma_1dcl},
in which a profile-likelihood method, denoted as \textsc{Prob}, is also applied.
Results from the \textsc{Plugin} and \textsc{Prob} methods are found to be identical.
Figure~\ref{fig:2dcl} shows the two-dimensional confidence levels of the $(\gamma, \rb)$ and $(\db, \rb)$ pairs.

Additional studies are performed to evaluate uncertainties of $\gamma$ from different sources and different approaches. 
The statistical and systematic uncertainties from \lhcb are found to be $4.9{\degrees}$ and $0.8{\degrees}$, respectively,
which are compatible with the previous binned measurement~\cite{LHCb-PAPER-2020-019}.
The \besiii data has an impact of $0.6{\degrees}$ on $\gamma$, which is smaller than that in Ref.~\cite{LHCb-PAPER-2020-019}, owing to the increase in the \besiii sample size.
The \lhcb data is refitted using the binned approach with updated strong-phase inputs from $\decay{D}{\KSpp}$ decays~\cite{BESIII:2025nsp} and using the unbinned approach with higher orders.
Results are listed in Table~\ref{tab:parameters_lhcb}, where the uncertainties are from \lhcb statistics only.
It is shown that the statistical uncertainty of $\gamma$ is improved by 5\% in the unbinned method.
The precision of $\gamma$ is found to be improved when employing higher Fourier orders, 
indicating that more information can be extracted from higher-order observables.
This is because the model to compute weights does not perfectly describe the data.
In future unbinned measurement of $\gamma$, better precision can be achieved by incorporating higher-order observables in the fits. 

\begin{table}[!tb]
    \caption{Results of the extracted parameters $(\gamma, \rdk, \ddk, \rdpi, \ddpi)$, where the uncertainties are statistical only, deriving from \lhcb data.}
    \label{tab:parameters_lhcb}
    {\small
    \renewcommand{\arraystretch}{1.1}
    \begin{center}
        \begin{tabular}{lr@{$\;$}c@{$\;$}lr@{$\;$}c@{$\;$}lcr@{$\;$}c@{$\;$}lc}
            \hline
 Configuration & \multicolumn{3}{c}{$\gamma~({\degrees})$} & \multicolumn{3}{c}{$\rdk~(10^{-2})$} &$\ddk~({\degrees})$ & \multicolumn{3}{c}{$\rdpi~(10^{-2})$} & $\ddpi~({\degrees})$ \\
            \hline
Binned (refitted)         &$67.7$&$\pm$& $5.1$  &$9.61$&$^{+}_{-}$&$_{0.77}^{0.78}$ &$118.6^{+5.2}_{-5.6}$ & $0.53$& $\pm$& $0.19$ & $286^{+24}_{-28}$ \\
$M_\pi=2$, $M_K=1$ (baseline)    &$71.3$&$\pm$& $4.9$  &$9.48$&$\pm$& $0.81$ &$121.6_{-5.5}^{+5.3}$ & $0.64$& $\pm$& $0.20$ & $311^{+16}_{-20}$ \\
$M_\pi=2$, $M_K=2$    &$71.6$&$^{+}_{-}$&$^{4.7}_{4.8}$  &$9.58$&$^{+}_{-}$&$^{0.81}_{0.80}$ &$122.5^{+5.1}_{-5.4}$ & $0.62$&$^{+}_{-}$&$^{0.20}_{0.19}$ & $309^{+17}_{-20}$ \\
$M_\pi=3$, $M_K=1$    &$71.0$&$\pm$& $4.9$  &$9.36$&$^{+}_{-}$&$_{0.79}^{0.80}$ &$120.7_{-5.6}^{+5.2}$ & $0.64$&$^{+}_{-}$&$^{0.20}_{0.19}$ & $311^{+16}_{-20}$ \\
$M_\pi=3$, $M_K=2$    &$71.3$&$^{+}_{-}$&$_{4.9}^{4.8}$  &$9.46$&$^{+}_{-}$&$_{0.79}^{0.80}$ &$121.5^{+5.1}_{-5.5}$ & $0.63$&$^{+}_{-}$&$^{0.19}_{0.20}$ & $309^{+17}_{-20}$ \\
            \hline
        \end{tabular}
    \end{center}
    }
\end{table}

\section{Conclusions}
\label{sec:conclusions}

In this paper and an accompanying Letter~\cite{LHCb-PAPER-2025-064}, a novel approach developed from Ref.~\cite{Poluektov:2017zxp} is adopted to determine the \CP observables that lead to a precise measurement of the CKM angle $\gamma$.
The approach uses per-event weights that account for the strong-phase variation, magnitude of the decay amplitude and background level to optimize the sensitivity to~$\gamma$.
The approach is employed in datasets collected by the \besiii and \lhcb detectors,
corresponding to integrated luminosities of 8\invfb and 9\invfb, respectively.
The CKM angle $\gamma$ is determined from $\decay{\Bpm}{D\Kpm}$ and $\decay{\Bpm}{D\pipm}$ decays followed by $\decay{D}{\KS\pip\pim}$ and $\decay{D}{\KS\Kp\Km}$ decays with input of the strong-phase information 
from the doubly-tagged $\decay{\psiprpr}{D\Db}$ process.
The \CP-violating parameters are determined from a joint fit to both datasets,
and an interpretation of the underlying physics parameters yields $\gamma = (71.3\pm5.0){\degrees}$.
The statistical uncertainty from \lhcb data is improved by 5\% with respect to the binned phase-space measurement~\cite{LHCb-PAPER-2020-019}, while the systematic uncertainty from \lhcb is at the same level
and the uncertainty introduced by the strong-phase input from \besiii is lower.
The hadronic parameters $\rb$ and $\db$ of the $\decay{\Bpm}{D\Kpm}$ and $\decay{\Bpm}{D\pipm}$ decays are also measured.
Finally, the strong-phase parameters of the novel approach are measured from \besiii data, and will serve as inputs to future $\gamma$ measurement.

\section*{Acknowledgements}

\noindent We acknowledge important input from Alex Bondar, which helped to shape the analysis reported here.
The BESIII collaboration thanks the staff of BEPCII (https://cstr.cn/31109.02.BEPC) and the IHEP computing center for their strong support. This work is supported in part by National Key R\&D Program of China under Contracts Nos. 2023YFA1606000, 2023YFA1606704, 2022YFA1601901; National Natural Science Foundation of China (NSFC) under Contracts Nos. 11635010, 11935015, 11935016, 11935018, 12025502, 12035009, 12035013, 12061131003, 12192260, 12192261, 12192262, 12192263, 12192264, 12192265, 12221005, 12225509, 12235017, 12342502, 12361141819, 12375087, 12405112; the Chinese Academy of Sciences (CAS) Large-Scale Scientific Facility Program; the Strategic Priority Research Program of Chinese Academy of Sciences under Contract No. XDA0480600; CAS under Contract No. YSBR-101; 100 Talents Program of CAS; The Institute of Nuclear and Particle Physics (INPAC) and Shanghai Key Laboratory for Particle Physics and Cosmology; ERC under Contract No. 758462; German Research Foundation DFG under Contract No. FOR5327; Istituto Nazionale di Fisica Nucleare, Italy; Knut and Alice Wallenberg Foundation under Contracts Nos. 2021.0174, 2021.0299, 2023.0315; Ministry of Development of Turkey under Contract No. DPT2006K-120470; National Research Foundation of Korea under Contract No. NRF-2022R1A2C1092335; National Science and Technology fund of Mongolia; Polish National Science Centre under Contract No. 2024/53/B/ST2/00975; STFC (United Kingdom); Swedish Research Council under Contract No. 2019.04595; U. S. Department of Energy under Contract No. DE-FG02-05ER41374.

%
%
The LHCb collaboration expresses gratitude to our colleagues in the CERN
accelerator departments for the excellent performance of the LHC. We
thank the technical and administrative staff at the LHCb
institutes.
We acknowledge support from CERN and from the national agencies:
ARC (Australia);
CAPES, CNPq, FAPERJ and FINEP (Brazil); 
MOST and NSFC (China); 
CNRS/IN2P3 and CEA (France);  
BMFTR, DFG and MPG (Germany);
INFN (Italy); 
NWO (Netherlands); 
MNiSW and NCN (Poland); 
MEC/IFA (Romania); 
MICIU and AEI (Spain);
SNSF and SER (Switzerland); 
NASU (Ukraine); 
STFC (United Kingdom); 
DOE NP and NSF (USA).
We acknowledge the computing resources that are provided by ARDC (Australia), 
CBPF (Brazil),
CERN, 
IHEP and LZU (China),
IN2P3 (France), 
KIT and DESY (Germany), 
INFN (Italy), 
SURF (Netherlands),
Polish WLCG (Poland),
IFIN-HH (Romania), 
PIC (Spain), CSCS (Switzerland), 
GridPP (United Kingdom),
and NSF (USA).  
We are indebted to the communities behind the multiple open-source
software packages on which we depend.
Individual groups or members have received support from
RTP (Australia), 
Key Research Program of Frontier Sciences of CAS, CAS PIFI, CAS CCEPP (China); 
Minciencias (Colombia);
EPLANET, Marie Sk\l{}odowska-Curie Actions, ERC and NextGenerationEU (European Union);
A*MIDEX, ANR, IPhU and Labex P2IO, and R\'{e}gion Auvergne-Rh\^{o}ne-Alpes (France);
Alexander-von-Humboldt Foundation (Germany);
ICSC (Italy); 
Severo Ochoa and Mar\'ia de Maeztu Units of Excellence, GVA, XuntaGal, GENCAT, InTalent-Inditex and Prog.~Atracci\'on Talento CM (Spain);
the Leverhulme Trust, the Royal Society and UKRI (United Kingdom).

\clearpage

\section*{Appendices}

\appendix

\section{$\boldsymbol{D}$-candidate mass fit at \besiii}
\label{app:bes_k0kk_datafit_result}

\begin{figure}[!h]
    \begin{center}
        \includegraphics[width=1.0\textwidth]{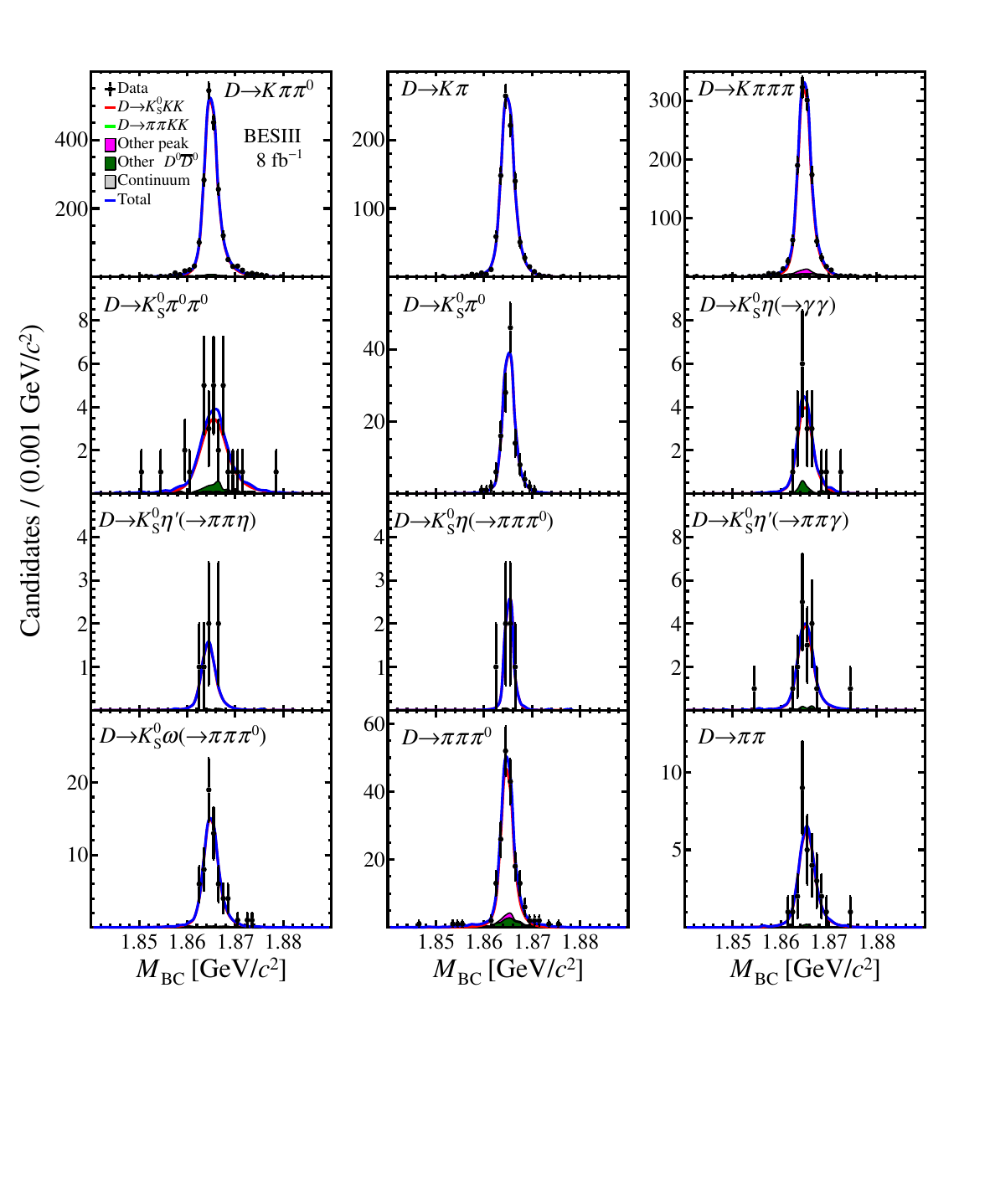}
    \end{center}
    \caption{One-dimensional mass-fit results for the $M_{\rm BC}$ distributions of $\decay{D}{\KS\Kp\Km}$ decays with the indicated fully reconstructed tags.}
    \label{fig:kskk_datafit_full}
\end{figure}

\begin{figure}[tb]
    \begin{center}
        \includegraphics[width=0.7\textwidth]{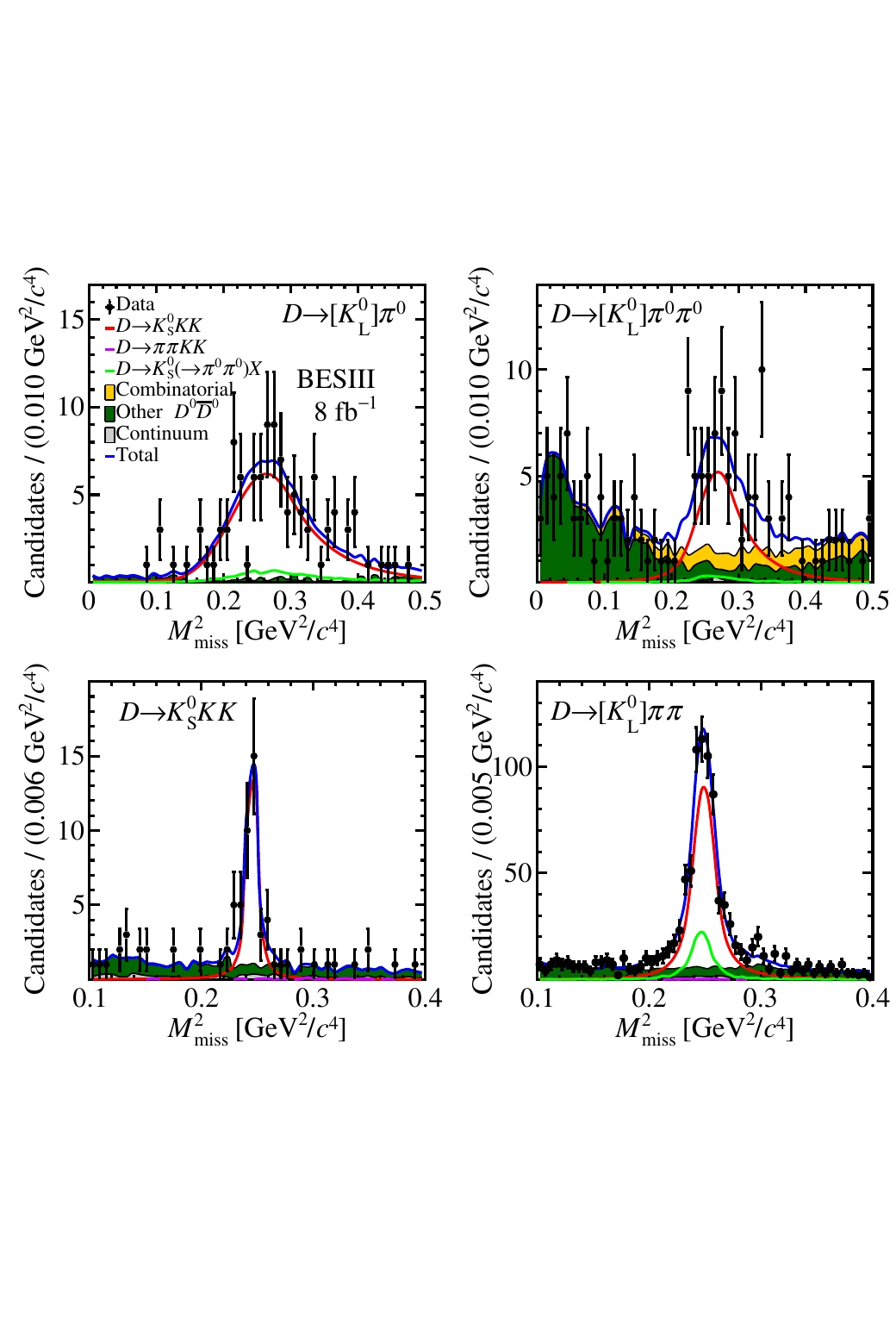}
    \end{center}
    \caption{One-dimensional mass-fit results for the $M_{\rm miss}^{2}$ distributions of $\decay{D}{\KS\Kp\Km}$ decays with partially reconstructed tags and $\KS \Kpm[K^{\mp}]$ with fully reconstructed tags. The decay-chain labels indicate the tag decays.} 
    \label{fig:kskk_datafit_partial}
\end{figure}

\begin{figure}[tb]
    \begin{center}
        \includegraphics[width=1\textwidth]{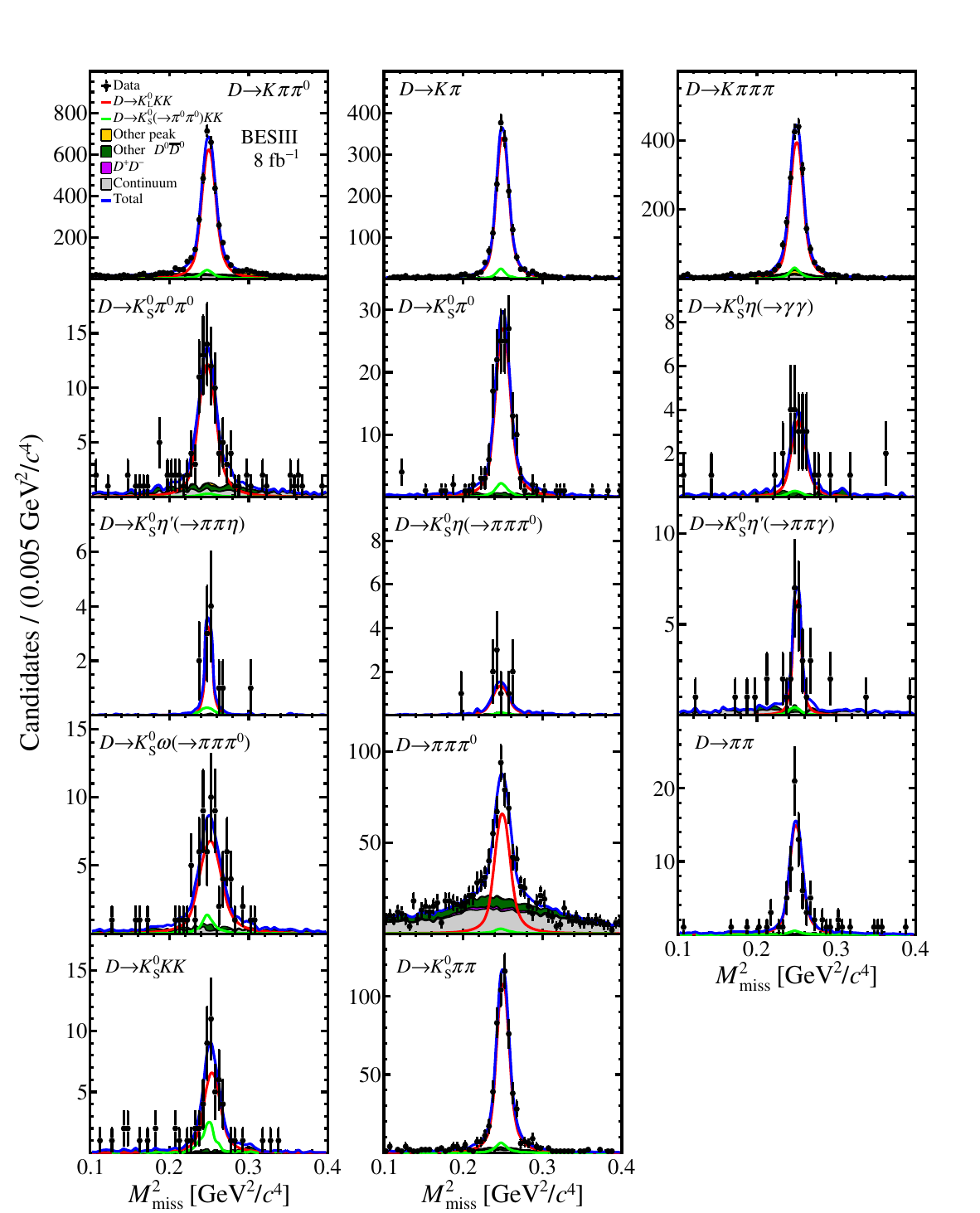}
    \end{center}
    \caption{One-dimensional mass-fit results for the $M_{\rm miss}^{2}$ distributions, featuring $\decay{D}{[\KL]\Kp\Km}$ decays against the indicated fully reconstructed tag modes.}
    \label{fig:klkk_datafit}
\end{figure}

Figures~\ref{fig:kskk_datafit_full}--\ref{fig:klkk_datafit} show the one-dimensional mass fits for the $\decay{D}{\KSL\Kp\Km}$ channel in \besiii, from which the signal yields and DT efficiencies are extracted. These results are summarized in Table~\ref{tab:bes_k0kk_massfit_res}, while those for the $\decay{D}{\KSL\pip\pim}$ channel can be found in Ref.~\cite{BESIII:2025nsp}.

\begin{table}[!h]
    \caption{Signal yields and DT efficiencies of the $\decay{D}{\KSL\Kp\Km}$ channel in \besiii.}
    \label{tab:bes_k0kk_massfit_res}
    \begin{center}
    \begin{tabular}{l|cccc}
        \hline
        \multirow{2}{*}{Tag mode} & \multicolumn{2}{c}{$\decay{D}{\KS\Kp\Km}$} & \multicolumn{2}{c}{$\decay{D}{\KL\Kp\Km}$} \\
         & Signal yield & $\epsilon$ (\%) & Signal yield & $\epsilon$ (\%) \\
        \hline
        $K\pi$                    & $\phantom{0}963\pm31$           & $13.22\pm0.07$           & $1594\pm42$                     & $15.15\pm0.07$ \\
        $K\pi\piz$                & $1946\pm45$                     & $\phantom{0}7.43\pm0.05$ & $3284\pm64$                     & $\phantom{0}8.69\pm0.06$  \\
        $K\pi\pi\pi$            & $1166\pm35$                     & $\phantom{0}7.74\pm0.05$ & $1949\pm48$                     & $\phantom{0}8.61\pm0.06$ \\
        $Ke\neue$                   & $\phantom{0}684\pm26$           & $10.46\pm0.04$           & --- & --- \\
        $\pi\pi\piz$                        & $\phantom{0}160\pm13$           & $\phantom{0}6.92\pm0.05$ & $\phantom{0}381\pm27$           & $\phantom{0}9.99\pm0.05$ \\
        $\pi\pi$                            & $\phantom{00}29\pm\phantom{0}5$ & $11.78\pm0.07$           & $\phantom{00}71\pm\phantom{0}9$ & $16.61\pm0.07$ \\
        $KK$                              & $\phantom{00}86\pm\phantom{0}9$ & $11.51\pm0.06$           & $\phantom{0}222\pm15$           & $16.28\pm0.07$ \\
        $\KS\piz\piz$                         & $\phantom{00}26\pm\phantom{0}5$ & $\phantom{0}2.82\pm0.03$ & $\phantom{00}79\pm10$           & $\phantom{0}4.33\pm0.04$ \\
        $\KL\piz$                             & $\phantom{00}94\pm11$           & $\phantom{0}5.91\pm0.05$ & --- & --- \\
        $\KS\piz$                             & $\phantom{0}128\pm11$           & $\phantom{0}8.10\pm0.05$ & $\phantom{0}147\pm13$           & $\phantom{0}7.04\pm0.06$ \\
        $\KS\eta(\to\gamma\gamma)$            & $\phantom{00}17\pm\phantom{0}4$ & $\phantom{0}6.66\pm0.05$ & $\phantom{00}20\pm\phantom{0}5$ & $\phantom{0}5.98\pm0.05$ \\
        $\KS\eta^{\prime}(\to\pi\pi\eta)$   & $\phantom{000}6\pm\phantom{0}2$ & $\phantom{0}2.71\pm0.03$ & $\phantom{000}9\pm\phantom{0}3$ & $\phantom{0}2.59\pm0.03$ \\
        $\KS\eta(\to\pi\pi\piz)$            & $\phantom{000}5\pm\phantom{0}2$ & $\phantom{0}3.67\pm0.04$ & $\phantom{000}7\pm\phantom{0}3$ & $\phantom{0}3.57\pm0.04$ \\
        $\KS\eta^{\prime}(\to\pi\pi\gamma)$ & $\phantom{00}15\pm\phantom{0}4$ & $\phantom{0}4.00\pm0.04$ & $\phantom{00}19\pm\phantom{0}5$ & $\phantom{0}3.92\pm0.04$  \\
        $\KS\omega(\to\pi\pi\piz)$          & $\phantom{00}62\pm\phantom{0}8$ & $\phantom{0}3.35\pm0.03$ & $\phantom{00}55\pm\phantom{0}8$ & $\phantom{0}3.30\pm0.04$ \\
        $\KL\piz\piz$                         & $\phantom{00}49\pm\phantom{0}7$ & $\phantom{0}3.65\pm0.03$ & --- & --- \\
        $\KS\Kp\Km$                           & $\phantom{00}35\pm\phantom{0}7$ & $12.72\pm0.03$           & $\phantom{00}40\pm\phantom{0}8$ & $\phantom{0}4.86\pm0.02$ \\
        $\KS\pip\pim$                         & $\phantom{0}564\pm27$           & $16.62\pm0.03$           & $\phantom{0}512\pm24$           & $\phantom{0}8.53\pm0.02$ \\
        $\KL\pip\pim$                         & $\phantom{0}562\pm29$           & $\phantom{0}9.61\pm0.03$ & --- & --- \\
        \hline
    \end{tabular}
    \end{center}
    \label{tab:datafit_yields_k0kk}
\end{table}

\clearpage

\section{Alternative fits to strong-phase parameters}
\label{app:strong_phase_alt}

Table~\ref{tab:strong_phase_res_kspp} lists the strong-phase parameters from fits to \besiii and \lhcb data or only to \besiii data.
Data from \lhcb are found to slightly improve the sensitivity to $S_n$ parameters.

\begin{table}[!b]
    \centering
    \caption{Strong-phase parameters of the $\decay{D}{\KSpp}$ decay with different setups under $M_{\pi} = 2$ and $M_K = 1$. 
    The uncertainties are statistical only.}
    \label{tab:strong_phase_res_kspp}
    \begin{tabular}{c|c|rr}
        \hline
        Parameter & Weight & \multicolumn{1}{c}{\besiii and \lhcb} & \multicolumn{1}{c}{\besiii only} \\
        \hline
        $C_0$ & $w^{\rm opt}$            &  $0.0701 \pm 0.0038$   & $0.0691 \pm 0.0038$  \\ 
        $C_1$ & $w^{\rm opt}\sin{\phi}$  &  $0.0102 \pm 0.0023$   & $0.0102 \pm 0.0023$  \\ 
        $C_2$ & $w^{\rm opt}\cos{\phi}$  &  $0.1729 \pm 0.0028$   & $0.1726 \pm 0.0028$  \\ 
        $C_3$ & $w^{\rm opt}\sin{2\phi}$ &  $0.0067 \pm 0.0021$   & $0.0064 \pm 0.0022$  \\ 
        $C_4$ & $w^{\rm opt}\cos{2\phi}$ &  $0.0437 \pm 0.0031$   & $0.0439 \pm 0.0032$  \\ 
        $S_0$ & $w^{\rm opt}$            & $-0.0213 \pm 0.0085$   & $-0.0185 \pm 0.0088$ \\
        $S_1$ & $w^{\rm opt}\sin{\phi}$  &  $0.0933 \pm 0.0052$   & $0.0939 \pm 0.0052$  \\
        $S_2$ & $w^{\rm opt}\cos{\phi}$  &  $0.0073 \pm 0.0067$   & $0.0071 \pm 0.0069$  \\
        $S_3$ & $w^{\rm opt}\sin{2\phi}$ &  $0.0198 \pm 0.0064$   & $0.0214 \pm 0.0067$  \\
        $S_4$ & $w^{\rm opt}\cos{2\phi}$ &  $0.0129 \pm 0.0069$   & $0.0116 \pm 0.0072$  \\
        \hline
    \end{tabular}
\end{table}

\section{Correlation matrices}
\label{app:correlations}

The statistical uncertainties and correlation matrices of the \CP observables from \lhcb are given in Table~\ref{tab:corr_cpv_lhcb_stat}, while those from \besiii are given in Table \ref{tab:corr_cpv_bes_stat}. The corresponding information pertaining to systematic uncertainties are then given in Tables~\ref{tab:corr_cpv_lhcb_syst}--\ref{tab:corr_cpv_bes_syst}.

\begin{table}[!b]
    \caption{Statistical uncertainties and correlation matrix of the \CP observables from \lhcb.}
    \label{tab:corr_cpv_lhcb_stat}
    {
    \renewcommand{\arraystretch}{1.1}
    \begin{center}
        \begin{tabular}{l|rrrrrr}
            \hline
            \multicolumn{7}{c}{Uncertainty~($10^{-2}$)} \\
            \hline
             & $\xmdk$ & $\ymdk$ & $\xpdk$ & $\ypdk$ & $\xxi$ & $\yxi$ \\
            \hline
            $\sigma$ & $0.96$ & $1.34$ & $1.01$ & $1.36$ & $2.06$ & $2.30$ \\
            \hline\hline
            \multicolumn{7}{c}{Correlation matrix} \\
            \hline
             & $\xmdk$ & $\ymdk$ & $\xpdk$ & $\ypdk$ & $\xxi$ & $\yxi$ \\
            \hline
            $\xmdk$ & $ 1.000$ & $-0.083$ & $-0.028$ & $ 0.021$ & $ 0.111$ & $-0.124$ \\
            $\ymdk$ &          & $ 1.000$ & $-0.018$ & $-0.041$ & $ 0.100$ & $ 0.192$ \\
            $\xpdk$ &          &          & $ 1.000$ & $ 0.227$ & $-0.240$ & $-0.046$ \\
            $\ypdk$ &          &          &          & $ 1.000$ & $-0.115$ & $-0.231$ \\
             $\xxi$ &          &          &          &          & $ 1.000$ & $ 0.302$ \\
             $\yxi$ &          &          &          &          &          & $ 1.000$ \\
            \hline
        \end{tabular}
    \end{center}
    }
\end{table}

\begin{table}[H]
    \caption{Statistical uncertainties and correlation matrix of the \CP observables from \besiii.}
    \label{tab:corr_cpv_bes_stat}
    {
    \renewcommand{\arraystretch}{1.1}
    \begin{center}
        \begin{tabular}{l|rrrrrr}
            \hline
            \multicolumn{7}{c}{Uncertainty~($10^{-2}$)} \\
            \hline
             & $\xmdk$ & $\ymdk$ & $\xpdk$ & $\ypdk$ & $\xxi$ & $\yxi$ \\
            \hline
            $\sigma$ & $0.20$ & $0.37$ & $0.15$ & $0.34$ & $0.38$ & $0.57$ \\
            \hline\hline
            \multicolumn{7}{c}{Correlation matrix} \\
            \hline
             & $\xmdk$ & $\ymdk$ & $\xpdk$ & $\ypdk$ & $\xxi$ & $\yxi$ \\
            \hline
            $\xmdk$ & $ 1.000$ & $ 0.147$ & $-0.572$ & $ 0.530$ & $-0.258$ & $-0.155$ \\
            $\ymdk$ &          & $ 1.000$ & $-0.022$ & $-0.234$ & $-0.213$ & $-0.043$ \\
            $\xpdk$ &          &          & $ 1.000$ & $-0.013$ & $-0.096$ & $-0.186$ \\
            $\ypdk$ &          &          &          & $ 1.000$ & $-0.307$ & $-0.395$ \\
             $\xxi$ &          &          &          &          & $ 1.000$ & $ 0.954$ \\
             $\yxi$ &          &          &          &          &          & $ 1.000$ \\
            \hline
        \end{tabular}
    \end{center}
    }
\end{table}

\begin{table}[H]
    \caption{Systematic uncertainties and correlation matrix of the \CP observables from \lhcb.}
    \label{tab:corr_cpv_lhcb_syst}
    {
    \renewcommand{\arraystretch}{1.1}
    \begin{center}
        \begin{tabular}{l|rrrrrr}
            \hline
            \multicolumn{7}{c}{Uncertainty~($10^{-2}$)} \\
            \hline
             & $\xmdk$ & $\ymdk$ & $\xpdk$ & $\ypdk$ & $\xxi$ & $\yxi$ \\
            \hline
            $\sigma$ & $0.18$ & $0.23$ & $0.24$ & $0.31$ & $0.30$ & $0.20$ \\
            \hline\hline
            \multicolumn{7}{c}{Correlation matrix} \\
            \hline
             & $\xmdk$ & $\ymdk$ & $\xpdk$ & $\ypdk$ & $\xxi$ & $\yxi$ \\
            \hline
            $\xmdk$ & $ 1.000$ & $ 0.322$ & $ 0.200$ & $ 0.167$ & $ 0.000$ & $ 0.277$ \\
            $\ymdk$ &          & $ 1.000$ & $-0.361$ & $-0.358$ & $-0.320$ & $ 0.283$ \\
            $\xpdk$ &          &          & $ 1.000$ & $ 0.423$ & $ 0.163$ & $-0.415$ \\
            $\ypdk$ &          &          &          & $ 1.000$ & $ 0.223$ & $ 0.008$ \\
             $\xxi$ &          &          &          &          & $ 1.000$ & $ 0.298$ \\
             $\yxi$ &          &          &          &          &          & $ 1.000$ \\
            \hline
        \end{tabular}
    \end{center}
    }
\end{table}

\begin{table}[H]
    \caption{Systematic uncertainties and correlation matrix of the \CP observables from \besiii.}
    \label{tab:corr_cpv_bes_syst}
    {
    \renewcommand{\arraystretch}{1.1}
    \begin{center}
        \begin{tabular}{l|rrrrrr}
            \hline
            \multicolumn{7}{c}{Uncertainty~($10^{-2}$)} \\
            \hline
             & $\xmdk$ & $\ymdk$ & $\xpdk$ & $\ypdk$ & $\xxi$ & $\yxi$ \\
            \hline
            $\sigma$ & $0.07$ & $0.09$ & $0.08$ & $0.06$ & $0.01$ & $0.02$ \\
            \hline\hline
            \multicolumn{7}{c}{Correlation matrix} \\
            \hline
             & $\xmdk$ & $\ymdk$ & $\xpdk$ & $\ypdk$ & $\xxi$ & $\yxi$ \\
            \hline
            $\xmdk$ & $ 1.000$ & $-0.008$ & $-0.701$ & $-0.288$ & $ 0.031$ & $-0.563$ \\
            $\ymdk$ &          & $ 1.000$ & $ 0.191$ & $-0.520$ & $ 0.528$ & $-0.235$ \\
            $\xpdk$ &          &          & $ 1.000$ & $ 0.321$ & $ 0.298$ & $ 0.427$ \\
            $\ypdk$ &          &          &          & $ 1.000$ & $ 0.058$ & $ 0.721$ \\
             $\xxi$ &          &          &          &          & $ 1.000$ & $ 0.073$ \\
             $\yxi$ &          &          &          &          &          & $ 1.000$ \\
            \hline
        \end{tabular}
    \end{center}
    }
\end{table}
\clearpage

\addcontentsline{toc}{section}{References}
\bibliographystyle{LHCb/LHCb}
\bibliography{main,LHCb/standard,LHCb/LHCb-PAPER,LHCb/LHCb-CONF,LHCb/LHCb-DP,LHCb/LHCb-TDR}

\newpage
\centerline
{\large\bf BESIII collaboration}
\vspace{0.2cm}
\small
\noindent M.~Ablikim$^{1}$\BESIIIorcid{0000-0002-3935-619X},
M.~N.~Achasov$^{4,d}$\BESIIIorcid{0000-0002-9400-8622},
P.~Adlarson$^{83}$\BESIIIorcid{0000-0001-6280-3851},
X.~C.~Ai$^{89}$\BESIIIorcid{0000-0003-3856-2415},
C.~S.~Akondi$^{31A,31B}$\BESIIIorcid{0000-0001-6303-5217},
R.~Aliberti$^{39}$\BESIIIorcid{0000-0003-3500-4012},
A.~Amoroso$^{82A,82C}$\BESIIIorcid{0000-0002-3095-8610},
Q.~An$^{79,65,\dagger}$,
Y.~H.~An$^{89}$\BESIIIorcid{0009-0008-3419-0849},
Y.~Bai$^{63}$\BESIIIorcid{0000-0001-6593-5665}
% ...

\vspace{0.2cm} {\footnotesize\it
\noindent $^{1}$ Institute of High Energy Physics, Beijing 100049, People's Republic of China\\
% ...
$^{4}$ Budker Institute of Nuclear Physics SB RAS (BINP), Novosibirsk 630090, Russia\\
% ...
$^{31}$ INFN Sezione di Ferrara, (A)INFN Sezione di Ferrara, I-44122, Ferrara, Italy; (B)University of Ferrara, I-44122, Ferrara, Italy\\
% ...
$^{39}$ Johannes Gutenberg University of Mainz, Johann-Joachim-Becher-Weg 45, D-55099 Mainz, Germany\\
% ...
$^{63}$ Southeast University, Nanjing 211100, People's Republic of China\\
% ...
$^{65}$ State Key Laboratory of Particle Detection and Electronics, Beijing 100049, Hefei 230026, People's Republic of China\\
% ...
$^{79}$ University of Science and Technology of China, Hefei 230026, People's Republic of China\\
% ...
$^{82}$ University of Turin and INFN, (A)University of Turin, I-10125, Turin, Italy; (B)University of Eastern Piedmont, I-15121, Alessandria, Italy; (C)INFN, I-10125, Turin, Italy\\
$^{83}$ Uppsala University, Box 516, SE-75120 Uppsala, Sweden\\
% ...
$^{89}$ Zhengzhou University, Zhengzhou 450001, People's Republic of China\\

\vspace{0.2cm}
\noindent $^{\dagger}$ Deceased\\
% ...
$^{d}$ Also at the Novosibirsk State University, Novosibirsk, 630090, Russia\\
% ...

}

\centerline
{\large\bf LHCb collaboration}
\begin
{flushleft}
\small
R.~Aaij$^{38}$\lhcborcid{0000-0003-0533-1952},
A.S.W.~Abdelmotteleb$^{58}$\lhcborcid{0000-0001-7905-0542},
C.~Abellan~Beteta$^{52}$\lhcborcid{0009-0009-0869-6798},
F.~Abudin\'en$^{60}$\lhcborcid{0000-0002-6737-3528},
T.~Ackernley$^{62}$\lhcborcid{0000-0002-5951-3498},
A. A. ~Adefisoye$^{70}$\lhcborcid{0000-0003-2448-1550},
B.~Adeva$^{48}$\lhcborcid{0000-0001-9756-3712},
M.~Adinolfi$^{56}$\lhcborcid{0000-0002-1326-1264},
P.~Adlarson$^{88}$\lhcborcid{0000-0001-6280-3851},
C.~Agapopoulou$^{14}$\lhcborcid{0000-0002-2368-0147}
% ...

{\footnotesize \it

$^{1}$School of Physics and Astronomy, Monash University, Melbourne, Australia\\
% ...
$^{4}$Department of Engineering Physics, Tsinghua University, Beijing, China\\
$^{5}$Institute Of High Energy Physics (IHEP), Beijing, China\\
% ...
$^{14}$Universit{\'e} Paris-Saclay, CNRS/IN2P3, IJCLab, Orsay, France\\
% ...
$^{20}$Physikalisches Institut, Albert-Ludwigs-Universit{\"a}t Freiburg, Freiburg, Germany\\
% ...
$^{31}$INFN Sezione di Milano-Bicocca, Milano, Italy\\
% ...
$^{38}$Nikhef National Institute for Subatomic Physics, Amsterdam, Netherlands\\
$^{39}$Nikhef National Institute for Subatomic Physics and VU University Amsterdam, Amsterdam, Netherlands\\
% ...
$^{48}$Instituto Galego de F{\'\i}sica de Altas Enerx{\'\i}as (IGFAE), Universidade de Santiago de Compostela, Santiago de Compostela, Spain\\
% ...
$^{52}$Physik-Institut, Universit{\"a}t Z{\"u}rich, Z{\"u}rich, Switzerland\\
% ...
$^{56}$H.H. Wills Physics Laboratory, University of Bristol, Bristol, United Kingdom\\
% ...
$^{58}$Department of Physics, University of Warwick, Coventry, United Kingdom\\
% ...
$^{60}$School of Physics and Astronomy, University of Edinburgh, Edinburgh, United Kingdom\\
$^{61}$School of Physics and Astronomy, University of Glasgow, Glasgow, United Kingdom\\
$^{62}$Oliver Lodge Laboratory, University of Liverpool, Liverpool, United Kingdom\\
$^{63}$Imperial College London, London, United Kingdom\\
$^{64}$Department of Physics and Astronomy, University of Manchester, Manchester, United Kingdom\\
$^{65}$Department of Physics, University of Oxford, Oxford, United Kingdom\\
% ...
$^{70}$Syracuse University, Syracuse, NY, United States\\
% ...
$^{79}$Institute of Physics of  the Czech Academy of Sciences, Prague, Czech Republic, associated to $^{64}$\\
% ...
$^{82}$Faculty of Physics, Vilnius University, Vilnius, Lithuania, associated to $^{20}$\\
$^{83}$Institute of Physics and Technology, Ulan Bator, Mongolia, associated to $^{5}$\\
% ...
$^{88}$Department of Physics and Astronomy, Uppsala University, Uppsala, Sweden, associated to $^{61}$\\
$^{89}$Taras Schevchenko University of Kyiv, Faculty of Physics, Kyiv, Ukraine, associated to $^{14}$\\
% ...
$^{d}$Hangzhou Institute for Advanced Study, UCAS, Hangzhou, China\\
% ...
$ ^{\dagger}$Deceased
}
\end{flushleft}

\end{document}